\documentclass[examplefnt]{nowfnt} 
\usepackage[utf8]{inputenc}
\usepackage{amsmath,amsfonts,bm, amssymb, fixmath,dsfont}

\def\circlen#1{%
        \raisebox{.9pt}{\textcircled{\raisebox{-.9pt}{#1}}}%
}

\def\eqref#1{equation~\ref{#1}}

\def\1{\bm{1}}

\DeclareMathAlphabet{\mathsfit}{\encodingdefault}{\sfdefault}{m}{sl}
\SetMathAlphabet{\mathsfit}{bold}{\encodingdefault}{\sfdefault}{bx}{n}
\newcommand{\tens}[1]{\bm{\mathsfit{#1}}}

\def\tM{{\tens{M}}}

\def\vpi{{\boldsymbol{\pi}}}
\def\vpi{{\boldsymbol{\pi}}}

\definecolor{CeruleanRef}{RGB}{12,127,172}

\DeclareUnicodeCharacter{0308}{HERE!HERE!} 
\DeclareUnicodeCharacter{1EF3}{HERE!HERE!}
\usepackage[english,french]{babel} 
\usepackage{csquotes} 
\usepackage{url}
\usepackage{longtable}
\usepackage{tabularx}
\usepackage{multirow}
\usepackage[format=hang,font=small,labelfont=bf]{caption}
\usepackage{etoolbox}
\usepackage{graphicx}
\usepackage{booktabs} 
\usepackage{siunitx}
\usepackage{array}
\usepackage{subfigure}
\usepackage{enumitem,balance}
\usepackage{algorithm,algpseudocode}
\usepackage{tabulary}
\usepackage{tabularx}
\usepackage{nicefrac}
\usepackage{xcolor}
\usepackage{algorithm}
\usepackage[T1]{fontenc}
\usepackage{lmodern}
\DeclareUnicodeCharacter{00A0}{~}
\usepackage{amsmath}
\usepackage{newtxtext}

\usepackage{todonotes}
\newtheorem{assumption}{Assumption}

\usepackage{pbox}

\newcolumntype{P}[1]{>{\centering\arraybackslash}p{#1}}
\DeclareUnicodeCharacter{0301}{*************************************}

\title{Game-Theoretic Multiagent Reinforcement Learning}

\maintitleauthorlist{
Yaodong Yang \\
Peking University, China \\
yaodong.yang@pku.edu.cn
\and
Chengdong Ma \\
Peking University, China \\
chengdong.ma@stu.pku.edu.cn
\and
Zihan Ding \\
Princeton University, USA \\
zihand@princeton.edu
\and
Stephen McAleer \\
Carnegie Mellon University, USA \\
smcaleer@cs.cmu.edu
\and
Chi Jin \\
Princeton University, USA \\
chij@princeton.edu
\and
Jun Wang \\
University College London, UK \\
jun.wang@cs.ucl.ac.uk
\and
Tuomas Sandholm \\
Carnegie Mellon University, USA \\
sandholm@cs.cmu.edu
}

\issuesetup
{%
 copyrightowner={Yaodong Yang et al. * Equal contribution. $\dag$ Corresponding author.},
 volume        = xx,
 issue         = xx,
 pubyear       = 2025,
 isbn          = xxx-x-xxxxx-xxx-x,
 eisbn         = xxx-x-xxxxx-xxx-x,
 doi           = 10.1561/XXXXXXXXX,
 firstpage     = 1, 
 lastpage      = 18,
 }

\usepackage{mwe}

\author[]{Yang $^{1, *, \dag}$,Yaodong}
\author[]{Ma $^{2, *}$,Chengdong}
\author[3]{Ding,Zihan}
\author[4]{McAleer,Stephen}
\author[5]{Jin,Chi}
\author[6]{Wang,Jun}
\author[7]{Sandholm,Tuomas}

\affil[1]{Institute for Artificial Intelligence, Peking University, Beijing, China, yaodong.yang@pku.edu.cn}
\affil[2]{Institute for Artificial Intelligence, Peking University, Beijing, China, chengdong.ma@stu.pku.edu.cn}
\affil[3]{Department of Electrical and Computer Engineering, Princeton University, Princeton, NJ, USA, zihand@princeton.edu}
\affil[4]{Carnegie Mellon University, PA, USA, smcaleer@cs.cmu.edu}
\affil[5]{Department of Electrical and Computer Engineering, Princeton University, Princeton, NJ, USA, chij@princeton.edu}
\affil[6]{UCL Centre for Artificial Intelligence, University College London, UK, jun.wang@cs.ucl.ac.uk}
\affil[7]{Carnegie Mellon University, PA, USA, sandholm@cs.cmu.edu}

\articledatabox{\nowfntstandardcitation}

\begin{document}
\selectlanguage{english}
\makeabstracttitle

\begin{abstract}
Tremendous advances have been made in \textit{multiagent reinforcement learning (MARL)}. MARL corresponds to the learning problem in a multiagent system in which multiple agents learn simultaneously. It is an interdisciplinary field of study with a long history that includes game theory, machine learning, stochastic control, psychology, and optimization. Despite great successes in MARL, there is a lack of a self-contained overview of the literature that covers game-theoretic foundations of modern MARL methods and summarizes the recent advances. The majority of existing surveys are outdated and do not fully cover the recent developments since 2010. In this work, we provide a monograph on MARL that covers both the fundamentals and the latest developments on the research frontier. The goal of this monograph is to provide a self-contained assessment of the current state-of-the-art MARL techniques from a game-theoretic perspective. We expect this work to serve as a stepping stone for both new researchers who are about to enter this fast-growing field and experts in the field who want to obtain a panoramic view and identify new directions based on recent advances.
\end{abstract}

\definecolor{amber}{rgb}{0.81, 0.71, 0.23}
\definecolor{turq}{rgb}{0.28, 0.82, 0.8}

\chapter{Introduction}
\label{sec:intro}

\section{Deep Reinforcement Learning}
Deep Reinforcement Learning (RL) is a subfield of artificial intelligence that combines deep learning\cite{lecun2015deep} and reinforcement learning\cite{sutton1998reinforcement}, enabling machines to make decisions and solve problems. It has seen significant advancements in recent years and has been applied to various domains including video games\cite{mnih2015human, silver2016mastering,silver2017mastering, silver2018general}, robotics\cite{kober2013reinforcement, akkaya2019solving}, finance\cite{deng2016deep, liu2018practical}, energy\cite{degrave2022magnetic}, transportation\cite{wei2018intellilight, haydari2020deep}, etc. In RL, an agent learns to maximize the expected cumulative reward over time by interacting with an environment and receiving feedback in the form of rewards.

\paragraph{A Car Driving Example.} To illustrate the key components of decision-making process in RL, let us consider the real-world example of controlling a car to drive safely through an intersection.
At each time step, a robot car can move by steering, accelerating, and braking. 
The goal is to safely exit the intersection and reach the destination (with possible decisions of going straight or turning left/right into another lane). 
Therefore,  in addition to being able to detect objects, such as traffic lights, lane markings, and other cars (by converting data to knowledge), we aim to find a steering policy that can control the car to make a sequence of maneuvers to achieve the goal (making decisions based on the knowledge gained). 
In a decision-making setting such as this, two additional challenges arise: 

 \begin{figure}[t]
\centering
\includegraphics[width=.75\linewidth]{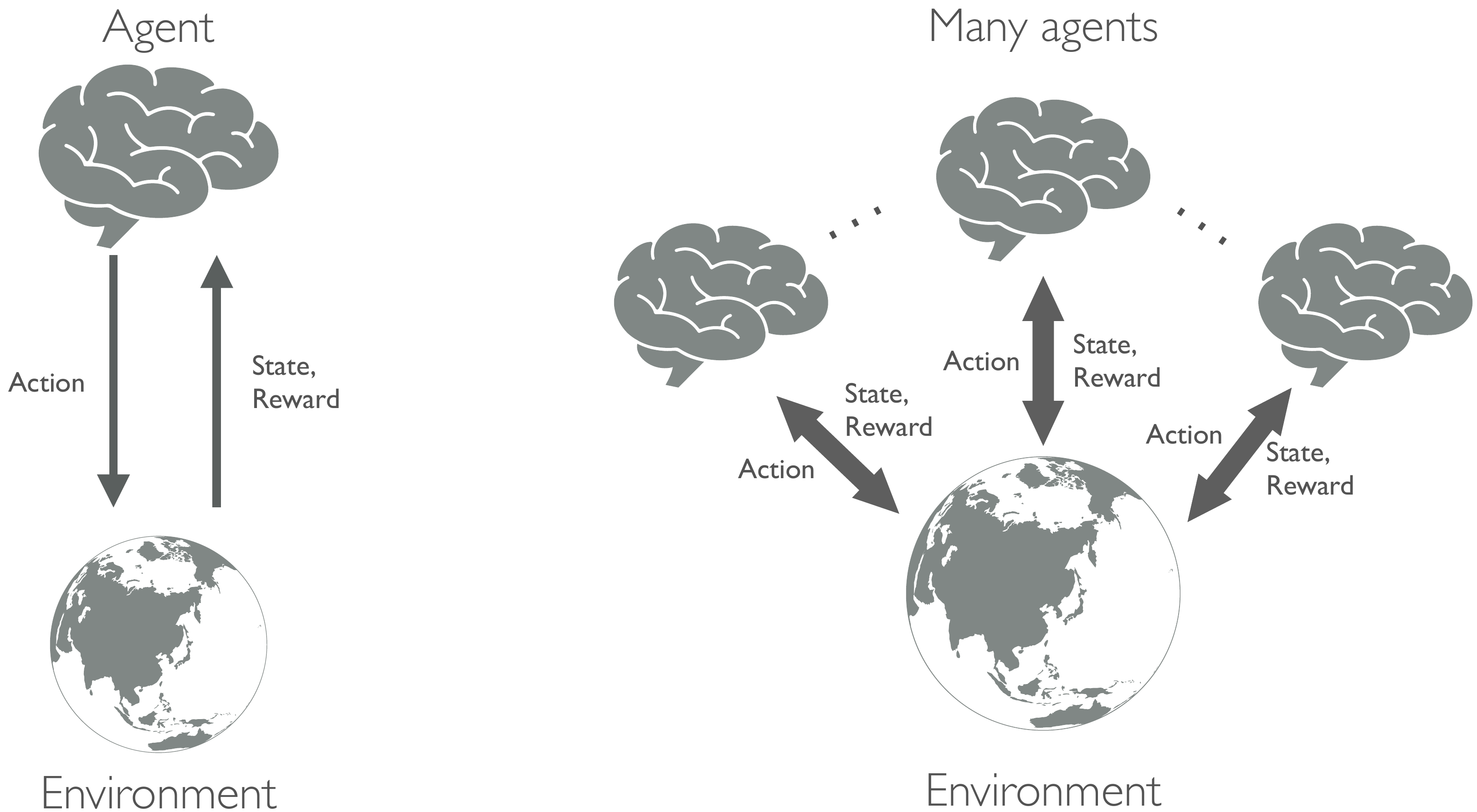}
\caption{Diagram of a single-agent MDP (left) and a multiagent MDP  (right). }
\label{fig:single_multi}
\end{figure}

\begin{enumerate}
\item First, during the decision-making process, at each time step, the robot car should consider not only the immediate value of its current action but also the consequences of its current action in the future.
For example, in the case of driving through an intersection, it would be detrimental to have a policy that chooses to steer in a ``safe'' direction at the beginning of the process if it would eventually lead to a car crash later on.  
 
\item 
Second, to make each decision correctly and safely, the car must also consider other cars' behavior and act accordingly.  Human drivers, for example, often predict in advance other cars' movements and then take strategic moves in response (like giving way to an oncoming car or accelerating to merge into another lane). 
\end{enumerate}

The need for an adaptive decision-making framework, together with the complexity of addressing multiple interacting learners, has led to the development of multiagent RL (Fig.~\ref{fig:single_multi}), which will be discussed later~\ref{sec:intro_marl}.

\paragraph{A Short History of RL.}

RL is a sub-field of machine learning, where agents learn how to behave optimally based on a trial-and-error procedure during their interaction with the environment. 
Unlike supervised learning, which takes labeled data as the input (for example, an image labeled with cats), RL is goal-oriented: it constructs a learning model that learns to achieve the optimal long-term goal by improvement through trial and error, with the learner having no labeled data to obtain knowledge from. 
The word ``reinforcement'' refers to the learning mechanism since the actions that lead to satisfactory outcomes are reinforced in the learner's set of behaviours.  

Historically, the RL mechanism was originally developed based on studying cats' behaviour in a puzzle box \cite{thorndike1898animal}.  
\cite{minsky1954theory} first proposed the computational model of RL in his Ph.D. thesis and named his resulting analog machine the \emph{stochastic neural-analog reinforcement calculator}. 
Several years later, he first suggested the connection between dynamic programming  \cite{bellman1952theory} and RL \cite{minsky1961steps}. 
In 1972, \cite{klopf1972brain} integrated the trial-and-error learning process with the finding of \emph{temporal difference (TD)} learning from psychology.  
TD learning quickly became indispensable in scaling RL for larger systems.  
Based on dynamic programming and TD learning, \cite{watkins1992q} laid the foundations for present-day RL using the Markov decision process (MDP) and proposing the famous Q-learning method as the solver. 
As a dynamic programming method, the original Q-learning process inherits Bellman's ``curse of dimensionality'' \cite{bellman1952theory}, which strongly limits its applications when the number of state variables is large.   
To overcome such a bottleneck, \cite{bertsekas1996neuro} proposed approximate dynamic programming methods based on neural networks. 
More recently, \cite{mnih2015human} from DeepMind made a significant breakthrough by introducing the deep Q-learning (DQN) architecture, which leverages the representation power of DNNs for approximate dynamic programming methods. 
DQN has demonstrated human-level performance on $49$ Atari games.
Since then, deep RL techniques have become common in machine learning/AI and have attracted considerable attention from the research community.

RL originates from an understanding of animal behaviour where animals use trial-and-error to reinforce beneficial behaviours, which they then perform more frequently. During its development, computational RL incorporated ideas such as optimal control theory and other findings from psychology that help mimic the way humans make decisions to maximise the long-term profit of decision-making tasks.
As a result, RL methods can naturally be used to train a computer program (an agent) to a performance level comparable to that of a human on certain tasks.
The earliest success of RL methods against human players can be traced back to the game of backgammon \cite{tesauro1995temporal}. 
More recently, the advancement of applying RL to solve sequential decision-making problems was marked by the remarkable success of the AlphaGo series \cite{silver2016mastering,silver2017mastering, silver2018general}, a self-taught RL agent that beats top professional players of the game Go, a game whose search space ($10^{761}$ possible games) is even greater than the number of atoms in the universe\footnote{There are an estimated $10^{82}$ atoms in the universe. If one had one trillion computers, each processing one trillion states per second for one trillion years, one could only reach $10^{43}$ states.}.

In fact, the majority of successful RL applications, such as those for the game Go\footnote{AlphaGo can also be treated as a multiagent technique if we consider the opponent in self-play as another agent.} and Stratego\cite{perolat2022mastering}, robotic control \cite{kober2013reinforcement}, and autonomous driving \cite{shalev2016safe}, naturally involve the participation of multiple AI agents, which probe into the realm of multiagent RL. 

\section{Multiagent Reinforcement Learning}
\label{sec:intro_marl}

\paragraph{A Short History of MARL.}
The development of MARL from the 1980s to the early 2000s laid the foundation for modern methodologies, primarily driven by the extension of RL techniques to multi-agent settings. Early works were strongly rooted in game theory, particularly stochastic games, which extended Markov decision processes (MDPs) to settings involving multiple interacting agents. This shift established the groundwork for MARL by introducing key concepts such as equilibrium-based strategies, coordination, and scalability. A significant milestone in this period was the introduction of Minimax Q-learning by Littman \cite{littman1994markov}, which formalized the application of Q-learning to zero-sum games, offering a way to model competitive interactions between agents. Hu and Wellman \cite{hu1998multiagent,hu2003nash} expanded this framework to Nash Q-learning, a general-sum extension that accounted for both competitive and cooperative agents by incorporating Nash equilibria as the solution concept. This approach, however, faced challenges, particularly with convergence in environments with multiple equilibria, which necessitated the development of advanced solution techniques. In parallel, joint-action learning frameworks like those proposed by Claus and Boutilier \cite{claus1998dynamics} focused on cooperation in multi-agent environments, advancing methods such as joint-action value functions and optimistic exploration for coordination in shared-reward settings. These frameworks addressed the need for agents to effectively collaborate while managing the complexities of dynamic environments. The period also saw key algorithmic advancements aimed at tackling the distinct challenges of non-stationarity and scalability. The problem of non-stationarity, where the environment continually changes due to simultaneous policy learning by multiple agents, was addressed by methods such as WoLF (Win or Learn Fast) Policy Hillclimbing by Bowling and Veloso \cite{bowling2001rational}, which adapted learning rates to balance rationality and convergence in mixed cooperative-competitive environments. Additionally, strategies like Brownian bandits and Replicator Dynamics \cite{tuyls2003selection} were introduced to explore dynamic multi-agent interactions and their effects on learning in non-stationary environments. Scalability remained a pressing challenge, as the joint-action space in MARL grew exponentially with the number of agents, making the use of full state-action tables impractical. Sparse Q-learning provided a partial solution by reducing the complexity of value function representations \cite{kok2004sparse}.

Despite these advances, several core challenges persisted. Non-stationarity remained a significant hurdle due to the continuously changing policies of other agents, often requiring adaptations such as opponent modeling \cite{hu1999learning} and recency-based exploration \cite{ballard2002overcoming}. Moreover, the presence of multiple Nash equilibria in general-sum games further complicated convergence, requiring sophisticated coordination strategies among agents to select appropriate equilibria. Key works from this era, such as Minimax Q-learning (Littman, 1994) \cite{littman1994markov}, Nash Q-learning \cite{hu1998multiagent,hu2003nash}, and joint-action RL \cite{claus1998dynamics}, played pivotal roles in shaping the field. These foundational contributions provided the theoretical groundwork for MARL, highlighting its potential in both competitive and cooperative scenarios, while also identifying the persistent challenges that would drive subsequent research. Early empirical evaluations, conducted in controlled environments such as grid-world, predator-prey scenarios, and cooperative tasks, demonstrated the practical strengths of these techniques but also revealed the theoretical gaps in convergence and scalability.
These foundational works paved the way for future advancements, as they not only contributed to the theoretical development of MARL but also highlighted persistent issues that continue to shape the direction of research in the field.

Multiagent Reinforcement Learning (MARL) extends the concept of single-agent RL to situations where multiple agents are present and interact with each other and their environment. In these scenarios, each agent must take into account the actions of other agents when making decisions, making MARL a challenging area of research. The goal in MARL is for each agent to learn its own policy while adapting to the changing policies of other agents. This requires the development of algorithms that can handle the complexity and non-stationarity of multiagent systems. MARL has been applied in various fields such as autonomous robots, networked systems, and cyber-physical systems, and has seen successful results in cooperative navigation, formation control, traffic control, and energy management.

When designing MARL algorithms, it is critical to consider equilibria, which are the outcomes of interactions between agents that are stable, meaning no single agent has the incentive to change its behaviour. The mathematical concept of equilibria traces back to the origin of \textit{game theory}, which is widely considered to start from the founding book ``\textit{Theory of Games and Economic behaviour}'' in 1944, by a mathematician and physicist named John von Neumann and economist Oskar Morgenstern. Their approach to game theory involved defining games as mathematical models consisting of players, actions, and payoffs. The concept of mixed strategies was introduced, which allows players to randomize their actions, and the \textit{minimax theorem}, shows that in any two-player zero-sum game, there is always a strategy that minimizes the maximum possible loss. Another famous concept, \textit{Nash equilibrium}, was introduced by John Nash in his 1950 paper ``\textit{Non-Cooperative Games}'', for which he was later awarded the Nobel Prize in Economics. Nash developed the concept as a generalization of the classical equilibrium concept used in economics, which is based on the assumption of perfect competition and rational behaviour. A Nash equilibrium is a set of strategies where no player can improve their outcome by changing their strategy, assuming that the other players' strategies remain the same. It serves as a fundamental concept in game theory and is widely used to analyze a wide range of multiagent interaction scenarios, boosting the development of theoretical game theory.
Notably, in some cases, the focus of study shifts from player strategies to programs. For example, one approach to achieving cooperation in the one-shot Prisoner's Dilemma is program equilibrium \cite{tennenholtz2004program}, where players submit programs instead of strategies. These programs are then allowed to read each other's source code in order to decide on their actions.
As a mathematical framework for analyzing interactions between agents, game theory provides a solution to this challenge by allowing us to develop MARL algorithms with theoretical equilibrium properties and convergence guarantees. For example, game theoretic algorithms for two-player zero-sum games have provable convergence guarantees and have been shown to achieve astonishing practical achievements, like beating top human players in various games such as Go\cite{silver_2017}, poker\cite{brown2018superhuman}, and Stratego\cite{perolat2022mastering}. Multiplayer video games such as StarCraft II\cite{alphastar} and Dota2\cite{dota} have also been tackled by MARL. 

\paragraph{Recent Advances in MARL}



One popular testbed of MARL is StarCraft II \cite{vinyals2017starcraft}, a multi-player real-time strategy computer game that has its own professional league. 
In this game, each player has only limited information about the game state, and the dimension of the search space is orders of magnitude larger than that of go ($10^{26}$ possible choices for every move). The design of effective RL methods for StarCraft II was once believed to be a long-term challenge for AI \cite{vinyals2017starcraft}. 
However, a breakthrough was accomplished by AlphaStar in $2019$ \cite{vinyals2019grandmaster}, which has exhibited grandmaster-level skills by ranking above $99.8\%$ of human players.

Another prominent video game-based test-bed for MARL is Dota2, a zero-sum game played by two teams, each composed of five players. 
From each agent's perspective, in addition to the difficulty of incomplete information (similar to StarCraft II), Dota2 is more challenging, in the sense that both cooperation among team members and competition against the opponents must be considered.
The OpenAI Five AI system \cite{pachocki2018openai} demonstrated superhuman performance in Dota2 by defeating world champions in a public e-sports competition.

In addition to StarCraft II and Dota2, \cite{jaderberg2019human} and \cite{baker2019emergent} showed human-level performance in capture-the-flag and hide-and-seek games, respectively. Although the games themselves are less sophisticated than either StarCraft II or Dota2, it is still non-trivial for AI agents to master their tactics, so the agents' impressive performance again demonstrates the efficacy of MARL. Interestingly, both authors reported emergent behaviours induced by their proposed MARL methods that humans can understand and are grounded in physical theory.

Another achievement of MARL worth mentioning is its application to the poker game Texas hold' em, which is a multi-player extensive-form game with incomplete information accessible to the player. 
Heads-up (namely, two-player) no-limit hold'em has more than $6\times10^{161}$ information states. Only recently have ground-breaking achievements in the game been made, thanks to MARL.  Two independent programs, \emph{DeepStack} \cite{moravvcik2017deepstack} and \emph{Libratus} \cite{brown2018superhuman}, are able to beat professional human players. 
Even more recently, Libratus was upgraded to Pluribus \cite{brown2019superhuman} and showed remarkable performance by winning over one million dollars from five elite human professionals in a no-limit setting. 

More recently, MARL techniques based on Neural Replicator Dynamics\cite{neurd} have achieved expert-level performance in the very large game of Stratego\cite{perolat2022mastering}. Also in 2022, Cicero achieved human-level performance in the press version of Diplomacy by using game-theoretic MARL combined with large language models\cite{meta2022human}. MARL techniques have also been applied to large language models, where red-team and blue-team language models engage in a game-theoretic interaction to enhance the security detection of large language models \cite{ma2023red} and align them with human preferences \cite{wang2024magnetic}.

In conclusion, MARL requires a deep understanding of equilibria and the interactions between agents. Game theory provides a powerful tool for studying multiagent systems and developing MARL algorithms with nice theoretical properties and convergence guarantees. By combining game theory and deep RL, it is possible to scale MARL algorithms to complex multiagent systems and achieve superhuman performance.

For a deeper understanding of RL and MARL, mathematical notation and deconstruction of the concepts are needed. In the next section, we provide mathematical formulations for these concepts, starting from single-agent RL and progressing to multiagent RL methods.  

\chapter{Single-Agent RL}

Through trial and error, an RL agent attempts to find the optimal policy to maximise its long-term reward. 
This process is formulated by Markov Decision Processes. 

\section{Problem Formulation: Markov Decision Process}
\begin{definition}[Markov Decision Process]
An MDP can be described by a tuple of key elements $ \langle \mathbb{S}, \mathbb{A}, P, R, \gamma  \rangle$.
\begin{itemize}
	\item $\mathbb{S}$: the set of environmental states.
		\item $\mathbb{A}$: the set of agent's possible actions.
	\item $P: \mathbb{S} \times \mathbb{A} \rightarrow \Delta(\mathbb{S})$: for each time step $t \in \mathbb{N}$, given agent's action $a\in \mathbb{A}$, the transition probability from a state $s \in \mathbb{S}$ to  the state in the next time step $ s' \in \mathbb{S}$.
	\item $R: \mathbb{S} \times \mathbb{A} \times \mathbb{S} \rightarrow \mathbb{R}$: the reward function that returns a scalar value to the agent for a transition from $s$ to $s'$ as a result of action $a$. The rewards have absolute values uniformly bounded by $R_{\text{max}}$. 

	\item $\gamma \in [0, 1]$ is the discount factor that represents the value of time.  
\end{itemize}	
\label{def:mdp}
\end{definition}
  At each time step $t$, the environment has a state $s_t$. The learning agent observes this state\footnote{The agent can only observe part of the full environment state. The partially observable setting is introduced in Definition \ref{def:decpomdp} as a special case of Dec-PODMP.  } and executes an action $a_t$. The action makes the environment transition into the next state $s_{t+1} \sim P(\cdot | s_t, a_t)$, and the new environment returns an immediate reward $R(s_t, a_t, s_{t+1})$ to the agent.  
  The reward function  can  be also written as $R:\mathbb{S}\times \mathbb{A}\rightarrow \mathbb{R}$, which is interchangeable with $R: \mathbb{S} \times \mathbb{A} \times \mathbb{S} \rightarrow \mathbb{R}$ (see \cite{van2012reinforcement}, page 10). 
  The goal of the agent is to solve the MDP: to find the optimal policy that maximises the reward over time.
Mathematically, one common objective is for the agent to find a Markovian (i.e., the input depends on only the current state) and stationary (i.e., function form is time-independent) policy function\footnote{Such an optimal policy exists  as long as the transition function and the reward function are both Markovian and stationary \cite{feinberg2010total}.} $\pi: \mathbb{S} \rightarrow \Delta(\mathbb{A})$, with $\Delta(\cdot)$ denoting the probability simplex, which  can guide it to take sequential actions such that the discounted cumulative reward is maximised:  
 \begin{equation}
\mathbb{E}_{s_{t+1} \sim P(\cdot | s_t, a_t)}\left[\sum_{t \geq 0} \gamma^{t} R\left(s_{t}, a_{t}, s_{t+1}\right) \Big| a_{t} \sim \pi\left(\cdot \mid s_{t}\right), s_{0}\right].	
\label{eq:rlobj}
 \end{equation}
 Another common mathematical objective of an MDP is to maximise the time-average reward: \begin{equation}
 	 \lim _{T \rightarrow \infty} \mathbb{E}_{s_{t+1}\sim P(\cdot | s_t, a_t)}\left[\frac{1}{T} \sum_{t=0}^{T-1} R(s_t, a_t, s_{t+1}) \Big| a_{t} \sim \pi\left(\cdot \mid s_{t}\right), s_{0}\right],
 	 \label{eq:time_average_reward}
 \end{equation}
  which we do not consider in this work and refer to \cite{mahadevan1996average} for a full analysis of the objective of time-average reward.

 Based on the objective function of Eq. (\ref{eq:rlobj}), under a given policy $\pi$, we can define the state-action function (namely, the Q-function, which determines the expected return from undertaking action $a$ in state $s$) and the value function (which determines the return associated with the policy in state $s$) as:
  \begin{align}
Q^{\pi}(s, a) &=  \mathbb{E}^{\pi} \left[\sum_{t \geq 0} \gamma^{t} R\left(s_{t}, a_{t}, s_{t+1}\right) \Big| a_{0}=a, s_{0}=s\right] 	, \forall s \in \mathbb{S}, a\in \mathbb{A}
\label{eq:q_func} \\
V^{\pi}(s) &=  \mathbb{E}^{\pi} \left[\sum_{t \geq 0} \gamma^{t} R\left(s_{t}, a_{t}, s_{t+1}\right) \Big| s_{0}=s\right], \forall s \in \mathbb{S} 	
\label{eq:v_func} 
 \end{align}
where $\mathbb{E}^{\pi}$ is the expectation under the probability measure $\mathbb{P}^{\pi}$  over the set of infinitely long state-action trajectories $\tau=(s_0, a_0, s_1, a_1, ...)$ and  where $\mathbb{P}^{\pi}$  is induced by  state transition probability $P$, the policy $\pi$,  the initial state $s$ and initial action $a$ (in the case of the Q-function). 
The connection between the Q-function and value function is $V^{\pi}(s) = \mathbb{E}_{a\sim \pi(\cdot|s)}[Q^{\pi}(s, a)]$ and $Q^{\pi} = \mathbb{E}_{s'\sim P(\cdot| s, a)}[R(s, a, s') + V^{\pi}(s')]$.

\section{Justification of Reward Maximisation}

The current model for RL, as given by Eq. (\ref{eq:rlobj}), suggests that the expected value of a single reward function is sufficient for any problem we want our ``intelligent agents'' to solve.
The justification for this idea is deeply rooted in the \emph{von Neumann-Morgenstern (VNM) utility theory} \cite{von2007theory}. This theory essentially proves that an agent is \emph{VNM-rational} if and only if there exists a real-valued utility (or, reward) function such that every preference of the agent is characterised by maximising the single expected reward.  
 The VNM utility theorem is the basis for the well-known \emph{expected utility theory} \cite{schoemaker2013experiments}, which essentially states that \emph{rationality} can be modelled as maximising an expected value. 
  Specifically, the VNM utility theorem provides both necessary and sufficient conditions under which the expected utility hypothesis holds. 
  In other words, rationality is equivalent to VNM-rationality, and it is safe to assume an intelligent entity will always choose the action with the highest expected utility in any complex scenario.

Admittedly, it was accepted long before that some of the assumptions on rationality could be violated by real decision-makers in practice \cite{gigerenzer2002bounded}. In fact, those conditions are rather taken as the ``axioms'' of rational decision-making. 
In the case of the multi-objective MDP, we are still able to convert multiple objectives into a single-objective MDP with the help of a \emph{scalarisation function} through a two-timescale process; we refer to \cite{roijers2013survey} for more details.

\section{Solving Markov Decision Processes}

One commonly used notion in MDPs is the (discounted-normalised) occupancy measure $\mu^{\pi}(s, a)$, which uniquely corresponds to a   given policy $\pi$ and vice versa \cite[Theorem 2]{syed2008apprenticeship}, defined by  
\begin{align}
	\mu^{\pi}(s, a) &= \mathbb{E}_{s_t\sim P, a_t \sim \pi}\left[{ (1-\gamma)}\sum_{t \ge 0} \gamma^t \mathds{1}_{(s_t=s \wedge a_t=a)} \right]. \nonumber \\
&= (1-\gamma) \sum_{t\ge 0}\gamma^t\mathbb{P}^{\pi}(s_t = s, a_t=a), 
	\label{eq:occupancy}
\end{align}
where $\mathds{1}$ is an indicator function. Note that in Eq. (\ref{eq:occupancy}), $P$ is the state transitional probability, and $\mathbb{P}^{\pi}$ is the probability of specific state-action pairs when following stationary policy $\pi$. 
The physical meaning of  $\mu^{\pi}(s, a)$ is a probability measure that counts the expected discounted number of visits to the individual admissible state-action pairs. 
Correspondingly,  $\mu^{\pi}(s)  = \sum_a \mu^{\pi}(s, a)$ is the discounted state visitation frequency, i.e., the stationary distribution of the Markov process induced by $\pi$. 
With the occupancy measure, we can write Eq. (\ref{eq:v_func}) as an inner product of $V^{\pi}(s) = \frac{1}{1-\gamma} \big\langle \mu^{\pi}(s, a), R(s, a) \big\rangle $. This implies that solving an MDP
can be regarded as solving a linear program (LP) of $\max_{\mu}\big\langle \mu(s, a), R(s, a) \big  \rangle $, and the optimal policy  is then
\begin{equation}
	\pi^*(a|s) = \mu^{*}(s, a)/{\mu^{*}(s)}
	\label{eq:occupancy_policy}
\end{equation}
 However, this method is not practical in the case of a large-scale LP with millions of variables   \cite{papadimitriou1987complexity}. When the state-action space of an MDP is continuous, LP formulation cannot help solve either. 

In the context of optimal control \cite{bertsekas2005dynamic}, the classical dynamic-programming approaches, such as policy iteration and value iteration, can also be applied to solve for the optimal policy that maximises Eq. (\ref{eq:q_func}) and Eq. (\ref{eq:v_func}), but these approaches require knowledge of the exact form of the model: the transition function $P(\cdot | s, a)$, and the reward function $R(s, a, s')$. 

On the other hand, in the setting of RL, the agent learns the optimal policy by a trial-and-error process during its interaction with the environment rather than using prior knowledge of the model. 
RL algorithms can be categorised into two types: value-based methods and policy-based methods.

\subsubsection{Value-Based Methods}
For all MDPs with finite states and actions, there exists at least one deterministic stationary optimal policy \cite{szepesvari2010algorithms,sutton1998reinforcement}. 
Value-based methods are introduced to find the optimal Q-function $Q^*$ that maximises Eq. (\ref{eq:q_func}). Correspondingly, the optimal policy can be derived from the Q-function by taking the greedy action of $\pi^* = \arg\max_a Q^*(s, a)$. 
The classic Q-learning algorithm \cite{watkins1992q} approximates $Q^*$ by $\hat{Q}$ and updates its value via temporal-difference learning \cite{sutton1988learning}. 
{
\begin{equation}
	{\displaystyle \underbrace{\hat{Q}(s_{t},a_{t})}_{\text{new value}}\leftarrow \underbrace {\hat{Q}(s_{t},a_{t})} _{\text{old value}}+\underbrace {\alpha } _{\text{learning rate}}\cdot \overbrace {{\bigg (}\underbrace {{R_t} + {\gamma } \cdot {\max _{a\in\mathbb{A}}\hat{Q}(s_{t+1},a)}} _{\text{temporal difference target}}- \underbrace {\hat{Q}(s_{t},a_{t})} _{\text{old value}}{\bigg )}} ^{\text{temporal difference error }}}
	\label{eq:q_learning}
\end{equation}}
Theoretically, given the Bellman optimality operator $\mathbf{H}^*$,  defined by 
\begin{equation}
	(\mathbf{H}^*Q)(s, a) = \sum_{s'}P(s'| s, a)\left[R(s, a, s') + \gamma \max_{b\in \mathbb{A}}Q(s, b) \right], 
	\label{eq:operator_q}
\end{equation}
which is a contraction mapping and the optimal Q-function is the unique\footnote{Note that although the optimal Q-function is unique, its corresponding optimal policies may have multiple candidates. } fixed point, i.e., $\mathbf{H}^*(Q^*) = Q^*$.
The Q-learning algorithm draws random samples of $(s, a, R, s')$ in Eq. (\ref{eq:q_learning}) to approximate Eq. (\ref{eq:operator_q}), but is still guaranteed to converge to the optimal Q-function \cite{szepesvari1999unified} under the assumptions that the state-action sets are discrete and finite and are visited an infinite number of times.
\cite{munos2008finite} extended the convergence result to a more realistic setting by deriving the high probability error bound for an infinite state space with a finite number of samples.

Recently, \cite{mnih2015human} applied neural networks as a function approximator for the Q-function in updating Eq. (\ref{eq:q_learning}). Specifically, DQN optimises the following equation: 
\begin{equation}
	\min_\theta \mathbb{E}_{(s_t, a_t, R_t, s_{t+1})\sim \mathcal{D}}\left[\left(R_t + \gamma \max_{a\in\mathbb{A}}Q_{\theta^{-}}\left(s_{t+1}, a \right) - Q_\theta \left(s_t, a_t\right) \right)^2\right].
	\label{eq:dqn}
\end{equation}
The neural network parameters $\theta$ are fitted by drawing i.i.d. samples from the replay buffer  $\mathcal{D}$ and then updating in a supervised learning fashion. $Q_{\theta^{-}}$ is a slowly updated target network that helps stabilise training.  
The convergence property and finite sample analysis of DQN have been studied by  \cite{yang2019theoretical}.

\subsubsection{Policy-Based  Methods}
\label{sec:single_pg}
Policy-based methods are designed to directly search over the policy space to find the optimal policy $\pi^*$. 
One can parameterise the policy expression $\pi^* \approx \pi_\theta(\cdot | s)$ and update the parameter $\theta$ in the direction that maximises the cumulative reward $\theta \leftarrow \theta + \alpha \nabla_\theta V^{\pi_\theta}(s)$ to find the optimal policy.  However, the gradient will depend on the unknown effects of policy changes on the state distribution. The famous policy gradient (PG) theorem \cite{sutton2000policy} derives an analytical solution that does not involve the state distribution, that is:     
\begin{equation}
 \nabla_\theta V^{\pi_\theta}(s) =	\mathbb{E}_{s\sim \mu^{\pi_{\theta}}(\cdot), a\sim \pi_{\theta}(\cdot| s)}\Big[ \nabla_\theta \log \pi_\theta (a|s) \cdot Q^{\pi_\theta}(s, a) \Big]
 \label{eq:pg}
\end{equation}
where $\mu^{\pi_{\theta}}$ is the state occupancy measure under policy $\pi_\theta$ and $\nabla\log \pi_\theta (a|s)$ is the updating score of the policy. When the policy is deterministic and the action set is continuous, one obtains the deterministic policy gradient (DPG) theorem \cite{silver2014deterministic}  as 
\begin{equation}
 \nabla_\theta V^{\pi_\theta}(s) =	\mathbb{E}_{s\sim \mu^{\pi_{\theta}}(\cdot)}\Big[ \nabla_\theta \pi_\theta (a|s) \cdot \nabla_a Q^{\pi_\theta}(s, a) \big|_{a=\pi_\theta(s)} \Big].
 \label{eq:dpg}
\end{equation}

A classic implementation of the PG theorem is REINFORCE \cite{williams1992simple}, which uses a sample return $R_t = \sum_{i=t}^{T}\gamma^{i-t}r_i$ to estimate $Q^{\pi_\theta}$. 
Alternatively, one can use a model of $Q_\omega$ (also called \emph{critic}) to approximate the true $Q^{\pi_\theta}$ and update the parameter $\omega$ via TD learning. This approach gives rise to the famous actor-critic methods \cite{konda2000actor,peters2008natural}. Important variants of actor-critic methods include trust-region methods \cite{schulman2015trust, schulman2017proximal}, PG with optimal baselines \cite{weaver2001optimal,zhao2011analysis},  soft actor-critic methods \cite{haarnoja2018soft}, and deep deterministic policy gradient (DDPG) methods  \cite{lillicrap2015continuous}.

\chapter{Multi-Agent RL}
\label{sec:marl}

In the multiagent scenario, much like in the single-agent scenario, each agent is still trying to solve the sequential decision-making problem through a trial-and-error procedure. 
The difference is that the evolution of the environmental state and the reward function that each agent receives is now determined by all agents' joint actions (see Figure \ref{fig:single_multi}). 
As a result, agents need to take into account and interact with not only the environment but also other learning agents. 
A decision-making process that involves multiple agents is usually modeled through stochastic games (covered below in section~\ref{stochastic}) or extensive-form games (covered in section~\ref{extensive}).

\section{Stochastic Game}\label{stochastic}

\subsection{Problem Formulation}

\begin{figure}[t]
\centering
\includegraphics[width=.95\linewidth]{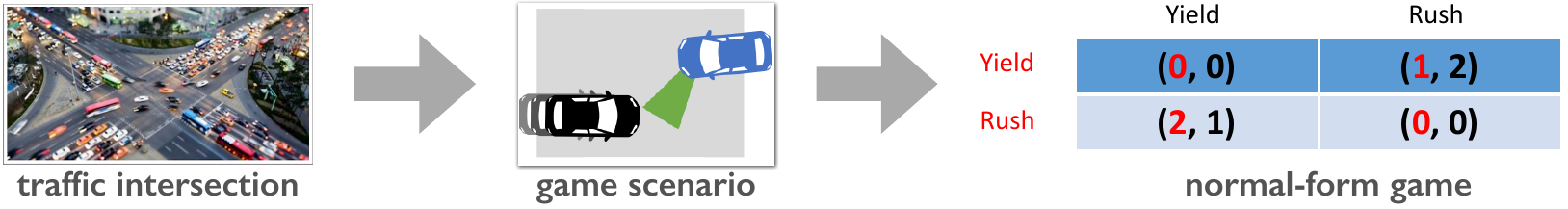}
\caption{A snapshot of stochastic time in the intersection example. The scenario is abstracted such that there are two cars, with each car taking one of two possible actions: to yield or to rush. The outcome of each joint action pair is represented by a normal-form game, with the reward value for the row player denoted in red and that for the column player denoted in black. The Nash equilibria (NE) of this game are (rush, yield) and (yield, rush). If both cars maximize their own reward selfishly without considering the others, they will end up in an accident.}
\label{fig:intersection}
\end{figure}

\begin{definition}[Stochastic Game]
A stochastic game\cite{shapley1953stochastic, littman1994markov} can be regarded as a multi-player\footnote{Player is a common word used in game theory; the agent is more commonly used in machine learning. We do not discriminate between their usages in this work. The same holds for strategy vs policy and utility/payoff vs reward. Each pair refers to the game theory usage vs machine learning usage.} extension to the MDP in Definition \ref{def:mdp}. Therefore, it is also defined by a set of key elements $\langle N, \mathbb{S}, \{\mathbb{A}^i\}_{i\in\{1,...,N\}}, P,  \{R^i\}_{i\in\{1,...,N\}}, \gamma \rangle $.
\begin{itemize}
\item $N$: the number of agents, $N=1$ degenerates to a single-agent MDP, $N \gg 2$ is referred as many-agent cases in this paper. 
	\item $\mathbb{S}$: the set of environmental states shared by all agents.
	\item $\mathbb{A}^i$: the set of  actions of agent $i$. We denote $\pmb{\mathbb{A}}:=\mathbb{A}^1\times \cdots \times \mathbb{A}^N$.
	\item $P: \mathbb{S} \times \pmb{\mathbb{A}} \rightarrow \Delta(\mathbb{S})$: for each time step $t \in \mathbb{N}$, given  agents' joint actions $\bm{a}\in \pmb{\mathbb{A}}$, the transition probability from state $s \in \mathbb{S}$ to  state $ s' \in \mathbb{S}$ in the next time step.
	\item $R^i: \mathbb{S} \times \pmb{\mathbb{A}} \times \mathbb{S} \rightarrow \mathbb{R}$: the reward function that returns a scalar value to the $i-th$ agent for a transition from $(s, \bm{a})$ to $s'$. The rewards have absolute values uniformly bounded by $R_{\text{max}}$. 
	\item $\gamma \in [0, 1]$ is the discount factor that represents the value of time.  
\end{itemize}
	\label{def:sg}
\end{definition}
We use the superscript of $(\cdot^i, \cdot^{-i})$ (for example, $\bm{a}=(a^i, a^{-i})$), when it is necessary to distinguish between agent $i$ and all  other $N-1$ opponents.  

Ultimately, the stochastic game (SG) acts as a framework that allows simultaneous moves from agents in a decision-making scenario\footnote{Extensive-form games allow agents to take sequential moves; the full description can be found in  \cite[Chapter 5]{shoham2008multiagent}.}. The game can be described sequentially, as follows: 
 At each time step $t$, the environment has a state $s_t$, and given $s_t$, each agent executes its action $a^i_t$, simultaneously with all other agents. The joint action from all agents makes the environment transition into the next state $s_{t+1} \sim P(\cdot | s_t, \bm{a}_t)$; then, the environment determines an immediate reward $R^i(s_t, \bm{a}_t, s_{t+1})$ for each agent. 
 As seen in the single-agent MDP scenario, the goal of each agent $i$ is to solve the SG. In other words, each agent aims to find a behavioral policy (or,  a mixed strategy\footnote{A behavioral policy refers to a function map from the history  $(s_0, a_0^i, s_1, a_1^i, ..., s_{t-1})$ to an action. The policy is typically assumed to be Markovian such that it depends on only the current state $s_t$ rather than the entire history. A mixed strategy refers to a randomization over pure strategies (for example, the actions). In SGs, the behavioral policy and mixed policy are exactly the same. In extensive-form games, they are different, but if the agent retains the history of previous actions and states (has perfect recall), each behavioral strategy has a realization-equivalent mixed strategy, and vice versa \cite{kuhn1950extensive}.} in game theory terminology \cite{osborne1994course}), that is,  $\pi^i \in \Pi^i: \mathbb{S} \rightarrow \Delta(\mathbb{A}^i)$ that can guide the agent to take sequential actions such that the discounted cumulative reward\footnote{Similar to single-agent MDP, we can adopt the objective of time-average rewards.} in Eq. (\ref{eq:marl_vfunc}) is maximized.   
Here, $\Delta(\cdot)$ is the probability simplex on a set.  In game theory, $\pi^i$ is also called a pure strategy (vs a mixed strategy) if  $\Delta(\cdot)$ is replaced by a Dirac measure. 
 \begin{equation}
 \scriptstyle
 V^{\pi^i, \pi^{-i}}(s) = \mathbb{E}_{s_{t+1} \sim P(\cdot | s_t, a_t), a^i_{t} \sim \pi^i\left(\cdot \mid s_{t}\right), a^{-i} \sim {\pi}^{-i}(\cdot | s_t)}\left[\sum_{t \geq 0} \gamma^{t} R^i_t\left(s_{t}, \bm{a}_{t}, s_{t+1}\right) \Big| a^i_{t}, s_{0}\right]. 	
\label{eq:marl_vfunc}
 \end{equation}
 Comparison of Eq. (\ref{eq:marl_vfunc}) with Eq. (\ref{eq:v_func}) indicates that the optimal policy of each agent is influenced by not only its own policy but also the policies of the other agents in the game. 
 This scenario leads to fundamental differences in the \emph{solution concept} between single-agent RL and multiagent RL.

 \subsection{Solving Stochastic Games}
 Given the definitions of SGs defined above, we will primarily introduce some solutions for SGs in this section, especially with MARL methods. More detailed discussions about advanced algorithms for solving SGs are provided later in Sec.~\ref{sec:sg_alg}.
 An SG can be considered as a sequence of normal-form games, which are games that can be represented in a matrix. 
 Take the original intersection scenario as an example (see Figure \ref{fig:intersection}).  A snapshot of the SG at time $t$ (stage game) can be represented as a normal-form game in a matrix format. The rows correspond to the action set $\mathbb{A}^1$ for agent $1$, and the columns correspond to the action set $\mathbb{A}^2$ for agent $2$. The values of the matrix are the rewards given for each of the joint action pairs.
 In this scenario, if both agents care only about maximizing their own possible reward with no consideration of other agents (the solution concept in a single-agent RL problem) and choose the action to rush, they will reach the outcome of crashing into each other. Clearly, this state is unsafe and is thus sub-optimal for each agent, despite the fact that the possible reward was the highest for each agent when rushing. Therefore, to solve an SG and truly maximize the cumulative reward, each agent must take strategic actions with consideration of others when determining their policies.
 
 Unfortunately, in contrast to MDPs, which have polynomial time-solvable linear programming formulations, solving SGs usually involves applying Newton's method for solving nonlinear programs. However, there are two special cases of two-player general-sum discounted-reward SGs that can still be written as LPs \cite [Chapter 6.2]{shoham2008multiagent}\footnote{According to \cite{filar2012competitive} [Section 3.5],  single-controller SG is solvable in polynomial time only under zero-sum cases rather than general-sum cases, which contradicts the result in \cite{shoham2008multiagent}  [Chapter 6.2], and we believe \cite{shoham2008multiagent} made a typo.}. They are as follows:
 \begin{itemize}
 \item 	\emph{single-controller SG}:  the transition dynamics are determined by a single player, i.e., $P(\cdot| \bm{a}, s) = P(\cdot| a^i, s)$ if the i-th index in the vector $\bm{a}$ is $\bm{a}[i]=a^i, \forall s \in \mathbb{S}, \forall \bm{a} \in \pmb{\mathbb{A}}$.
 \item \emph{separable reward state independent transition (SR-SIT) SG}: the states and the actions have independent effects on the reward function and the transition function depends on only the joint actions, i.e., $\exists \alpha:\mathbb{S}\rightarrow \mathbb{R},   \beta:\pmb{\mathbb{A}}\rightarrow \mathbb{R}$ such that these two conditions hold:  $1) \   R^i(s, \bm{a})=\alpha(s) + \beta (\bm{a}), \forall i \in \{1,..., N\},  \forall s \in \mathbb{S}, \forall \bm{a} \in \pmb{\mathbb{A}},  $ and $ 2) \  P(\cdot| s', \bm{a}) = P(\cdot| s, \bm{a}), \forall \bm{a} \in \pmb{\mathbb{A}}, \forall s, s' \in \mathbb{S}$. 
 \end{itemize}


 \subsubsection{Value-Based MARL Methods } 
 The single-agent Q-learning update in Eq. (\ref{eq:q_learning}) still holds in the multiagent case.
 In the $t$-th iteration, for each agent $i$, 
 given the transition data $\big\{(s_t, \bm{a}_t, R^i, s_{t+1}) \big\}_{t\ge0}$ sampled from the replay buffer, it updates only the value of $Q(s_t, \bm{a}_t)$ and keeps the other entries of the Q-function unchanged. Specifically, we have 
\begin{equation}
\scriptstyle
	{ Q^i(s_{t},\bm{a}_{t})\leftarrow {Q^i(s_{t}, \bm{a}_{t})} +{\alpha } \cdot } \bigg( R^i + \gamma \cdot \mathbf{eval}^i \Big( \big\{Q^i(s_{t+1}, \cdot) \big\}_{i\in\{1,...,N\}} \Big)- Q^i(s_t, \bm{a}_t )  \bigg).
	\label{eq:ma_q_learning}
\end{equation}
 Compared to Eq. (\ref{eq:q_learning}), the $\max$ operator is changed to another operator $\mathbf{eval}^i \big(\{Q^i(s_{t+1}, \cdot) \}_{i\in\{1,...,N\}} \big)$ in Eq. (\ref{eq:ma_q_learning}) to reflect the fact that each agent can no longer consider only itself but must \textbf{eval}uate the situation of the stage game at time step $t+1$ by considering all agents'  interests, as represented by the set of their Q-functions. Then, the optimal policy can be solved by $\textbf{solve}^i \big(\{Q^i(s_{t+1}, \cdot) \}_{i\in\{1,...,N\}} \big) =\pi^{i, *} $. 
Therefore, we can further write the evaluation operator as 
\begin{equation}
\scriptstyle
	\mathbf{eval}^i \Big( \big\{Q^i(s_{t+1}, \cdot) \big\}_{i\in\{1,...,N\}} \Big) = V^{i}\Big(s_{t+1}, \Big\{\mathbf{solve}^i \big(\{Q^i(s_{t+1}, \cdot) \}_{i\in\{1,...,N\}}\big) \Big\} \Big).
		\label{eq:ma_solve_sg}
\end{equation}

In summary, $\mathbf{solve}^i$ returns agent $i'$s part of the optimal policy at  some equilibrium point (not necessarily corresponding to its largest possible reward),   
and $\mathbf{eval}^i$ gives agent $i$'s expected long-term reward under this equilibrium, assuming all other agents agree to play the same equilibrium.  
In Chapter \ref{chaper:general_sum}, we will further demonstrate how to leverage eval and solve when using value-based and policy-based algorithms, illustrating their practical applications in equilibrium computation and policy optimization.
 
 \subsubsection{Policy-Based MARL Methods} 
 The value-based approach suffers from the curse of dimensionality due to the combinatorial nature of multiagent systems (for further discussion, see Section \ref{sec:combinatorial}). 
  This characteristic necessitates the development of policy-based algorithms with function approximations.  
 Specifically, each agent learns its own optimal policy  $\pi^i_{\theta^i}: \mathbb{S} \rightarrow  \Delta(\mathbb{A}^i)$ by updating the parameter $\theta^i$ of, for example, a neural network.  
 Let $\theta=(\theta^i)_{i \in \{1,..., N\}}$ represent the collection of policy parameters for all agents, and let $\bm{\pi}_\theta:=\prod_{i\in\{1,...,N\}}\pi^i_{\theta^i}(a^i|s)$ be the joint policy. 
To optimize the parameter $\theta^i$, 
 the policy gradient theorem in Section \ref{sec:single_pg} can be extended to the multiagent context. 
 Given agent $i$'s objective function $J^i(\theta) = \mathbb{E}_{s\sim P, \bm{a}\sim{\bm{\pi}}_{\theta}}\big[\sum_{t\ge0} \gamma_tR_t^i\big]$, we have:
 \begin{equation}
 \nabla_{\theta^i} J^i(\theta) =	\mathbb{E}_{s\sim \mu^{\bm{\pi}_{\theta}}(\cdot),  \bm{ a}\sim \bm{\pi}_{\theta}(\cdot| s)}\Big[ \nabla_{\theta^i} \log \pi_{\theta^i} (a^i|s) \cdot Q^{i, \bm{\pi}_\theta}(s, \bm{a}) \Big].
 \label{eq:mas_pg}
\end{equation}
Considering a continuous action set with a deterministic policy, we have the multiagent deterministic policy gradient (MADDPG) \cite{lowe2017multi} written as  
\begin{equation}
 \nabla_{\theta^i} J^i(\theta) =	\mathbb{E}_{s\sim \mu^{\bm{\pi}_{\theta}}(\cdot)}\Big[ \nabla_{\theta^i} \log \pi_{\theta^i} (a^i|s) \cdot  \nabla_{a_i} Q^{i, \bm{\pi}_\theta}(s, \bm{a}) \big|_{\bm{a} = \bm{\pi}_{\theta}(s)}  \Big].
 \label{eq:ma_dpg}
\end{equation}
Note that in both Eqs. (\ref{eq:mas_pg}) \& (\ref{eq:ma_dpg}), the expectation over the joint policy $\bm{\pi}_\theta$ implies that other agents' policies must be observed; this is often a strong assumption for many real-world applications. 

 \subsubsection{Solution Concept--Nash Equilibrium}
 \label{sec:nash_solution}
 
 Game theory plays an essential role in multiagent learning by offering so-called \emph{solution concepts} that describe the outcomes of a game by showing which strategies will finally be adopted by players.  
Many types of solution concepts exist for MARL (see  Section \ref{sec:many_obj}), among which the most famous is probably the Nash equilibrium (NE)  in non-cooperative game theory \cite{nash1951non}. The word ``non-cooperative'' does not mean agents cannot collaborate or have to fight against each other all the time, it merely means that each agent maximizes its own reward independently and that agents cannot group into coalitions to make collective decisions. 

In a normal-form game, the NE characterizes an equilibrium point of the joint strategy profile $(\pi^{1,*}, ..., \pi^{N,*})$, where each agent acts according to their \textbf{best response} to the others. The best response produces the optimal outcome for the player once all other players' strategies have been considered. 
Player $i$'s best response\footnote{Best responses may not be unique; if a mixed-strategy best response exists, there must be at least one best response that is also a pure strategy.} to $\pi^{-i}$ is a set of policies in which the following condition is satisfied. 
\begin{equation}
\pi^{i,*} \in   \mathbf{Br}(\pi^{-i}) := \Big\{ \arg\max_{\hat{\pi}\in\Delta(\mathbb{A}^i)} \mathbb{E}_{\hat{\pi}^{i}, \pi^{-i}}\big[R^i (a^i, a^{-i})\big] \Big\}.
\label{eq:best_response}
\end{equation}
NE states that if all players are perfectly rational, none of them will have the motivation to deviate from their best response $\pi^{i,*}$ given others are playing $\pi^{-i,*}$.
Note that NE is defined in terms of the best response, which relies on relative reward values, suggesting that the exact values of rewards are not required for identifying NE. In fact, NE is invariant under positive affine transformations of a player's reward functions. 
By applying Brouwer's fixed point theorem, \cite{nash1951non} proved that a mixed-strategy NE always exists for any games with a finite set of actions. 
In the example of driving through an intersection in Figure \ref{fig:intersection}, the NE are $(yield, rush)$ and $(rush, yield)$.

For an SG, one commonly used equilibrium is a stronger version of the NE, called the Markov perfect NE \cite{maskin2001markov}, which is defined by:
\begin{definition}[Nash Equilibrium for Stochastic Game]
A Markovian strategy profile $\bm{\pi}^*=(\pi^{i,*}, \pi^{-i,*})$ is a Markov perfect NE of a SG defined in Definition \ref{def:sg} if the following condition holds, with the requirement that the randomness in all players' strategies remains independent.  
\begin{equation}
V^{\pi^{i,*}, \pi^{-i,*}}(s) \ge  	V^{\pi^{i}, \pi^{-i,*}}(s), \ \ \ \  \forall s\in\mathbb{S}, \forall \pi^i \in \Pi^i, \forall i \in \{1,...,N\}.
\end{equation}	
\end{definition}
``Markovian'' means the Nash policies are measurable with respect to a particular partition of possible histories (usually referring to the last state). 
The word ``perfect'' means that the equilibrium is also subgame-perfect \cite{selten1965spieltheoretische} regardless of the starting state. Considering the sequential nature of SGs, these assumptions are necessary, while still maintaining generality. Hereafter, the Markov perfect NE will be referred to as NE. 

A mixed-strategy NE\footnote{Note that this is different from a single-agent MDP, where a single, ``pure'' strategy optimal policy always exists. A simple example is the rock-paper-scissors game, where none of the pure strategies is the NE and the only NE is to mix between the three equally.} always exists for both discounted and average-reward\footnote{Average-reward SGs entail more subtleties because the limit of Eq. (\ref{eq:time_average_reward}) in the multiagent setting may be a cycle and thus not exist. Instead, NE is proved to exist on a special class of irreducible SGs, where every stage game can be reached regardless of the adopted policy.} SGs \cite{filar2012competitive}, though they may not be unique. In fact, checking for uniqueness is $NP$-hard \cite{conitzer2008new}.
With the NE as the solution concept of optimality, we can re-write Eq. (\ref{eq:ma_solve_sg}) as:  \par
 {\small
\begin{equation}
	\mathbf{eval}_{\text{Nash}}^i \Big( \big\{Q^i(s_{t+1}, \cdot) \big\}_{i\in\{1,...,N\}} \Big) = V^{i}\Big(s_{t+1}, \Big\{\mathbf{Nash}^i \big(\{Q^i(s_{t+1}, \cdot) \}_{i\in\{1,...,N\}}\big) \Big\} \Big).
		\label{eq:ma_nash_q}
\end{equation}}
In the above equation, $\mathbf{Nash}^i(\cdot) = \pi^{i,*}$ computes the NE of agent $i$'s strategy, and $V^i\big(s, \{\mathbf{Nash^i}\}_{i\in\{1,...,N\}}\big)$ is the expected payoff for agent $i$ from state $s$ onwards under this equilibrium. 
Eq. (\ref{eq:ma_nash_q}) and Eq. (\ref{eq:ma_q_learning}) form the learning steps of Nash Q-learning \cite{hu1998multiagent}. This process essentially leads to the outcome of a learned set of optimal policies that reach NE for every single-stage game encountered. In the case when NE is not unique, Nash-Q adopts hand-crafted rules for equilibrium selection (e.g., all players choose the first NE). 
Furthermore, similar to normal Q-learning, the Nash-Q operator defined in Eq. (\ref{eq:operator_nash}) is also proved to be a contraction mapping, and the stochastic updating rule provably converges to the NE for all states when the NE is unique:
\begin{equation}
\scriptstyle
	(\mathbf{H}^{\text{Nash}}Q)(s, a) = \sum_{s'}P(s'| s, a)\bigg[R(s, a, s') + \gamma\cdot \mathbf{eval}_{\text{Nash}}^i \Big( \big\{Q^i(s_{t+1}, \cdot) \big\}_{i\in\{1,...,N\}} \Big)  \bigg]. 
	\label{eq:operator_nash}
 \end{equation}

\begin{figure}[t]
\centering
\includegraphics[width=.85\linewidth]{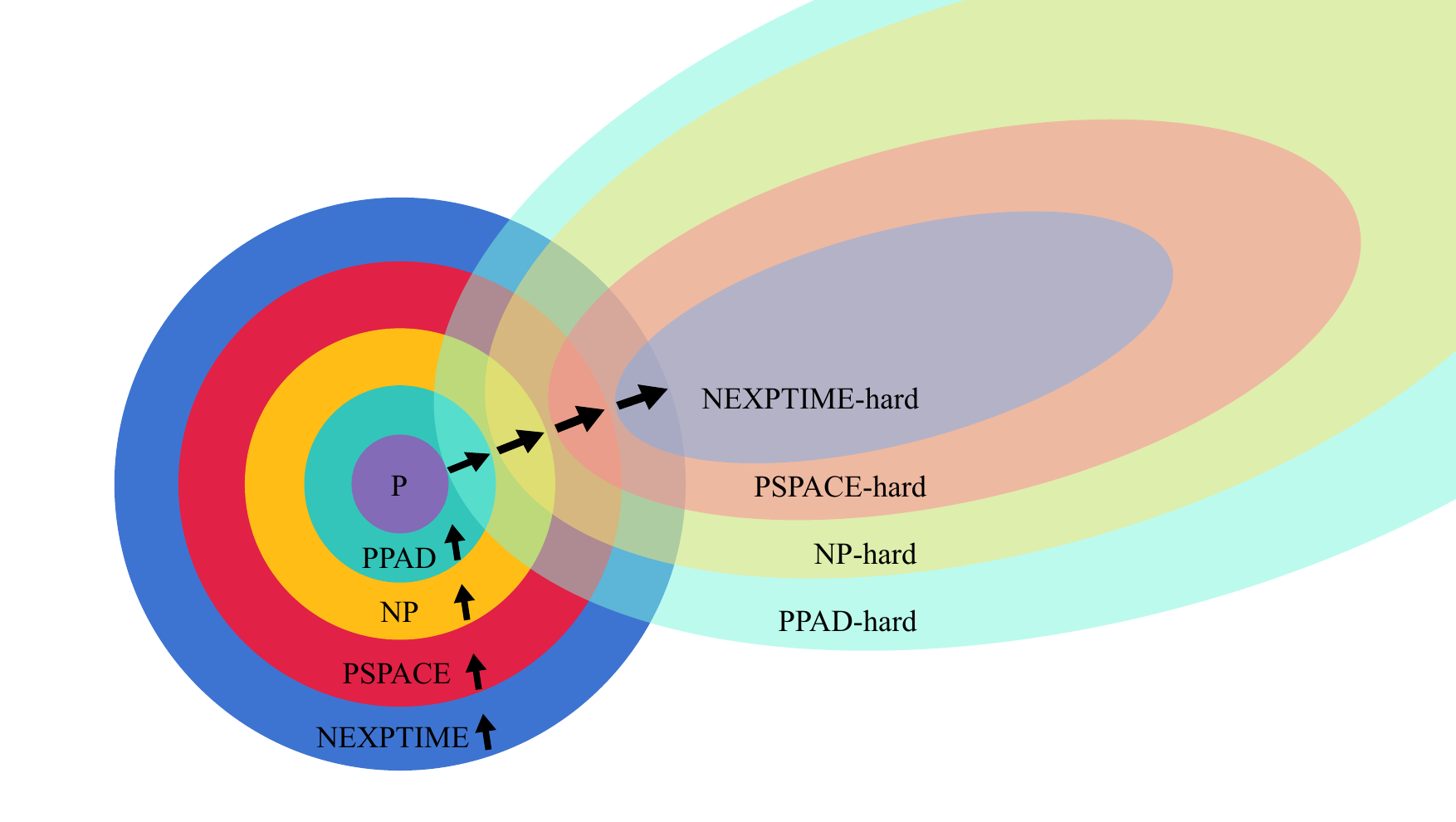}
\caption{The landscape of different complexity classes. Relevant examples are 1) solving the NE in a two-player zero-sum game, $P$-complete  \cite{neumann1928theorie}, 2) solving the NE in a general-sum game, $PPAD$-hard \cite{daskalakis2009complexity}, 3) checking the uniqueness of the NE, $NP$-hard \cite{conitzer2008new}, 4) checking whether a pure-strategy NE exists in a stochastic game, $PSPACE$-hard \cite{conitzer2008new}, and 5) solving Dec-POMDP, $NEXPTIME$-hard \cite{bernstein2002complexity}. }
\label{fig:complexity}
\end{figure}
The process of finding a NE in a two-player general-sum game can be formulated as a linear complementarity problem (LCP), which can then be solved using the \emph{Lemke-Howson} algorithm \cite{shapley1974note}. 
However, the exact solution for games with more than three players is unknown. 
The known theoretical result is that solving three-player zero-sum games is believed to be $PPAD$-hard \cite{daskalakis2005three}. 
In fact, the process of finding the NE is computationally demanding.
Even in the case of two-player games, the complexity of solving the NE is $PPAD$-hard (polynomial parity arguments on directed graphs) \cite{daskalakis2009complexity, chen2006settling}; therefore, in the worst-case scenario, the solution could take time that is exponential in relation to the game size. This complexity\footnote{The class of $NP$-complete is not suitable to describe the complexity of solving the NE because the NE is proven to always exist \cite{nash1951non}, while a typical $NP$-complete problem -- the traveling salesman problem (TSP), for example --  searches for the solution to the question: ``Given a distance matrix and a budget B, find a tour that is cheaper than B, or report that none exists \cite{daskalakis2009complexity}.'' }  prohibits any brute force or exhaustive search solutions unless $P=NP$ (see Figure \ref{fig:complexity}). 
As we would expect, the NE is much more difficult to solve for general SGs, where determining whether a pure-strategy NE exists is $PSPACE$-hard. Even if the SG has a finite-time horizon, the calculation remains $NP$-hard \cite{conitzer2008new}. 
When it comes to approximation methods to $\epsilon$-NE, the best known polynomially computable algorithm can achieve $\epsilon=0.3393$ on bimatrix games \cite{tsaknakis2007optimization}; its approach is to turn the problem of finding NE into an optimization problem that searches for a stationary point.
Furthermore, communication complexity can be used as a lower bound for the required learning time or cost \cite{Conitzer04:Communication}, which is general for any learning algorithm and provides communication complexity for various solution concepts in game theory, including Nash equilibrium, iterated dominant strategies (both strict and weak), and backward induction \cite{Conitzer04:Communication}. This is a very important principle that many works have followed to generate lower bounds for equilibrium finding and MARL.

\subsection{Special Types of Stochastic Games}
\label{sec:st-sg}

 \begin{figure}[htbp]
\centering
\includegraphics[width=.55\linewidth]{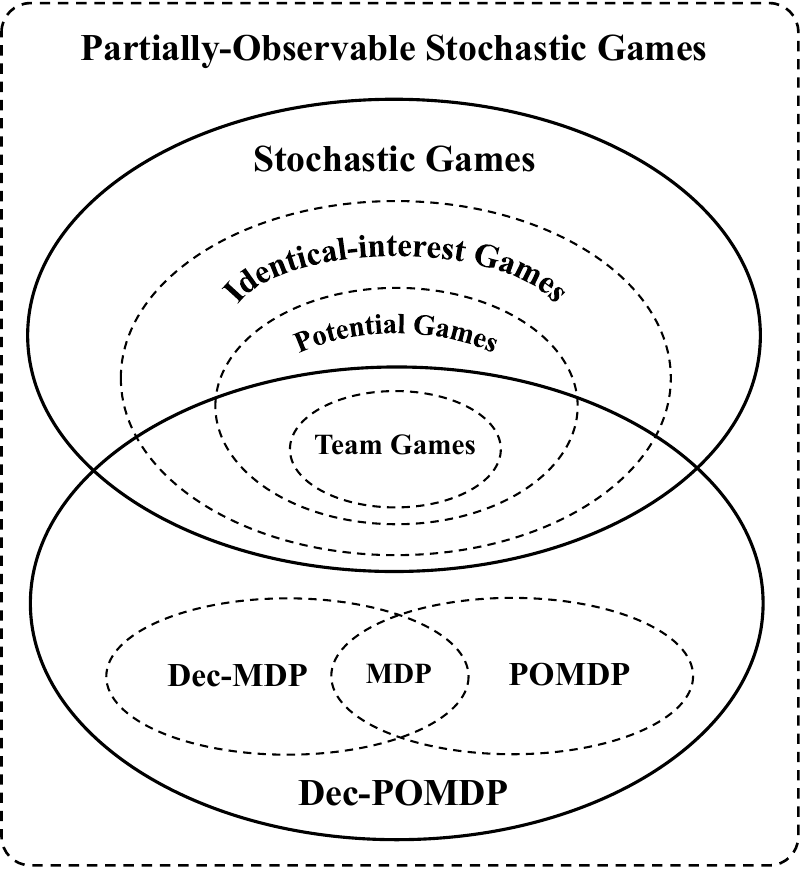}
\caption{Venn diagram of different types of games in the context of POSGs. The intersection of SG and Dec-POMDP is the team game. In the upper-half  SG, we have MDP $\subset$ team games $\subset$ potential games $\subset$ identical-interest games $\subset$ SGs. In the bottom-half Dec-POMDP, we have MDP $\subset$ team games $\subset$ Dec-MDP $\subset$ Dec-POMDPs, and MDP $\subset$ POMDP $\subset$ Dec-POMDP. We refer to Sections (\ref{sec:st-sg} \& \ref{sec:st-pomdp}) for detailed definitions of these games. }
\label{fig:games_types}
\end{figure}
To summarise the solutions to SGs,
one can think of the ``master'' equation: $$\textbf{Normal-form game solver} + \textbf{MDP solver} = \textbf{SG solver},$$
which was first summarised by \cite{bowling2000analysis} (in Table 4). 
The first term refers to solving an equilibrium (NE) for the stage game encountered at every time step.  
It assumes the transition and reward function is known.
 The second term refers to applying a RL technique (such as Q-learning) to model the temporal structure in the sequential decision-making process. 
It assumes to only receive observations of the transition and reward function.
 The combination of the two gives a solution to SGs, where agents reach a certain type of equilibrium at each and every time step during the game.

Since solving general SGs with NE as the solution concept for the normal-form game is computationally challenging, researchers instead aim to study special types of SGs that have tractable solution concepts. In this section, we provide a brief summary of these special types of games.

\begin{definition}[Special Types of Stochastic Games] Given the general form of SG in Definition \ref{def:sg}, we have the following special cases:
\begin{itemize} 
		\item \textbf{normal-form game/repeated game}: $|S|=1$, see the example in Figure \ref{fig:intersection}.  These games have only a single state. Though not theoretically grounded, it is practically easier to solve a small-scale SG. 
	\item \textbf{identical-interest setting\footnote{In some of the literature on this topic, identical-interest games are equivalent to team games. Here, we refer to this type of game as a more general class of games that involve a shared objective function that all agents collectively optimize, although their individual reward functions can still be different.}}: agents share the same learning objective, which we denote as $\mathsf{R}$. Since all agents are treated independently, each agent can safely choose the action that maximizes its own reward. As a result, single-agent RL algorithms can be applied safely, and a decentralized method is developed. 
	 Several types of SGs fall into this category. 
	\begin{itemize}
	\item \textbf{team games/fully cooperative games/multiagent MDP (MMDP)}: agents are assumed to be homogeneous and interchangeable, so importantly, they share the same reward function\footnote{In some of the literature on this topic (for example, \cite{wang2003reinforcement}), agents are assumed to receive the same expected reward in a team game, which means in the presence of noise, different agents may receive different reward values at a particular moment.},  $\mathsf{R}=R^1=R^2=\cdots=R^N.$ 
	\item \textbf{team-average reward games/networked multiagent MDP (M-MDP)}: agents can have different reward functions, but they share the same objective, $\mathsf{R}=\frac{1}{N}\sum_{i=1}^{N}R^i $.
	\item \textbf{stochastic potential games}: agents can have different reward functions, but their mutual interests are described by a shared potential function $\mathsf{R}= \phi$, defined as $\phi: \mathbb{S}\times \pmb{\mathbb{A}}\rightarrow \mathbb{R}$ such that $\forall (a^i, a^{-i}), (b^{i}, a^{-i}) \in \pmb{\mathbb{A}}, \forall i \in \{1,...,N\}, \forall s \in \mathbb{S}$ and the following equation holds:
	  \begin{equation}
   \scriptstyle
	  R^{i}\left(s, \left(a^{i}, a^{-i}\right)\right)-R^{i}\left(s,\left(b^{ i}, a^{-i}\right)\right)=\phi\left(s,\left(a^{i}, a^{-i}\right)\right)-\phi\left(s,\left(b^{i}, a^{-i}\right)\right). 
	  \label{eq:potential}
	  \end{equation} 
	  Games of this type are guaranteed to have a pure-strategy NE \cite{mguni2020stochastic}.
	  	  Moreover, potential games degenerate into team games if one chooses the reward function to be a potential function. 
	\end{itemize}
	\item \textbf{zero-sum setting}: agents share opposite interests and act competitively, and each agent optimizes against the worst-case scenario. The NE in a zero-sum setting can be solved using a linear program (LP) in polynomial time because of the minimax theorem developed by \cite{neumann1928theorie}. The idea of min-max values is also related to robustness in machine learning. We can subdivide the zero-sum setting as follows:
	\begin{itemize}
	\item \textbf{two-player constant-sum games}: $R^1(s, a, s') + R^2(s, a, s') = c, \forall (s, a, s')$, where $c$ is a constant and usually $c=0$. For cases when $c \neq 0$, one can always subtract the constant $c$ for all payoff entries to make the game zero-sum.   
	\item \textbf{two-team competitive games}: two teams compete against each other, with team sizes $N_1$ and $N_2$.  Their reward functions are: 
	\[\{R^{1,1},..., R^{1, N_1}, R^{2, 1}, ..., R^{2, N_2} \}.\] Team members within a team share the same objective of either $$\mathsf{R}^1 = \sum_{i\in\{1,...,N_1\}}R^{1,i}/N_1,$$ or  $$\mathsf{R}^2 = \sum_{j\in\{1,...,N_2\}}R^{2,j}/N_2, $$ and $\mathsf{R}^1 + \mathsf{R}^2 = 0$. 
	\item \textbf{harmonic games}: Any normal-form game can be decomposed into a potential game plus a harmonic game \cite{candogan2011flows}.  A harmonic game (for example, rock-paper-scissors) can be regarded as a general class of zero-sum games with a harmonic property. Let $\forall \bm{p} \in \pmb{\mathbb{A}}$ be a joint pure-strategy profile, and let $ \pmb{\mathbb{A}}^{ [-i]}=\{\bm{q} \in \pmb{\mathbb{A}}:\bm{q}^i \neq \bm{p}^i, \bm{q}^{-i} = \bm{p}^{-i}\}$ be the set of strategies that differ from $\bm{p}$ on agent $i$; then, the harmonic property is:  
	\[
	\sum_{i\in\{1,...,N\}}\sum_{\bm{q} \in  \pmb{\mathbb{A}}^{[-i]}}\big(R^i(\bm{p}) - R^i(\bm{q})\big) = 0,  \ \ \ \  \forall \bm{p} \in \pmb{\mathbb{A}}.
	\]
	\end{itemize}
	\item \textbf{linear-quadratic (LQ) setting}: the transition model follows linear dynamics, and the reward function is quadratic with respect to the states and actions. Compared to a black-box reward function, LQ games offer a simple setting. For example, actor-critic methods are known to facilitate convergence to the NE of zero-sum LQ games \cite{al2007model}. Again, the LQ setting can be subdivided as follows:
		\begin{itemize}
	\item \textbf{two-player zero-sum LQ games}:  $Q\in \mathbb{R}^{| \mathbb{S}|}, U^1\in \mathbb{R}^{| \mathbb{A}^1|}$ and $W^2\in \mathbb{R}^{| \mathbb{A}^2|}$ are the known cost matrices for the state and action spaces, respectively, while the matrices $A\in \mathbb{R}^{| \mathbb{S}|\times | \mathbb{S}|}, B\in \mathbb{R}^{| \mathbb{S}|\times | \mathbb{A}^1|}, C\in \mathbb{R}^{| \mathbb{S}|\times | \mathbb{A}^2|}$ are usually unknown to the agent:
	\begin{align}
			s_{t+1} &= As_t +Ba^1_t + Ca^2_t, \ \ \ \ \   s_0 \sim P_0 \nonumber, \\
			R^1(a^1_t , a^2_t ) &= -R^2(a^1_t , a^2_t ) \nonumber \\
    &= -\mathbb{E}_{s_0 \sim P_0} \left[\sum_{t\ge0} s_t^TQs_t + {a^1_t}^TU^1a^1_t  - {a^2_t}^TW^2a^2_t \right].
			\label{eq:lq_zerosum}
	\end{align}
	\item \textbf{multi-player general-sum LQ games}: the difference with respect to a two-player game is that the summation of the agents' rewards does not necessarily equal zero: 
	\begin{align}
			s_{t+1} &= As_t +B\bm{a}_t, \ \ \ \ \   s_0 \sim P_0 \nonumber, \\
			R^i(\bm{a}) &= -\mathbb{E}_{s_0 \sim P_0} \left[\sum_{t\ge0} s_t^TQ^is_t + {a^i_t}^TU^ia^i_t \right].
			\label{eq:lq_general}
	\end{align}

	\end{itemize}
	\end{itemize}
\label{def:sgs_types}	
\end{definition}

 \subsection{Partially Observable Settings} 
 \label{sec:st-pomdp}
 
 A partially observable stochastic game (POSG) assumes that
  agents have no access to the exact environmental state but only an observation of the true state through an observation function. Formally, this scenario is defined by:
  \begin{definition}[partially-observable stochastic games] A POSG is defined by the set $\langle N, \mathbb{S}, \{\mathbb{A}^i\}_{i\in\{1,...,N\}}, P,  \{R^i\}_{i\in\{1,...,N\}}, \gamma,  \underbrace{\{\mathbb{O}^i\}_{i\in\{1,...,N\}}, O}_{\text{newly added}}\rangle $. In addition to the SG defined in Definition \ref{def:sg},  POSGs add the following terms:
  	 \begin{itemize}
  	 	\item   	 
  $\mathbb{O}^i$: 	 an observation set for each agent $i$. The joint observation set is defined as $\pmb{\mathbb{O}}:= \mathbb{O}^1\times \cdots \times \mathbb{O}^N$. 
  	 \item $O: S \times \pmb{\mathbb{A}} \rightarrow \Delta(\pmb{\mathbb{O}})$: an observation function $O(\bm{o}| \bm{a}, s')$ denotes the probability of observing  $\bm{o} \in \pmb{\mathbb{O}}$ given the action $\bm{a} \in   \pmb{\mathbb{A}}$, and the new state $s' \in \mathbb{S}$ from the environment transition. 
  	 \end{itemize}
Each agent's policy now changes to $\pi^i \in \Pi^i: \mathbb{O} \rightarrow \Delta(\mathbb{A}^i)$. 
\label{def:posg}
  \end{definition}

Although the added partial-observability constraint is common in practice for many real-world applications, theoretically it exacerbates the difficulty of solving SGs. Even in the simplest setting of a two-player fully cooperative finite-horizon game, solving a POSG is $NEXP$-hard (see Figure \ref{fig:complexity}), which means it requires super-exponential time to solve in the worst-case scenario \cite{bernstein2002complexity}.  However, the benefits of studying games in the partially observable setting come from the algorithmic advantages. 
Notably, recent advancements in weakly-revealing POSGs \cite{liu2022sample} show that even when different players have their own local observations, these games can still be provably learned with polynomial sample complexity. This mitigates some of the computational challenges traditionally associated with POSGs, making learning more tractable in certain structured settings.
Centralized-training-with-decentralized-execution methods \cite{oliehoek2016concise, lowe2017multi, foerster2017counterfactual,  rashid2018qmix, yang2020multi} have achieved many empirical successes, and together with DNNs, they hold great promise.  

 A POSG is one of the most general classes of games. An important subclass of POSGs is decentralized partially observable MDP (Dec-POMDP), where all agents share the same reward. Formally, this scenario is defined as follows:
  \begin{definition}[Dec-POMDP]
A Dec-POMDP is a special type of POSG defined in Definition \ref{def:posg} with $R^1=R^2=\cdots=R^N$.
\label{def:short-decpomdp}
\end{definition}

Dec-POMDPs are related to single-agent MDPs through the partial observability condition, and they are also related to stochastic team games through the assumption of identical rewards. In other words, versions of both single-agent MDPs and stochastic team games are particular types of Dec-POMDPs (see Figure \ref{fig:games_types}). 

\begin{definition}[Special types of Dec-POMDPs]
The following games are special types of Dec-POMDPs. 
\begin{itemize}
\item \textbf{partially observable MDP (POMDP)}: there is only one agent of interest, $N=1$. This scenario is equivalent to a single-agent MDP in Definition \ref{def:mdp} with a partial-observability constraint. 
\item \textbf{decentralized MDP (Dec-MDP)}: the agents in a Dec-MDP have joint full observability. That is, if all agents share their observations, they can recover the state of the Dec-MDP unanimously. Mathematically, we have $\forall \bm{o} \in \pmb{\mathbb{O}}, \exists s \in \mathbb{S}$  such that $\mathbb{P}(S_t=s | \pmb{\mathbb{O}}_t= \bm{o})=1$. 
 \item \textbf{fully cooperative stochastic games}: assuming each agent has full observability, $\forall i=\{1,...,N\},  \forall o^i \in {O}^i, \exists s \in \mathbb{S}$  such that $\mathbb{P}(S_t=s  | \mathbb{O}_t= o^i)=1$. The fully-cooperative SG from Definition \ref{def:sgs_types} is a type of Dec-POMDP. 
\end{itemize}
\label{def:decpomdp}
\end{definition}

We conclude  Section \ref{sec:marl} by presenting the relationships between the many different types of POSGs through a Venn diagram in Figure \ref{fig:games_types}.

\section{Extensive-Form Game}\label{extensive}
\subsection{Problem Formulation}
An SG assumes that a game is represented as a large table in each stage where the rows and columns of the table correspond to the actions of the two players\footnote{A multi-player game is represented as a high-dimensional tensor in an SG.}. 
Based on the big table, SGs model the situations in which agents act simultaneously and then receive their rewards.
Nonetheless, for many real-world games, players take actions alternately. 
Poker is one class of games in which who plays first has a critical role in the players' decision-making process. Games with alternating actions are naturally described by an extensive-form game (EFG) \cite{osborne1994course,von1945theory} through a tree structure. 
Recently, \citet{kovavrik2019rethinking} has made a significant contribution in unifying the framework of EFGs and the framework of POSGs.

\begin{figure}[htbp]
\centering
\includegraphics[width=.9\linewidth]{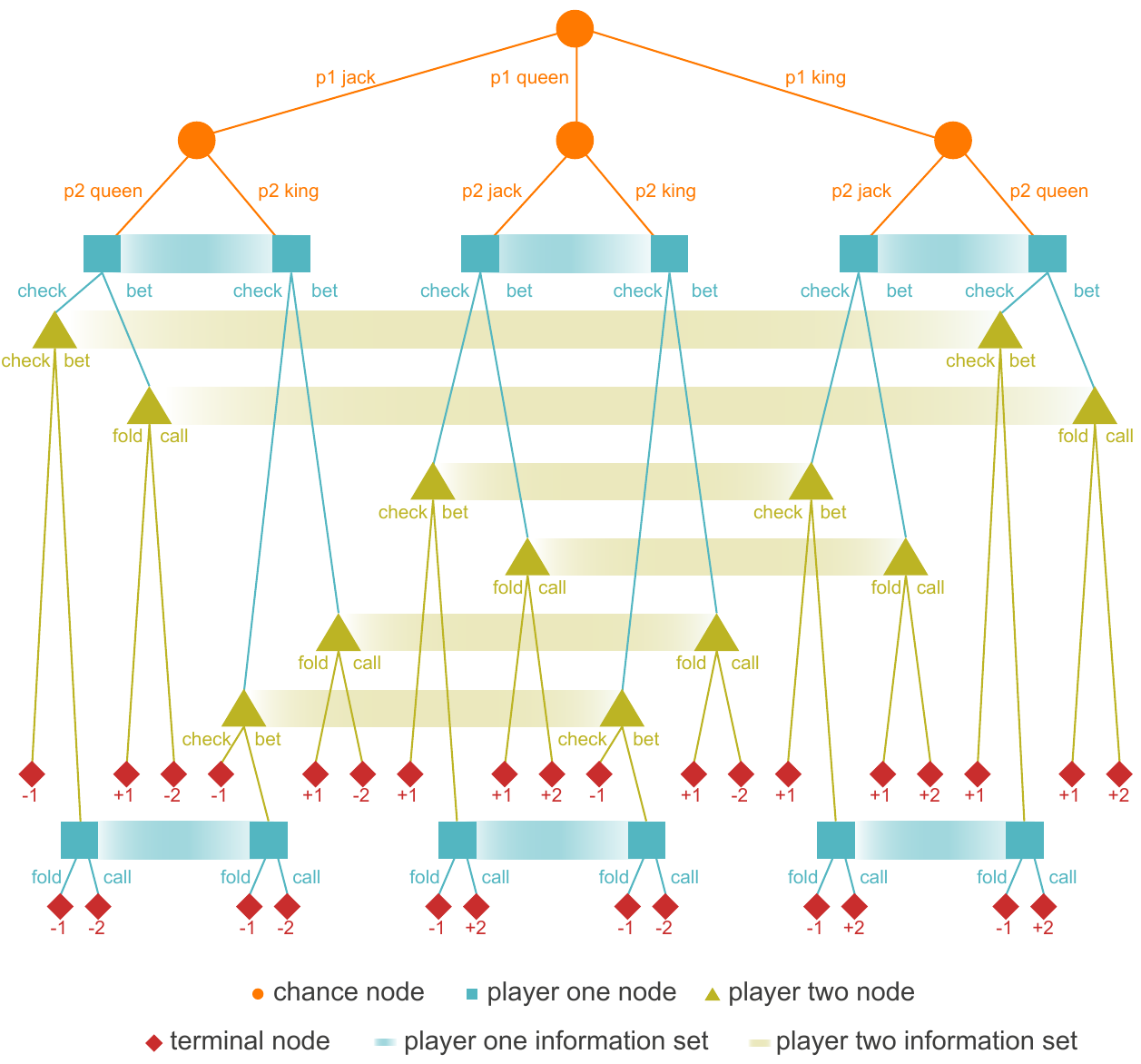}
\caption{Game tree of two-player Kuhn poker. Each node (i.e., circles, squares and rectangles) represents the choice of one player, each edge represents a possible action, and the leaves (i.e., diamond) represent final outcomes over which each player has a reward function (only player one's reward is shown in the graph since Kuhn poker is a zero-sum game). Each player can observe only their own card; for example, when player one holds a Jack, it cannot tell whether player two is holding a Queen or a King, so the choice nodes of player one in each of the two scenarios stay within the same information set.  }
\label{fig:kuhn}
\end{figure}

Figure \ref{fig:kuhn} shows the game tree of two-player Kuhn poker \cite{kuhn1950simplified}. In Kuhn poker, the dealer has three cards, a King, Queen, and Jack (King$>$Queen$>$Jack),  each player is dealt one card (the orange nodes in Figure \ref{fig:kuhn}), and the third card is put aside unseen. The game then develops as follows.    
\begin{itemize}
\item {\color{turq}Player one} acts first; he/she can \emph{check} or \emph{bet}.   
\item \quad If {\color{turq}player one} \emph{checks}, then {\color{amber}{player two}} decides to \emph{check} or \emph{bet}. 	
\item \qquad If {\color{amber}{player two}} \emph{checks}, then the higher card wins $1\$$  from the other player. 
\item \qquad If {\color{amber}{player two}} \emph{bets}, then {\color{turq}player one} can \emph{fold} or \emph{call}.
\item \qquad \quad If {\color{turq}player one} \emph{folds}, then {\color{amber}{player two}} wins $1\$$ from {\color{turq}{player one}}. 
\item  \qquad \quad If {\color{turq}player one} \emph{calls}, then the higher card wins $2\$$  from the other player. 
\item \quad If {\color{turq}player one} \emph{bets}, then {\color{amber}{player two}} decides to \emph{fold} or \emph{call}. 	
\item \qquad If {\color{amber}{player two}} \emph{folds}, then {\color{turq}player one} wins $1\$$ from {\color{amber}{player two}}. 
\item \qquad If {\color{amber}{player two}} calls, then the higher card wins $2\$$  from the other player.  
\end{itemize}

An important feature of EFGs is that they can handle imperfect information for multi-player decision-making. 
In the example of Kuhn poker, the players do not know which card the opponent holds. 
However, unlike Dec-POMDP, which also models imperfect information in the SG setting but is intractable to solve,   EFG, represented in an equivalent sequence form, can be solved by an LP in polynomial time in terms of game states \cite{koller1992complexity}. In the next section, we first introduce EFG and then consider the sequence form of EFG.

\begin{definition}[Extensive-form Game]
An (imperfect-information) EFG can be described by a tuple of key elements, which is written as

$ \langle N, \mathbb{A}, \mathbb{H}, \mathbb{T},  \{R^i\}_{i\in\{1,...,N\}}, \chi, \rho, P, \{\mathbb{S}^i\}_{i\in\{1,...,N\}} \rangle$.
\begin{itemize}
\item $N$: the number of players. Some EFGs involve a special player called ``chance'', which has a fixed stochastic policy that represents the randomness of the environment. For example, the chance player in Kuhn poker is the dealer, who distributes cards to the players at the beginning.  
	\item $\mathbb{A}$: the (finite) set of all agents' possible actions.
	\item $\mathbb{H}$: the (finite) set of non-terminal choice nodes. 
	\item $\mathbb{T}$: the (finite) set of terminal choice nodes, disjoint from $\mathbb{H}$.
	\item $\chi: \mathbb{H} \rightarrow 2^{|\mathbb{A}|}$ is the action function that assigns a set of valid actions to each choice node.
	\item $\rho: \mathbb{H} \rightarrow \{1,...,N\} $ is the player indicating function that assigns, to each non-terminal node, a player who is due to choose an action at that node.
	\item $P: \mathbb{H}\times \mathbb{A} \rightarrow \mathbb{H} \cup \mathbb{T} $ is the transition function that maps a choice node and an action to a new choice/terminal node such that $\forall h_1, h_2 \in \mathbb{H}$ and $\forall a_1, a_2 \in \mathbb{A}$,  if $P(h_1, a_1) = P(h_2, a_2)$, then $h_1 = h_2$ and $a_1= a_2$.
	\item $R^i: \mathbb{T} \rightarrow \mathbb{R}$ is a real-valued reward function for player $i$ on the terminal node. Kuhn poker is a zero-sum game since $R^1 + R^2 = 0$.
	\item $\mathbb{S}^i$:  a set of equivalence classes/partitions  $\mathbb{S}^i=(S^i_1, ... , S^i_{k^i})$ for agent $i$  on $\{h \in \mathbb{H}: \rho(h)=i\}$ with the property that $\forall j \in \{1,...,k^i\}, \forall h, h' \in S^i_j$, we have $\chi(h) = \chi(h')$ and $\rho(h) = \rho(h')$. The set $S^i_j$ is also called an \textbf{information state}. The physical meaning of the information state is that the choice nodes of an information state are indistinguishable. In other words, the set of valid actions and agent identities for the choice nodes within an information state are the same; one can thus use $\chi(S^i_j), \rho(S^i_j)$ to denote $\chi(h), \rho(h), \forall h \in S^i_j$. 
	
\end{itemize}	
\label{def:efg}
\end{definition}

	The inclusion of the information sets in EFG helps to model the imperfect information cases in which players have only partial or no knowledge about their opponents. In the case of Kuhn poker, each player can only observe their own card. For example, when player one holds a Jack, it cannot tell whether player two is holding a Queen or a King, so the choice nodes of player one under each of the two scenarios (Queen or King) stay within the same information set. 
	 Perfect-information EFGs (e.g., go or chess) are a special case where the information set is a singleton, i.e., $|S^i_j|=1, \forall j$, 	so a choice node can be equated to the unique history that leads to it. 
	Imperfect-information EFGs (e.g., Kuhn poker or Texas hold'em) are those in which there exists $i, j$ such that $|S^i_j| \ge 1$,  so the information state can represent more than one possible history. However, with the assumption of perfect recall (described later), the history that leads to an information state is still unique.

 Recent studies have explored the problem of steering no-regret learning agents towards desirable equilibria in EFG. Some works address the challenge of using nonnegative payments to guide players toward predetermined outcomes. A study shows that steering is impossible if the mediator's total budget remains finite, regardless of the game format, but demonstrates that vanishing average payments can still facilitate steering. Additionally, constant per-round payments allow steering when players' full strategies are observable at each round. In contrast, when only trajectories through the game tree are observable, steering is generally infeasible with constant per-round payments in EFG, though it remains possible in normal-form games or when the per-round budget can grow with time, maintaining vanishing average payments. \cite{zhang2023steering,zhang2023steering2}

\subsection{Normal-Form Representation}

A (simultaneous-move) NFG can be equivalently transformed into an imperfect-information EFG\footnote{This transformation is not unique, but they share the same equilibria as the original game. 
This transformation makes the payoff matrix very large, and the actual scalability of many optimization algorithms for large-scale games is often limited by the large payoff matrix of the game. Some researchers have proposed acceleration methods to improve solving scalability \cite{farina2022fast}.
Moreover, this transformation from NFG to EFG does not hold for perfect-information EFGs. } \cite{shoham2008multiagent} [Chapter 5].
Specifically, since the choices of actions by other agents are unknown to the central agent, this could potentially lead to different histories (triggered by other agents) that can be aggregated into one information state for the central agent.   

In the other direction, an imperfect-information EFG can also be transformed into an equivalent NFG in which the pure strategies  of each agent $i$ are defined by the Cartesian product  
$\prod_{S^i_j\in\mathbb{S}^i}\chi(S^i_j)$, 
which is a complete specification\footnote{One subtlety of the pure strategy is that it designates a decision at each choice node, regardless of whether it is possible to reach that node given the other choice nodes.} of which action to take at every information state of that agent. 
In the Kuhn poker example, one pure strategy for player one can be check-bet-check-fold-call-fold;  altogether, player one has $2^6=64$ pure strategies, corresponding to $3 \times 2^3=24$ pure strategies for the chance node and $2^6=64$ pure strategies for player two. 
The mixed strategy of each player is then a distribution over all its pure strategies. 
In this way, the NE in NFG in Eq. (\ref{eq:best_response}) can still be applied to the EFG, and the NE of an EFG can be solved in two steps: first, convert the EFG into an NFG; second, solve the NE of the induced NFG by means of the Lemke-Howson algorithm \cite{shapley1974note}. 
 If one further restricts the action space to be state-dependent and adopts the discounted accumulated reward at the terminal node, then the EFG recovers to an SG.
 While the NE of an EFG can be solved through its equivalent normal form, the computational benefit can be achieved by dealing with the extensive form directly; this motivates the adoption of the sequence-form representation of EFGs. 
   
\subsection{Sequence-Form Representation}
\label{sec:sequence-form}
Solving EFGs via the NFG representation, though universal, is inefficient because the size of the induced NFG is exponential in the number of information states. 
In addition, the NFG representation does not consider the temporal structure of games. 
One way to address these problems is to operate on the sequence form of the EFG, also known as the realization-plan representation, the size of which is only linear in the number of game states and is thus exponentially smaller than that of the NFG. 
Importantly, this approach enables polynomial-time solutions to EFGs \cite{koller1992complexity}. 

In the sequence form of EFGs, the main focus shifts from mixed strategies to  \emph{behavioral strategies} in which, rather than randomizing over complete pure strategies, the agents randomize independently at each information state $S^i \in \mathbb{S}^i$, i.e., $\pi^i: \mathbb{S}^i \rightarrow \Delta \big(\chi( S^i)\big)$.   
With the help of behavioral strategies, the key insight of the sequence form is that rather than building a player's strategy around the notion of pure strategies that can be exponentially many, one can build the strategy based on the paths in the game tree from the root to each node. 

In general, the expressive power of behavioral strategy and mixed strategy are non-comparable. 
However, if the game has \emph{perfect recall}, which intuitively\footnote{More formally, on the path from the root node to a decision node $h \in S^i_t$ of player $i$, list in chronological order which information sets of $i$ were encountered, i.e.,  $S^i_t \in \mathbb{S}^i$,  and what action player $i$ took at that information set, i.e., $a^i_t \in \chi(S^i_t)$. If one calls this list of $(S^i_0, a^i_0, ..., S^i_{t-1}, a^i_{t-1}, S^i_{t})$ the \emph{experience} of player $i$ in reaching node $h \in S^i_t$, then the game has perfect recall if and only if, for all players, any nodes in the same information set have the same experience. In other words, there exists one and only one experience that leads to each information state and the decision nodes in that information state; because of this,  all perfect-information EFGs are games of perfect recall.} means that each agent remembers all his historical moves in different information states precisely, then the behavioral strategy and mixed strategy are somehow equivalent. 
Specifically, suppose all choice nodes in an information state share the same history that led to them (otherwise the agent can distinguish between the choice nodes). In that case, the well-known Kuhn's theorem \cite{kuhn1950extensive} guarantees that the expressive power of behavioral strategies and that of mixed strategies coincide in the sense that they induce the same probability of outcomes for games of perfect recall. 
As a result, the set of NE does not change if one considers only behavioral strategies. In fact, 
the sequence-form representation is primarily useful for describing  imperfect-information EFGs of perfect recall, written as:

\begin{definition}[Sequence-form Representation]
The sequence-form representation of an imperfect-information EFG, defined in Definition \ref{def:efg},  of perfect recall  is described by a tuple of elements written as:
\begin{align}
(N, \boldsymbol\Sigma, \{G^i\}_{i\in\{1,...,N\}},  \{\pi^i\}_{i\in\{1,...,N\}}, \{\mu^{\pi^i}\}_{i\in\{1,...,N\}}, \{C^i\}_{i\in\{1,...,N\}})
\end{align}
where 
\begin{itemize}
	\item $N$: the number of agents, including the chance node, if any, denoted by $c$.
	\item $\boldsymbol{\Sigma} = \prod_{i=1}^{N} \Sigma^i$: where $\Sigma^i$ is the set of sequences available to agent $i$. A sequence of actions of player $i$, $\sigma^i \in \Sigma^i$, defined by a choice node $h \in \mathbb{H} \cup \mathbb{T}$, is the ordered set of player $i$'s actions that have been taken from the root to node $h$. Let $\varnothing$ be the sequence that corresponds to the root node.  
	 
	Note that other players' actions are not part of agent $i$'s sequence.  In the example of Kuhn poker,  $\Sigma^c=\{\varnothing, \ $Jack, Queen, King, Jack-Queen, Jack-King, Queen-Jack, Queen-King, King-Jack, King-Queen$\}$, $\Sigma^1=\{\varnothing,$ check, bet, check-fold, check-bet$\}$, and $\Sigma^2=\{\varnothing, \ \text{check, bet, fold, call}\}$. 
	\item $\pi^i$: $\mathbb{S}^i \rightarrow \Delta \big(\chi( S^i)\big)$ is the behavioral policy that assigns a probability of taking a valid action $a^i \in \chi(S^i)$ at an information state $S^i \in \mathbb{S}^i$.  This policy randomizes independently over different information states.  In the example of Kuhn poker, each player has six information states; their behavioral strategy is therefore a list of six independent probability distributions. 
	
	\item $\mu^{\pi^i}$: $\Sigma^i \rightarrow [0, 1]$ is the realization plan that provides the realization probability, i.e., $\mu^{\pi^i}(\sigma^i)=\prod_{c\in\sigma^i}\pi^i(c)$,  that a sequence $\sigma^i \in \Sigma^i$ would arise under a given behavioral policy $\pi^i$ of player $i$. In the Kuhn poker case, the realization probability that player one chooses the sequence of check and then fold is $ \mu^{\pi^1}(\text{check-fold})= \pi^1(\text{check}) \times \pi^1(\text{fold}) $. 
	
	Based on the realization plan, one can recover the underlying behavioral strategy\footnote{Empirically, it is often the case that working on the realization plan of a behavioral strategy is more computationally friendly than working on the behavioral strategy directly.} (an idea similar to Eq. (\ref{eq:occupancy_policy})). To do so, we need three additional pieces of notation.  Let $\mathsf{Seq}: \mathbb{S}^i \rightarrow \Sigma_i$ return the sequence $\sigma^i \in \Sigma^i$ that leads to a given information state $S^i \in \mathbb{S}^i$. Since the game assumes perfect recall, $\mathsf{Seq}(S^i)$ is known to be unique. Let $\sigma^i a^i$ denote a sequence that consists of the sequence $\sigma^i$ followed by the single action $a^i$. Since there are many possible actions $a^i$ to choose, let $\mathsf{Ext}: \Sigma^i \rightarrow 2^{\Sigma^i}$ denote the set of all possible sequences that extend the given sequence by taking one additional action. It is trivial to see that sequences that include a terminal node cannot be extended, i.e., $\mathsf{Ext}(T)=\emptyset$. Finally, we can write the behavioral policy $\pi^i$ for an information state $S^i$ as
	\begin{align}
		&\pi^i\big(a^i \in \chi(S^i)\big) = \dfrac{\mu^{\pi^i}\big(\mathsf{Seq}(S^i)a^i\big)}{\mu^{\pi^i}\big(\mathsf{Seq}(S^i)\big)}, \nonumber \\ 
        &\forall S^i \in \mathbb{S}^i, \ \forall  \big(\mathsf{Seq}(S^i)a^i\big) \in \mathsf{Ext}\big(\mathsf{Seq}(S^i)\big).
		\label{eq:behavorial_policy}
	\end{align}

	\item $G^i: \boldsymbol{\Sigma} \rightarrow \mathbb{R}$ is the reward function for agent $i$ given by $G^i(\boldsymbol{\sigma}) = R^i(T)$ if a terminal node $T \in \mathbb{T}$ is reached when each player plays their part of the sequence in $\boldsymbol{\sigma} \in \boldsymbol{\Sigma}$, and $G^i({\boldsymbol\sigma})=0$ if non-terminal nodes are reached. 
	
	Note that since each payoff that corresponds to a terminal node is stored only once in the sequence-form representation (due to the perfect recall, each terminal node has only one sequence that leads to it), compared to the normal-form representation, which is a  Cartesian product overall information sets for each agent and is thus exponential in size, the sequence form is only linear in the size of the EFG. In the example of Kuhn poker, the normal-form representation is a tensor with $64 \times 64 \times 32$ elements, while in the sequence-form representation, since there are $30$ terminal nodes and each node has only one unique sequence leading to it, the payoff tensor has only $30$ elements (plus $\varnothing$ for each player).  
	\item $C^i$: is a set of linear constraints on the realization probability of $\mu^{\pi^i}$. Under the notations of $\mathsf{Seq}$ and  $\mathsf{Ext}$ defined in the bullet points of $\mu^{\pi^i}$, we know the realization plan must meet the condition that 
	\begin{align}
	  \mu^{\pi^i}\big(\varnothing\big) &= 1, \ \ \  \mu^{\pi^i}\big(\sigma^i\big) \ge 0, \ \ \ \forall \sigma^i \in \Sigma^i \nonumber \\
		 \mu^{\pi^i}\Big(\mathsf{Seq}(S^i)\Big) &= \sum_{\sigma^{i} \in \mathsf{Ext}\big(\mathsf{Seq}(S^i)\big)} \mu^{\pi^i}\big(\sigma^i\big), \ \  \forall S^i \in \mathbb{S}^i. 
		\label{eq:realizaiton_constraints}
	\end{align}
	The first constraint requires that $\mu^{\pi^i}$ is a proper probability distribution. 
	In addition, the second constraint in Eq. (\ref{eq:realizaiton_constraints}) indicates that in order for a realization plan to be valid to recover a behavioral strategy, at each information state of agent $i$, the probability of reaching that information state must equal the summation of the realization probabilities of all the extended sequences. In the example of Kuhn poker, we have $C^1$ for player one by $\mu^{\pi^1}(\text{check}) = \mu^{\pi^1}(\text{check-fold}) + \mu^{\pi^1}(\text{check-call})$. 
\end{itemize}
\label{def:sequene-form}
\end{definition}
\subsection{Solving Extensive-Form Games}
\label{sec:efg-solve}
Based on the definition of EFG above and its different forms of representation,
we will primarily introduce some solutions for EFGs in this section. More detailed discussions about advanced algorithms for solving EFGs are provided later in Sec.~\ref{sec:efg_alg}.
In the sequence-form EFG,  
given a joint (behavioral) policy $\boldsymbol{\pi}=(\pi^1, ..., \pi^N)$, we can write the realization probability of agents reaching a terminal node $T\in \mathbb{T}$, assuming the sequence that leads to the node $T$ is $\boldsymbol{\sigma}_T$, in which each player, including the chance player, follows its own path $\sigma^i_T$  as 
\begin{equation}
\mu^{\boldsymbol{\pi}}\big(\boldsymbol{\sigma}_T\big) = \prod_{i \in \{1,...,N\}} \mu^{\pi^i}\big(\sigma^i_T\big).
\end{equation}
The expected reward for agent $i$, which covers all possible terminal nodes following the joint policy $\boldsymbol{\pi}$,  is thus given by Eq. (\ref{eq:efg-reward}). 
\begin{equation}
R^i(\boldsymbol \pi) = \sum_{T\in\mathbb{T}}\mu^{\boldsymbol{\pi}}\big(\boldsymbol{\sigma}_T\big)\cdot G^i(\boldsymbol \sigma_T) =  \sum_{T\in\mathbb{T}}\mu^{\boldsymbol{\pi}}\big(\boldsymbol{\sigma}_T\big)\cdot R^i(T).
\label{eq:efg-reward}
\end{equation}
 If we denote the expected reward by $R^i(\boldsymbol \pi)$ for simplicity, then the solution concept of NE for the EFG can be written as
\begin{equation}
	R^i(\pi^{i,*}, \pi^{-i,*}) \ge 	R^i(\pi^{i}, \pi^{-i,*}), \ \ \  \text{for any policy $\pi^i$ of agent $i$ and for all $i$}.
\end{equation}

\subsubsection{Perfect-Information Games}

Every finite perfect-information EFG has a pure-strategy NE \cite{zermelo1913application}. Since players take turns and every agent sees everything that has occurred thus far, it is unnecessary to introduce randomness or mixed strategies into the action selection. 
However, the NE can be too weak of a solution concept for the EFG. In contrast to that in NFGs, the NE in EFGs can represent \emph{non-credible threats}, which represent the situation where the Nash strategy is not executed as claimed if agents truly reach that decision node.  A refinement of the NE in the perfect-information EFG is a \emph{subgame-perfect equilibrium} (SPE). The SPE rules out non-credible threats by picking only the NE that is the best response at every subgame of the original game. 

The fundamental principle in solving the SPE is \emph{backward induction}, which identifies the NE from the bottom-most subgame and assumes those NE will be played as considered increasingly large trees. Specifically, 
backward induction can be implemented through a depth-first search algorithm on the game tree, which requires time that is only linear in the size of the EFG. 
In contrast, finding NE in NFG is known to be $PPAD$-hard, let alone the NFG representation is exponential in the size of an EFG.

In the case of two-player zero-sum EFGs, backward induction needs to propagate only a single payoff from the terminal node to the root node in the game tree. Furthermore,  due to the strictly opposing interests between players, one can further \emph{prune} the backward induction process by recognizing that certain subtrees will never be reached in NE, even without examining those subtree nodes\footnote{This occurs,  for example, in the case that the worst case of one player in one subgame is better than the best case of that player in another subgame.}, which leads to the well-known  Alpha-Beta-Pruning algorithm \cite[Chapter 5.1]{shoham2008multiagent}. For games with very deep game trees, such as Chess or GO, a common approach is to search only nodes up to certain depths and use an approximate value function to estimate those nodes' value without rolling outing to the end \cite{silver2016mastering}.  

Finally, backward induction can identify one NE in linear time; yet, it does not provide an effective way to find all NE. 
A theoretical result suggests that finding all NE in a two-player perfect-information EFG (not necessarily zero-sum) requires $\mathcal{O}(|\mathbb{T}|^3)$, which is still tractable \cite[Theorem 5.1.6]{shoham2008multiagent}. 

\subsubsection{Imperfect-Information  Games}

By means of the sequence-form representation, one can write the solution of a two-player EFG as an LP. Given a fixed behavioral strategy of player two, in the form of realization plan $\mu^{\pi^2}$, the best response for player one can be written as 
\[\max_{\mu^{\pi^1}} \sum_{\sigma^{1} \in \Sigma^{1}}\mu^{\pi^1}\left(\sigma^{1}\right) \left(\sum_{\sigma^{2} \in \Sigma^{2}} g^{1}\left(\sigma^{1}, \sigma^{2}\right) \mu^{\pi^2}\left(\sigma^{2}\right)\right)  \]
subject to the constraints in Eq. (\ref{eq:realizaiton_constraints}). In NE, player one and player two form a mutual best response. However, if we treat both $\mu^{\pi^1}$ and $\mu^{\pi^2}$ as variables, then the objective becomes nonlinear. The key to address this issue is to adopt the dual form of the LP \cite{koller1996finding}, which is written as
\begin{align}\min & \ \  v_{0} \nonumber \\ \text {s.t. } & v_{\mathcal{I}\left(\sigma^{1}\right)}-\sum_{I' \in \mathcal{I}\big(\mathsf{Ext}(\sigma^{1})\big)} v_{I'} \geq \sum_{\sigma^{2} \in \Sigma^{2}} g^{1}\left(\sigma^{1}, \sigma^{2}\right)  \mu^{\pi^2}\left(\sigma^{2}\right), \quad \forall \sigma^{1} \in \Sigma^{1}
\label{eq:dual_br}
\end{align}
where $\mathcal{I}: \Sigma^i \rightarrow \mathbb{S}^i$ is a mapping function that returns the information set\footnote{Recall that this information set is unique under the assumption of perfect recall.} encountered when the final action in $\sigma^i$ was taken. With slight abuse of notation, we let $\mathcal{I}\big(\mathsf{Ext}(\sigma^{1})\big)$\footnote{Recall that $\mathsf{Ext}(\sigma^{1})$ is the set of all possible sequences that extend $\sigma^1$ one step ahead.} denote the set of final information states encountered in the set of the extension of $\sigma^i$.  
The variable $v_0$ represents, given $\mu^{\pi^2}$, player one's expected reward under its own realization plan $\mu^{\pi^1}$ , and $v_{I'}$ can be considered as the part of this expected utility in the subgame starting from information state $I'$. Note that the constraint needs to hold for every sequence of player one. 

In the dual form of best response in Eq. (\ref{eq:dual_br}), if one treats $\mu^{\pi^2}$ as an optimizing variable rather than a constant, which means $\mu^{\pi^2}$ must meet the requirements in Eq. (\ref{eq:realizaiton_constraints})  to be a proper realization plan, then the LP formulation for a two-player zero-sum EFG can be written as follows. 
\begin{align}\min & \ \  v_{0} \label{eq:zero-sum_efg_obj} \\ \text {s.t. } & v_{\mathcal{I}\left(\sigma^{1}\right)}-\sum_{I' \in \mathcal{I}\big(\mathsf{Ext}(\sigma^{1})\big)} v_{I'} \geq \sum_{\sigma^{2} \in \Sigma^{2}} g^{1}\left(\sigma^{1}, \sigma^{2}\right)  \mu^{\pi^2}\left(\sigma^{2}\right), \quad \forall \sigma^{1} \in \Sigma^{1} \label{eq:zero-sum_efg_part1} \\
	 \mu^{\pi^2}& \big(\varnothing\big) = 1, \ \ \  \mu^{\pi^2}\big(\sigma^2\big) \ge 0, \ \ \ \forall \sigma^2 \in \Sigma^2  \\
		 \mu^{\pi^2} & \Big(\mathsf{Seq}(S^2)\Big) = \sum_{\sigma^{2} \in \mathsf{Ext}\big(\mathsf{Seq}(S^2)\big)} \mu^{\pi^2}\big(\sigma^2\big), \ \  \forall S^2 \in \mathbb{S}^2. 
\label{eq:zero-sum_efg}
\end{align}
Player two's realization plan is now selected to minimize player one's expected utility. 
Based on the minimax theorem \cite{von1945theory}, we know this process will lead to a NE. 
Notably, though the zero-sum EFG and zero-sum SG (see the formulation in Eq. (\ref{eq:opt_zero_sum_prime}))  both adopt the LP formulation to solve the NE and can be solved in polynomial time, the size of the representation for the game itself is very different.  
If one chooses first to transform the EFG into an NFG presentation and then solve it by LP, then the time complexity would in fact become exponential in the size of the original EFG. 

The solution to a two-player general-sum EFG can also be formulated using an approach similar to that used for the zero-sum EFG. 
The difference is that there will be no objective function such as Eq. (\ref{eq:zero-sum_efg_obj}) since in the general-sum context, one agent's reward can no longer be determined based on the other player's reward.
The LP with only Eqs. (\ref{eq:zero-sum_efg_part1} - \ref{eq:zero-sum_efg})  thus becomes a constraint satisfaction problem. Specifically, one would need to repeat Eqs. (\ref{eq:zero-sum_efg_part1} - \ref{eq:zero-sum_efg}) twice to consider each player independently. 
One final subtlety required in solving the two-player general-sum EFG is to ensure $v^1$ and $v^2$ are bounded\footnote{Since the constraints are linear, they remain satisfied when both $v^1$ and $v^2$ are increased by the same constant to any arbitrarily large values.},  a \emph{complementary slackness condition} must be further imposed; we have $\forall \sigma^{1} \in \Sigma^{1}$ (vice versa $ \forall \sigma^2 \in \Sigma^2$ for player two): 
\begin{equation}
\scriptstyle
\mu^{\pi^1}(\sigma^1) \bigg[\Big(v^1_{\mathcal{I}\left(\sigma^{1}\right)}-\sum_{I' \in \mathcal{I}\big(\mathsf{Ext}(\sigma^{1})\big)} v^1_{I'} \Big) - \Big( \sum_{\sigma^{2} \in \Sigma^{2}} g^{1}\left(\sigma^{1}, \sigma^{2}\right)  \mu^{\pi^2}\left(\sigma^{2}\right)\Big)\bigg]=0.  
\label{eq:efg-slack}
\end{equation}
The above condition indicates that for each player, either the sequence $\sigma^i$ is never played, i.e.,  $\mu^{\pi^i}(\sigma^i)=0$, or all sequences that are played by that player with positive probability must induce the same expected payoff such that $v^i$  takes arbitrarily large values, thus being bounded.  
Eqs. (\ref{eq:zero-sum_efg_part1} - \ref{eq:zero-sum_efg}), together with Eq. (\ref{eq:efg-slack}),  turns the solution to the NE into an LCP problem that can be solved by the generalized Lemke-Howson method \cite{lemke1964equilibrium}. Although in the worst case, polynomial time complexity cannot be achieved, as can for zero-sum games, this approach is still exponentially faster than running the Lemke-Howson method to solve the NE in a normal-form representation. 

 For a perfect-information EFG, recall that the SPE is a more informative solution concept than NE. Extending SPE to the imperfect-information scenario is therefore valuable.  
 However, such an extension is non-trivial because a well-defined notion of a subgame is lacking.
 However, 
for  EFGs with perfect recall, the intuition of subgame perfection can be effectively extended to a new solution concept, named the sequential equilibrium (SE) \cite{kreps1982reputation}, which is guaranteed to exist and coincides with the SPE if all players in the game have perfect information. We refer the readers to Chapter~\ref{chap:zero-sum} for discussions about advanced algorithms for solving EFGs.

Beyond defining equilibrium concepts, a crucial line of research has focused on optimizing the sample complexity required to learn these equilibria. Recent research has focused on optimizing the sample complexity of learning Nash equilibria (NE) in imperfect-information games (IIGs). \cite{kozuno2021learning} introduces the Implicit Exploration Online Mirror Descent (IXOMD) algorithm for two-player zero-sum IIGs, achieving a high-probability convergence rate of \( O(1/\sqrt{T}) \), where \( T \) is the number of games played. This model-free algorithm performs updates along sampled trajectories, ensuring computational efficiency. Building on this, \cite{bai2022near} presents the Balanced Online Mirror Descent (BOMD) and Balanced Counterfactual Regret Minimization (B-CFR) algorithms, which improve the sample complexity to \( \tilde{O}((XA+YB)/\epsilon^2) \), where \( X, Y \) are the number of information sets and \( A, B \) are the number of actions. This improves over previous bounds and matches the information-theoretic lower bound up to logarithmic factors. \cite{fiegel2023adapting} provides a problem-independent lower bound of \( \tilde{O}(H(A\mathcal{X} + B\mathcal{Y})/\epsilon^2) \) for learning optimal strategies in zero-sum IIGs. They propose two FTRL algorithms: Balanced FTRL, which matches this lower bound but requires prior knowledge of the information set structure, and Adaptive FTRL, which adapts the regularization dynamically, achieving \( \tilde{O}(H^2(A\mathcal{X} + B\mathcal{Y})/\epsilon^2) \). These work indicate how to improve the efficiency of IIG algorithms and promote their practical application.

Recent work has further expanded the scope of equilibrium concepts in EFGs, particularly by introducing new classes of equilibria such as communication and certification equilibria \cite{zhang2020small}. These models augment the game with a mediator capable of sending and receiving messages to and from the players, with the added capability of remembering the messages. In contrast to standard correlated equilibrium, the optimality of these new equilibrium notions can be computed in polynomial time, even in the context of extensive-form games. Some researchers explore these equilibrium concepts computationally, showing that the complexity of solving such equilibria is often driven by the mediator's imperfect recall \cite{zhang2022polynomial}. Their work also generalizes previous algorithms, including the polynomial-time solution for Bayes-Nash equilibria in automated mechanism design \cite{conitzer2004self}, and the correlation DAG algorithm for optimal correlation \cite{zhang2022optimal}. By leveraging a mediator-augmented game model, they provide a scalable framework for computing these equilibria, particularly in settings where players cannot lie about their information but may remain silent, which they define as full-certification equilibria. This framework not only enhances our understanding of the computational landscape of EFGs but also offers practical insights for both automated decision-making and game-theoretic analysis in multi-agent systems.

In contrast, the problem of finding an optimal extensive-form correlated equilibrium (CCE) remains computationally intractable. While CCE is a natural extension of correlated equilibrium to extensive-form games, it is NP-hard to compute an optimal CCE, particularly when aiming to maximize welfare or achieve other desirable outcomes. This hardness stems from the complex dependencies between players' strategies and the mediator's actions, which are difficult to manage in the context of imperfect recall and the intricate structure of extensive-form games. As a result, unlike the communication and certification equilibrium models, the computation of an optimal CCE requires significantly more sophisticated methods and remains a major challenge in game-theoretic research \cite{zhang2022optimal}. Thus, while recent advances have made significant strides in providing polynomial-time algorithms for some equilibrium concepts, the computational complexity of finding optimal correlated equilibria in extensive-form games continues to be a fundamental obstacle in the field.

\chapter{Grand Challenges of MARL}
\label{sec:challenge}
Compared to single-agent RL, 
multiagent RL is a general framework that better matches the broad scope of real-world AI applications. However, due to the existence of multiple agents that learn simultaneously, MARL methods pose more theoretical challenges, in addition to those already present in single-agent RL.  
Compared to classic MARL settings where there are usually two agents, solving a many-agent RL problem is even more challenging. 
As a matter of fact, \circlen1
\textbf{combinatorial complexity}, \circlen2 \textbf{multi-dimensional learning objectives}, and  \circlen3 \textbf{the issue of non-stationarity} all result in the majority of MARL algorithms being capable of solving games with  \circlen4 \textbf{only two players}, in particular, two-player zero-sum games.  
In this section, I will elaborate on each of the grand challenges in many-agent RL. 

\section{Combinatorial Complexity}
\label{sec:combinatorial}

In the context of multiagent learning, each agent has to consider the other opponents' actions when determining the best response; this characteristic is deeply rooted in each agent's reward function and for example is represented by the joint action $\bm{a}$ in their Q-function $Q^i(s, \bm{a})$ in Eq. (\ref{eq:ma_q_learning}). 
The size of the joint action space,  $|\mathbb{A}|^N$, grows exponentially with the number of agents and thus largely constrains the scalability of MARL methods. 
Furthermore, the combinatorial complexity is worsened by the fact that solving a NE in game theory is $PPAD$-hard, even for two-player games. Therefore, for multi-player general-sum games (neither team games nor zero-sum games), it is non-trivial to find an applicable solution concept.

One common way to address this issue is by assuming specific factorized structures on action dependency such that the reward function or Q-function can be significantly simplified. 
For example, a graphical game assumes an agent's reward is affected by only its neighboring agents, as defined by the graph from \cite{kearns2007graphical}. This assumption leads to a polynomial-time solution for the computation of a NE in specific tree graphs \cite{kearns2013graphical}, though the scope of applications is limited beyond this specific scenario. 

Recent progress has also been made toward leveraging particular neural network architectures for Q-function decomposition \cite{sunehag2018value, rashid2018qmix, yang2020multi}. 
In addition to the fact that these methods work only for the team-game setting, the majority of them lack theoretical backing. There remain open questions to answer, such as understanding the representational power (the approximation error) of the factorized Q-functions in a multiagent task and how factorization itself can be learned from scratch.

\section{Multi-Dimensional Learning Objectives}
\label{sec:many_obj}
Compared to single-agent RL, where the only goal is to maximize the learning agent's long-term reward, the learning goals in  MARL are naturally multi-dimensional, as the objective of all agents are not necessarily aligned by one metric.
\cite{bowling2001rational,bowling2002multiagent} proposed to classify the goals of the learning task into two types: \textbf{rationality} and \textbf{convergence}. Rationality ensures an agent takes the best possible response to the opponents when they are stationary, and convergence ensures the learning dynamics eventually lead to a stable policy against a given class of opponents.  Reaching both rationality and convergence gives rise to reaching the NE. 
 
In terms of rationality, 
 the NE characterizes a fixed point of a joint optimal strategy profile from which no agents would be motivated to deviate as long as they are all perfectly rational. 
However, in practice, an agent's rationality can easily be bound by either cognitive limitations and/or the tractability of the decision problem. 
In these scenarios, the rationality assumption can be relaxed to include other types of solution concepts, such as the recursive reasoning equilibrium, which results from modelling the reasoning process recursively among agents with finite levels of hierarchical thinking (for example, an agent may reason in the following way: I believe that you believe that I believe ...) \cite{wen2018probabilistic, wen2019modelling}; best response against a target type of opponent \cite{powers2005new}; the mean-field game equilibrium, which describes multiagent interactions as a two-agent interaction between each agent itself and the population mean \cite{guo2019learning, yang2018mean, yang2018learning}; evolutionary stable strategies, which describe an equilibrium strategy based on its evolutionary advantage of resisting invasion by rare emerging mutant strategies  \cite{maynard1972evolution,tuyls112005evolutionary, tuyls2007evolutionary, bloembergen2015evolutionary}; Stackelberg equilibrium \cite{zhang2019bi}, which assumes specific sequential order when agents take decisions; and the robust equilibrium  (also called the trembling-hand perfect equilibrium in game theory), which is stable against adversarial disturbance  \cite{li2019robust,goodfellow2014explaining, yabu2007multiagent}.

In terms of convergence, although most MARL algorithms are contrived to converge to the NE, the majority either lack a rigorous convergence guarantee \cite{zhang2019multi}, potentially converge only under strong assumptions such as the existence of a unique NE \cite{littman2001value, hu2003nash}, or are provably non-convergent in all cases \cite{mazumdar2019policy}. 
\cite{zinkevich2006cyclic} identified the non-convergent behavior of value-iteration methods in general-sum SGs and instead proposed an alternative solution concept to the NE -- \emph{cyclic equilibria} -- that value-based methods converge to.  
The concept of no regret (also called the Hannan consistency in game theory \cite{hansen2003reducing}), measures convergence by comparison against the best possible strategy in hindsight. This was also proposed as a new criterion to evaluate convergence in zero-sum self-plays  \cite{bowling2005convergence, hart2001reinforcement, zinkevich2008regret}. 
In two-player zero-sum games with a non-convex non-concave loss landscape (training GANs \cite{goodfellow2014generative}), gradient-descent-ascent methods are found to reach a Stackelberg equilibrium \cite{lin2019gradient,fiez2019convergence} or a local differential NE \cite{mazumdar2019finding} rather than the general NE.

Finally,  although the above solution concepts account for convergence, building a convergent objective for MARL methods with DNNs remains an uncharted area. This is partly because the global convergence result of a single-agent deep RL algorithm, for example, neural policy gradient methods \cite{wang2019neural,liu2019neural} and neural TD learning algorithms \cite{cai2019neural}, has not been extensively studied yet.

\section{The Challenges of Non-Stationarity and Adaptive Opponents}
The most well-known challenge of multiagent learning versus single-agent learning is probably the non-stationarity issue. 
Since multiple agents concurrently improve their policies according to their own interests, from each agent's perspective, the environmental dynamics become non-stationary and challenging to interpret when learning. This problem occurs because the agent itself cannot tell whether the state transition -- or the change in reward -- is an actual outcome due to its own action or if it is due to its opponent's explorations.
Although learning independently by completely ignoring the other agents can sometimes yield surprisingly powerful empirical performance \cite{papoudakis2020comparative, matignon2012independent},  
this approach essentially harms the stationarity assumption that supports the theoretical convergence guarantee of single-agent learning methods \cite{tan1993multi}. As a result, the Markovian property of the environment is lost, and the state occupancy measure of the stationary policy in Eq. (\ref{eq:occupancy}) no longer exists. For example, the convergence result of single-agent policy gradient methods in MARL is provably non-convergent in simple linear-quadratic games \cite{mazumdar2019finding}.

The non-stationarity issue can be further aggravated by TD learning, which occurs with the replay buffer that most deep RL methods currently adopt \cite{lin1992self,foerster2017stabilising}.    
In single-agent TD learning (see Eq. (\ref{eq:dqn})), the agent bootstraps the current estimate of the TD error, saves it in the replay buffer, and samples the data in the replay buffer to update the value function. 
In the context of multiagent learning, since the value function for one agent also depends on other agents' actions, the bootstrap process in TD learning also requires sampling other agents' actions, which leads to two problems.
First, the sampled actions barely represent the full behaviour of other agents' underlying policies across different states. Second, an agent's policy can change during training, so the samples in the replay buffer can quickly become outdated. Therefore, the dynamics that yielded the data in the agent's replay buffer must be constantly updated to reflect the current dynamics in which it is learning. 
This process further exacerbates the non-stationarity issue.

In general, the non-stationarity issue forbids the reuse of the same mathematical tool for analysing single-agent algorithms in the multiagent context.  However, one exception exists: the identical-interest game in Definition \ref{def:sgs_types}. 
In such settings, each agent can safely perform selfishly without considering other agents' policies since the agent knows the other agents will also act in their own interest. The stationarity is thus maintained, so single-agent RL algorithms can still be applied. 

Beyond non-stationarity, an additional challenge in competitive games arises due to the adaptivity of opponents \cite{he2016opponent}, who may actively exploit weaknesses in the learner’s policy. Unlike in cooperative settings, where all agents aim for shared goals, in competitive environments, adversarial opponents can adjust their strategies dynamically to counter the learner’s decisions \cite{ganzfried2011game, schadd2007opponent}. This means that a learning agent must not only improve its own policy but also anticipate and react to an ever-changing opponent strategy, making the learning process even more unstable. In extreme cases, naive learning algorithms can be systematically exploited, leading to poor performance or even divergence.


\section{\texorpdfstring{Scalability Issue when \( N \gg 2 \)}{Scalability Issue when N >> 2}}

Combinatorial complexity, multi-dimensional learning objectives, and the issue of non-stationarity all result in the majority of MARL algorithms being capable of solving games with only two players, in particular, two-player zero-sum games \cite{zhang2019multi}. 
As a result, solutions to general-sum settings with more than two agents (for example, the many-agent problem) remain an open challenge.
This challenge may be addressed from all three perspectives of multiagent intelligence: game theory, which provides realistic and tractable solution concepts to describe learning outcomes of a many-agent system; RL algorithms, which offer provably convergent learning algorithms that can reach stable and rational equilibria in the sequential decision-making process; and finally deep learning techniques, which provide the learning algorithms with expressive function approximators.

\begin{table}[t]
\caption{Common assumptions on the level of local knowledge made by MARL algorithms. }
\label{tb:information_set}
\centering
\resizebox{\columnwidth}{!}{ 
\begin{sc}
	\begin{tabular}{ c|l|l }
	\toprule
	Categories & \textbf{Assumptions} & \textbf{Examples} \\
	 \midrule
	 0 & Each agent observes the reward of his selected action. & Bandit \\
	 1 & Each agent observes the rewards of all possible actions. & Blackjack, Full-information Bandit \\
	 2 &  Each agent observes others' selected actions. & Chess, Texas Hold 'em \\
	 3 &  Each agent observes others' reward values. & Texas Hold 'em, Street Fighter \\
	 4  & Each agent knows others' exact policies. & ? \\
	 5  & Each agent knows others' exact reward functions.  & Tic-Tac-Toe, Rock-Paper-Scissor \\
	 6  & Each agent knows the equilibrium of the stage game. & Prisoner's Dilemma, Rock-Paper-Scissor \\
	\bottomrule
	\end{tabular}
	\end{sc}
 }
\end{table}

\chapter{A Survey of MARL Surveys}
\label{sec:survey}

In this section, we provide a non-comprehensive review of MARL algorithms. To begin, we introduce different taxonomies that can be applied to categorise prior approaches. Given multiple high-quality, comprehensive  surveys on MARL methods already exist, a survey of those surveys is provided. 
Based on the proposed taxonomy, we review related MARL algorithms, covering works on identical interest games, zero-sum games, and games with an infinite number of players.   
This section is written to be selective, focusing on the algorithms that have theoretical guarantees and less focus on those with only empirical success or those that are purely driven by specific applications.  

\section{Taxonomy of  MARL  Algorithms}
\label{sec:taxonomy}

One significant difference between the taxonomy of single-agent RL algorithms and MARL algorithms is that in the single-agent setting, since the objective is single and consistent (i.e. in terms of reward maximization), the taxonomy is driven mainly by the type of solution \cite{kaelbling1996reinforcement, li2017deep}, for example, model-free vs model-based, on-policy vs off-policy, TD learning vs Monte-Carlo methods. By contrast, in the multiagent setting, due to the existence of multiple learning objectives (see Section \ref{sec:many_obj}), the taxonomy is driven mainly by the type of problem rather than the solution. In fact, asking the right question for MARL algorithms is itself a research problem, which is referred to as the problem problem \cite{balduzzi2018re, shoham2007if}. 

\paragraph{Based on Stage Games Types.} Since the solution concept varies considerably according to the game type, 
one principal component of the MARL taxonomy is the nature of stage games. 
A common division\footnote{Such a division is complementary because any multi-player normal-form game can be decomposed into  a potential game \cite{monderer1996potential} plus a harmonic game \cite{candogan2011flows} (also see Definition \ref{def:sgs_types}); in the two-player case, it corresponds to a team game plus a zero-sum game.} includes team games (more generally, potential games), zero-sum games (more generally, harmonic games), and  a mixed setting of the two games, namely, general-sum games. 
Other types of ``exotic'' games, such as mean-field games \cite{lasry2007mean}, that originate from non-game-theoretical research domains exist and have recently attracted tremendous attention.  
Based on the type of stage game, the taxonomy can be further enriched by how many times they are played. A repeated game is where one stage game is played repeatedly without considering the state transition, and the learning cost in the learnability framework \cite{conitzer2003bl} is measured by the losses the learning agent accrues (rather than the number of rounds) \cite{conitzer2003bl}.
An SG is a sequence of stage games, which can be infinitely long, with the order of the games to play determined by the state-transition probability. 
Since solving a general-sum SG is at least $PSPACE$-hard \cite{conitzer2008new}, MARL algorithms usually have a clear boundary on what types of games they can solve. 
For general-sum games,  there are few MARL algorithms that have a provable convergence guarantee without strong, even unrealistic,   assumptions (e.g., the NE  is unique) \cite{shoham2007if, zhang2019multi}.    

\paragraph{Based on Level of Local Knowledge.}

The assumption on the level of local knowledge, i.e., what agents can and cannot know during training and execution time, is another major component to differentiate MARL algorithms.  
Having access to different levels of local knowledge leads to different local behaviours by agents and  various levels of difficulty in developing  theoretical analysis. 

One can think of some simplest cases by extending the bandit problems in online learning literature to multiagent settings, where each agent can observe the reward of its selected action or rewards of all possible actions. The latter one is more informative than the former. Beyond observing the rewards for selected actions, knowing the entire reward function can be even more informative (i.e., access to rewards for unselected actions and other agents' rewards). Apart from the reward functions, the actions of the other agents can be observable to the current agent in perfect information games, which provides additional information for it to make decisions. Knowledge of the agents' exact policy/reward function forms is a much stronger assumption than being able to observe their sampled actions/rewards. Therefore, being aware of other agents' policies are usually unlikely for complex games in practice, with examples like Texas Hold 'em and Chess. There are also extreme cases on the two ends. One is the case that the agent can observe nothing apart from itself, and another is that the agent knows the equilibrium point, i.e., the direct answer of the game. 
  
\paragraph{Based on Learning Paradigms.}

  \begin{figure}[t]
\centering
\includegraphics[width=.95\linewidth]{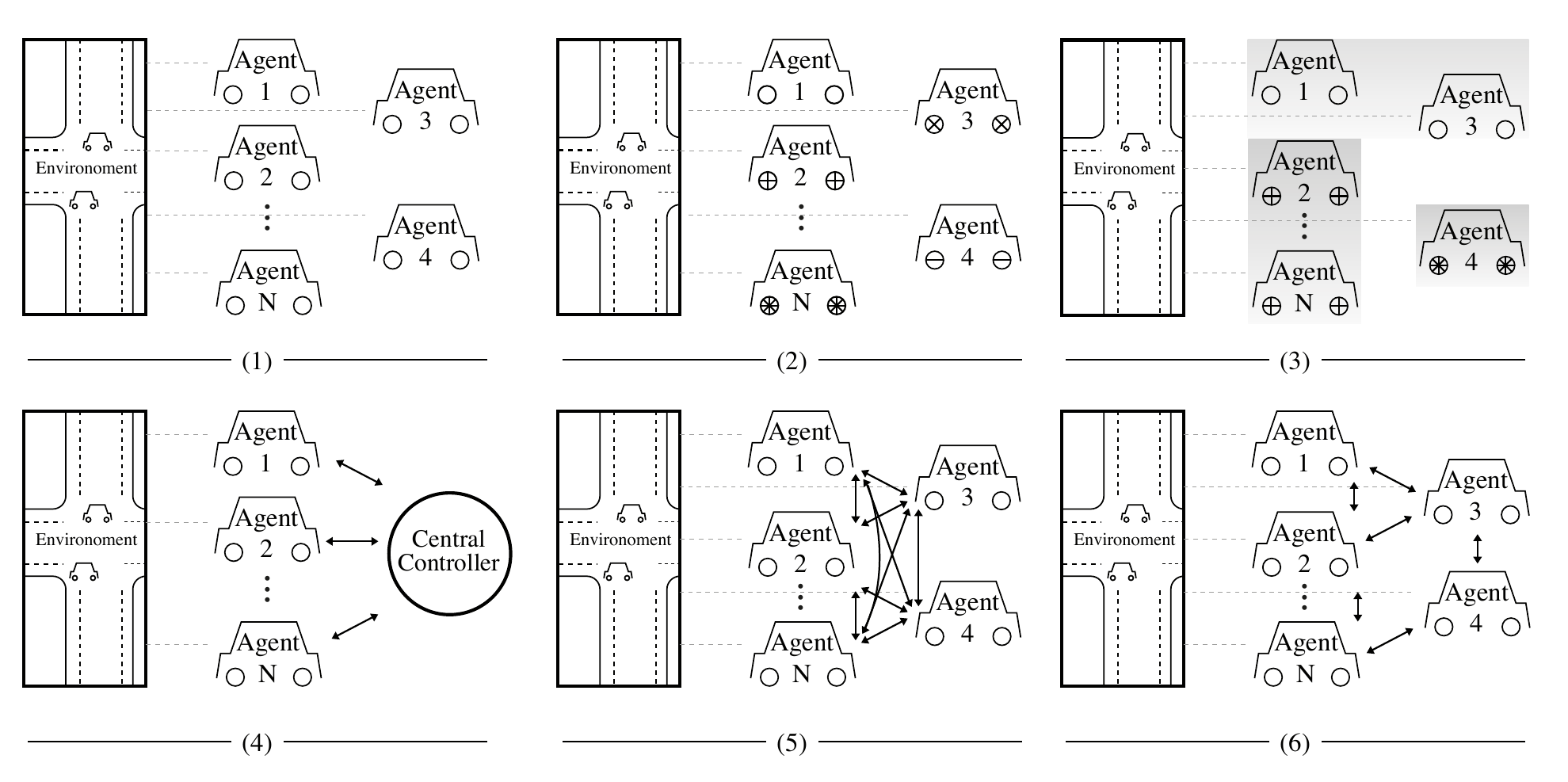}
\caption{Common learning paradigms of MARL algorithms. (1) Independent learners with shared policy. (2) Independent learners with independent policies (i.e., denoted by the difference in wheels). (3) Independent learners with shared policy within a group. (4) One central controller controls all agents: agents can exchange information with any other agents at any time. (5)  Centralised training with decentralised execution (CTDE): only during training, agents can exchange information with others; during execution, they act independently. (6) Decentralised training with networked agents: during training, agents can exchange information with their neighbours in the network; during execution, they act independently. }
\label{fig:mas_paradigm}
\end{figure}

In addition to various levels of local knowledge,  MARL algorithms can be classified based on the learning paradigm, as shown in Figure \ref{fig:mas_paradigm}. For example, 
the $4$th learning paradigm addresses multiagent problems by building  a single-agent controller, which takes the joint information from all agents as inputs and outputs the joint policies for all agents. In this paradigm, agents can exchange any information with any other opponent through the central controller. 
The information that can be exchanged depends on the assumptions about the  level of local knowledge describe in the last paragraph.
The $5$th paradigm allows agents to exchange information with  other agents only during training; during execution, each agent has to act in a decentralised manner, making decisions based on its own observations only. 
The $6$th category can be regarded as a particular case of paradigm $5$ in that agents are assumed to be interconnected via a (time-varying) network such that information can still spread across the whole network if agents communicate  with their neighbours.   
The most general case is paradigm $2$, where  agents are fully decentralised, with no information exchange of any kind allowed at any time, and each agent executes its own policy. 
Relaxation of paradigm $2$ yields the $1$st and the $3$rd paradigm, where the agents, although they cannot exchange  information, share a single set of policy parameters, or, within a pre-defined group, share a single set of policy parameters.

\paragraph{Based on Five AI Agendas.}

In order for MARL researchers to be specific about the problem being addressed and the associated evaluation criteria, \cite{shoham2007if, Sandholm07:Perspectives}  identified five coherent agendas for MARL studies, each of which has a clear motivation and success criterion.  
Though proposed more than a decade ago, these five distinct goals are still useful in evaluating and categorising  recent contributions.  
We, therefore, choose to incorporate the five agendas into the taxonomy of MARL algorithms, as described in Table~\ref{tb:five-agendas}. 

\begin{table}[htbp]
\caption{Summary of the five agendas for multiagent learning research  \cite{shoham2007if, Sandholm07:Perspectives}. }
\label{tb:five-agendas}
\centering
\small
	\begin{tabularx}{\linewidth}{ c|p{3.7cm} X }
	\toprule
	ID & \textbf{Agenda} & \textbf{Description}  \\
	 \midrule
	 1 & Computational  & To develop efficient methods that can find solutions (strategy profiles) to games according to given solution concepts. Examples: \cite{berger2007brown,leyton2005local}\\ \midrule 
	 2 &  Descriptive  & To develop formal models of learning that agree with the behaviors of people/animals/organizations. Examples: \cite{erev1998predicting,camerer2002sophisticated}\\ \midrule
	 3 &  Learning algorithms in equilibrium & To ensure that learning strategies themselves form an equilibrium, where no agent can benefit by unilaterally deviating. This addresses challenges such as unknown payoff matrices and the fragility of equilibria when agents fail to follow prescribed strategies. For example, in a repeated prisoner's dilemma, agents using learning algorithms that converge to a mutual cooperation equilibrium would achieve higher payoffs than those deviating to always-defect strategies.\\ \midrule
	 4  & Prescriptive, cooperative & To develop distributed learning algorithms for games with intra-team collaboration characteristics.  In this agenda, there is rarely a role for equilibrium analysis since the agents have no motivation to deviate from the prescribed algorithm. Examples: \cite{claus1998dynamics2} \\ \midrule
	 5  &  Prescriptive, noncooperative & To develop effective methods for obtaining a ``high reward'' in a given environment, for example, an environment with a  selected class of opponents. Examples: \cite{powers2005learning, powers2005new}\\ \midrule
	\bottomrule
	\end{tabularx}
\end{table}

The following sections are organized around the five agendas in Table~\ref{tb:five-agendas}. Specifically, the focus will be on investigating how game-theoretic models can be constructed in both cooperative and noncooperative scenarios, with an emphasis on designing equilibrium learning algorithms. These discussions aim to provide a comprehensive understanding of strategy selection and optimization in multi-agent systems, drawing on both theoretical foundations and practical applications.

\section{A  Survey of Surveys}

A multiagent system (MAS) is a generic concept that could refer to many different domains of research across different academic subjects;
general overviews are given by    \cite{weiss1999multiagent},  \cite{wooldridge2009introduction}, and \cite{shoham2008multiagent}. 
Due to the many possible ways of categorising multiagent (reinforcement) learning algorithms, it is impossible to have a single survey that includes all relevant works considering all directions of categorisations. 
In the past two decades, there has been no lack of survey papers that summarise the current progress of specific categories of multiagent learning research.
In fact, there are  so many that these surveys themselves deserve a comprehensive review. 
Before proceeding to review MARL algorithms based on the proposed taxonomy in Section \ref{sec:taxonomy}, in this section, I provide an overview of  relevant surveys that study multiagent systems from the machine learning, in particular, the RL, perspective.

 One of the earliest studies that surveyed MASs in the context of machine learning/AI was published by \cite{stone2000multiagent}: the research works up to that time were summarised into  four major scenarios considering whether agents were homogeneous or heterogeneous and whether or not agents were allowed to communicate with each other.  
 \cite{shoham2007if} considered the  game theory and RL perspective and introspectively asked the  question of ``if  multiagent learning is the answer, what is the question?''. Upon failing to find a single answer,    \cite{shoham2007if} proposed the famous five AI agendas for future research work to address.
\cite{stone2007multiagent}  tried to answer Shoham's question by emphasising that  MARL can be more broadly framed than through game theoretic terms, and he noted that how to apply the MARL technique remains an open question, rather than being an answer, in contrast to the suggestion of  \cite{shoham2007if}.
The survey  of \cite{tuyls2012multiagent} also reflected on Stone's viewpoint;  they believed that the entanglement of only RL and game theory is too narrow in its conceptual  scope, and MARL should  embrace other ideas, such as transfer learning \cite{taylor2009transfer}, swarm intelligence \cite{kennedy2006swarm}, and co-evolution \cite{tuyls2007evolutionary}.

\cite{panait2005cooperative} investigated  the cooperative MARL setting; instead of considering only reinforcement learners, they reviewed learning algorithms based on the division of \emph{team learning} (i.e., applying a single learner to search for the optimal joint behaviour for the whole team) and \emph{concurrent learning} (i.e., applying one learner per agent), which includes broader areas of evolutionary computation, complex systems, etc. 
\cite{matignon2012independent} surveyed the solutions for fully-cooperative games only; in particular, they  focused on evaluating independent RL solutions powered by Q-learning and its many variants. Due to the fact that Q-learning does not require an environmental model and can be used online, it is very suitable for repeated games with unknown opponents \cite{Sandholm96:Multiagent}.
\cite{jan2005overview} conducted an overview with a similar scope; moreover, they extended the work to include  fully competitive games in addition to fully cooperative games.
\cite{bucsoniu2010multi}, to the best of my knowledge, presented the first comprehensive survey on MARL  techniques, covering both value iteration-based and policy search-based methods, together with their strengths and weaknesses. In their survey, they considered not only fully cooperative or competitive games but also the effectiveness of different algorithms in the general-sum setting.    
\cite{nowe2012game}, in the $14$th chapter, addressed the same topic as \cite{bucsoniu2010multi} but with a much narrower coverage of multiagent RL algorithms.

\cite{tuyls112005evolutionary} and \cite{bloembergen2015evolutionary} both surveyed the dynamic models that have been derived for various MARL algorithms and revealed the deep connection between evolutionary game theory and MARL methods. We refer to Table 1 in \cite{tuyls112005evolutionary} for a summary of this connection.

\cite{hernandez2017survey} provided a different perspective on the taxonomy of how existing MARL algorithms cope with the issue of non-stationarity induced by opponents. On the basis of the opponent and environment characteristics, they  categorised the MARL algorithms according to the type of opponent modelling. 

\cite{da2019survey} introduced a new perspective of reviewing MARL algorithms based on how knowledge is reused, i.e., transfer learning. Specifically, they grouped the surveyed algorithms into \emph{intra-agent} and \emph{inter-agent} methods, which correspond to the reuse of knowledge from experience gathered from the agent itself and that acquired from other agents, respectively.

Most recently, deep MARL techniques  have received considerable attention. \cite{nguyen2020deep}  surveyed how deep  learning techniques were used to address the challenges in multiagent learning, such as partial observability, continuous state and action spaces, and transfer learning.  
\cite{oroojlooyjadid2019review} reviewed  the application of deep MARL techniques in fully cooperative games: the survey  on this setting is thorough. 
\cite{hernandez2019survey} summarised how the classic ideas from  traditional MAS research, such as emergent behaviour, learning communication, and opponent modelling, were incorporated into deep MARL domains, based on which they proposed a new categorisation for deep MARL methods. 
\cite{zhang2019multi} performed a selective survey on MARL algorithms that have theoretical convergence guarantees and complexity analysis. To the best of my knowledge, their review is the only one to cover  more advanced topics such as decentralised MARL with networked agents, mean-field MARL, and MARL for stochastic potential games.   

On the application side, \cite{muller2014application} surveyed $152$ real-world applications in various sectors powered by MAS techniques. 
\cite{campos2017multiagent} reviewed the application of multiagent techniques for automotive industry applications, such as traffic coordination and route balancing.  
\cite{derakhshan2019review} focused on real-world applications for wireless sensor networks,  
\cite{shakshuki2015multi} studied multiagent applications for the healthcare industry, and 
\cite{kober2013reinforcement} investigated the application of robotic control and summarised profitable RL approaches that can be applied to robots in  the real world.

\chapter{Learning in Identical-Interest Games}
The majority of MARL algorithms assume that agents collaborate with each other to achieve shared goals. 
In this setting, agents are usually considered homogeneous and play an interchangeable role in the environmental dynamics. In a two-player normal-form game or repeated game, for example, this means the payoff matrix is symmetrical.

\section{Stochastic Team Games}
One benefit of studying identical interest  games is that single-agent RL algorithms with a theoretical guarantee can be safely applied. 
For example, in the team game\footnote{The terms Markov team games, stochastic team games, and dynamic team games are interchangeably used across different domains of the literature.} setting, since all agents' rewards are always the same, the Q-functions are identical among all agents. As a result, one can simply apply the single-agent RL algorithms over the joint action space $\bm{a} \in \pmb{\mathbb{A}}$, equivalently, Eq. (\ref{eq:ma_solve_sg}) can be written as
\begin{equation}
		\mathbf{eval}^i \Big( \big\{Q^i(s_{t+1}, \cdot) \big\}_{i\in\{1,...,N\}} \Big) = V^{i}\Big(s_{t+1}, \arg \max_{\bm{a}\in{\pmb{\mathbb{A}}}  }Q^i\big( s_{t+1}, \bm{a} \big)\Big).
		\label{eq:coop_equilibrium}
\end{equation}
\cite{littman1994markov} first studied this approach in SGs. 
However, one issue with this approach is that when multiple equilibria exist (e.g., a normal-form game with reward ${ R = \big[ \begin{array}{cc} 0, 0 & 2,2 \\ 2,2 & 0,0 \end{array} \big]}$), unless the selection process is coordinated among agents, the agents' optimal policy can end up with a worse scenario even though their value functions have reached the optimal values. 
To address this issue, \cite{claus1998dynamics} proposed to build belief models about other agents' policies.  Similar to fictitious play  \cite{berger2007brown}, each agent chooses actions in accordance with its belief about the other agents. Empirical effectiveness, as well as convergence, have been reported for repeated games;
however, the convergent equilibrium may not be optimal. 
In solving this problem,  
\cite{wang2003reinforcement} proposed optimal adaptive learning (OAL) methods that provably converge to the optimal NE almost surely in any team SG. 
The main novelty of OAL is that it learns the game structure by building 
so-called \emph{weakly acyclic games} that eliminate all the joint actions with sub-optimal NE values and then applies adaptive play \cite{young1993evolution} to address the equilibrium selection problem for weakly acyclic games specifically.
Following this approach, 
\cite{arslan2016decentralized} proposed decentralised Q-learning algorithms that, under the help of two-timescale analysis \cite{leslie2003convergent},  converge to an equilibrium policy for weakly acyclic SGs. 
To avoid sub-optimal equilibria for weakly acyclic SGs, 
 \cite{yongacoglu2019learning} further  refined the decentralised Q-learners and derived theorems with stronger almost-surely convergence guarantees for optimal  policies.

\subsubsection{Solutions via Q-function Factorisation}
\label{sec:q-factor}

Another vital reason that team games have been repeatedly studied is that solving team games is a crucial step in building distributed AI (DAI) \cite{huhns2012distributed, gasser2014distributed}. 
The logic is that if each agent only needs to maintain the Q-function of $Q^i({s, a^i})$, which depends on the state and local action $a^i$, rather than joint action $\bm{a}$, then the combinatorial nature of multiagent problems can be avoided. 
Unfortunately, \cite{tan1993multi} previously noted that such independent Q-learning methods do not converge in team games.
\cite{lauer2000algorithm} reported similar negative results; however, when the state transition dynamics are deterministic, independent learning through distributed  Q-learning  can still obtain a convergence guarantee.  No additional expense is needed in comparison to the non-distributed case for computing the optimal policies.    

Factorised MDPs \cite{boutilier1999decision} are an effective way to avoid exponential blowups. 
For a coordination task, if the joint-Q function can be naturally written as \[Q=Q^1(a^1, a^2) + Q^2(a^2, a^4) + Q^3(a^1, a^3) + Q^4(a^3, a^4), \] then the nested structure can be exploited. For example, $Q^1$ and $Q^3$ are irrelevant in finding the optimal $a^4$; thus, given $a^4$, $Q^1$ becomes irrelevant for optimising $a^3$. Given $a^3, a^4$, one can then optimise $a^1, a^2$.     
Inspired by this result, 
\cite{guestrin2002coordinated,guestrin2002multiagent,kok2004sparse} studied the idea of \emph{coordination graphs}, which combine value function approximation with a message-passing scheme by which agents can efficiently find the globally optimal joint action.

However, the coordination graph may not always be available in real-world applications; thus, the ideal approach is to let agents  learn the Q-function factorisation  from the tasks automatically.  Deep neural networks are an effective way to learn such factorisations.  Specifically, the scope of the problem is then narrowed to the so-called \emph{decentralisable tasks} in the Dec-POMDP setting, that is, $\exists \big\{Q_i\big\}_{i \in \{1,...,N\}}$  $\forall  \bm{o} \in \pmb{\mathbb{O}}, \bm{a}\in \pmb{\mathbb{A}}$,  the following condition holds. 
\begin{equation}
\label{eq:igm}
\arg \max _{\bm{a}} Q^{\bm{\pi}}(\bm{o}, \bm{a})=\left[\begin{array}{c}{\arg \max _{a^{1}} Q^{1}(o^1, a^{1})} \\ {\vdots} \\ {\arg \max _{a^{N}} Q^{N}(o^N, a^{N})}\end{array} \right].  
\end{equation}
Eq. (\ref{eq:igm})   suggests that a task is  decentralisable only if  the
local maxima on the individual value function per every agent  amounts to the global maximum on the joint value function. 
Different structural constraints, enforced by particular neural architectures,  have been proposed to  satisfy this condition. For example, 
VDN \cite{sunehag2018value} maintains an additivity structure by making  $Q^{\bm{\pi}}(\bm{o}, \bm{a}):=\sum_{i=1}^{N}Q^i(o^i, a^i)$.   QMIX \cite{rashid2018qmix} adopts a monotonic structure  by means of a mixing network to ensure   $\frac{\partial Q^{\bm{\pi}}(\bm{o}, \bm{a})}{\partial Q^i(o^i, a^i)} \ge 0, \forall i \in \{1,...,N \}$.
QTRAN \cite{son2019qtran}  introduces a more rigorous  learning   objective on top of QMIX that proves to be a sufficient condition for Eq. (\ref{eq:igm}). 
However, these structure constraints heavily depend on specially designed neural architectures, which makes understanding the  representational power (i.e., the approximation error) of the above methods almost infeasible. 
Another drawback  is that the structure constraint also damages agents' efficient exploration during training.  
To mitigate these issues, \cite{yang2020multi} proposed Q-DPP, which eradicates the structure constraints by approximating the Q-function through a \emph{determinantal point process (DPP)} \cite{kulesza2012determinantal}. DPP pushes agents to explore and acquire diverse behaviours; consequently, it leads to natural decomposition of the joint Q-function with no need for \emph{a priori} structure constraints.   
In fact, VDN/QMIX/QTRAN  prove to be the exceptional cases of Q-DPP. 

\subsubsection{Solutions via Multiagent Soft Learning}
\label{sec:pr2}

\begin{figure}[t!]
     \centering
\includegraphics[width=.65\textwidth]{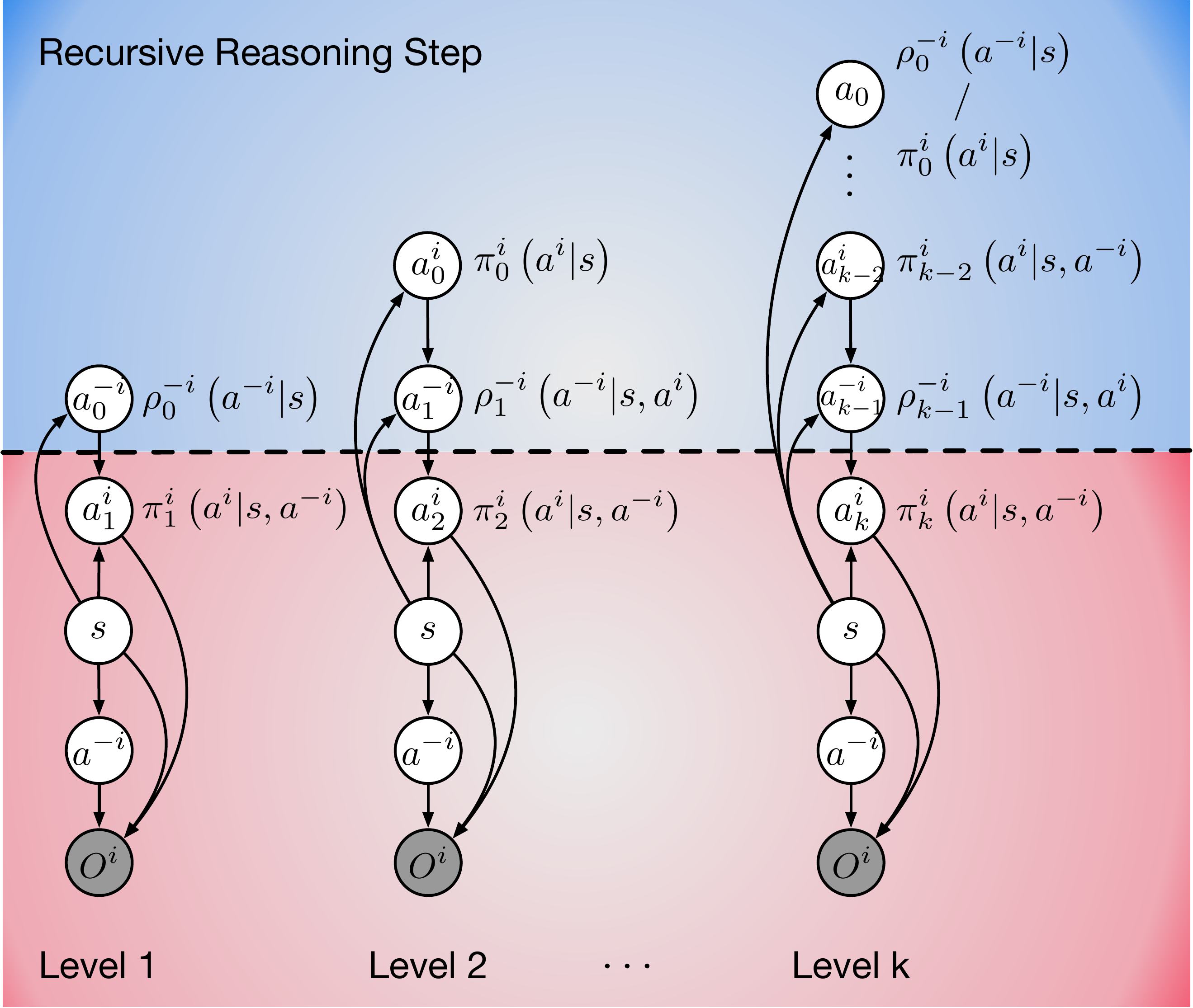}
     \caption{Graphical model of the $level$-$k$ reasoning model \cite{wen2019modelling}. The red part is the equivalent graphical model for the multiagent learning problem. The blue part corresponds to the recursive reasoning steps. 
     Subscript $a_*$  stands for the level of thinking, not the time step. The  opponent policies are approximated by $\rho^{-i}$. The omitted $level$-$0$ model  considers opponents that are fully randomised. Agent $i$ rolls out the recursive reasoning about opponents in its mind (blue area). In the recursion, agents with higher-level beliefs take the best response to the lower-level agents. The higher-level models conduct all the computations that the lower-level models have done, e.g., the $level$-$2$ model contains the $level$-$1$ model by integrating out $\pi_{0}^i(a^i|s)$. }
\label{fig:chk}
\end{figure}

In single-agent RL, the process of finding the optimal policy can be equivalently transformed into a probabilistic inference problem on a graphical model \cite{levine2018reinforcement}. 
The pivotal insight is that by introducing an additional binary  random variable $P(\mathcal{O}=1| s_t, a_t) \propto \exp(R(s_t, a_t))$, which denotes the \emph{optimality} of the state-action pair at time step $t$,  one can draw an equal connection between searching the optimal policies by RL methods and computing the marginal probability of $p(\mathcal{O}_{t}^{i}=1)$ by probabilistic inference methods, such as message passing or variational inference \cite{jordan1999introduction, beal2003variational, wainwright2008graphical}. 
This equivalence between optimal control and probabilistic inference also holds in the multiagent setting \cite{wen2018probabilistic,grau2018balancing, tian2019regularized, wen2019modelling, shi2019soft}.
In the context of SG (see the red part in Figure \ref{fig:chk}), the optimality variable for each agent $i$ is defined by $p\left(\mathcal{O}_{t}^{i}=1 | \mathcal{O}_{t}^{-i}=1, \tau_t^i \right) \propto \exp \big(r^i\left(s_{t}, a_{t}^{i}, a_{t}^{-i}\right)\big)$, which implies that the optimality of trajectory $\tau^i_t=(s_0, a^i_0, a^{-i}_0,...,  s_t, a^i_t, a^{-i}_t)$  depends on whether agent $i$ acts according to its best response against other agents, and $ \mathcal{O}_{t}^{-i}=1$ indicates that all other agents are perfectly rational and attempt to maximise their rewards.  
Therefore, from each agent's perspective, its objective becomes maximising   $p(\mathcal{O}_{1:T}^i = 1|\mathcal{O}_{1:T}^{-i}=1)$. 
As we assume no knowledge of the optimal policies  and the model of the environment, we treat states and actions as latent variables and apply variational inference \cite{blei2017variational} to approximate this objective, which leads to 
\begin{align}
    \label{eq:mas_soft}
    \max_{\theta^i} &  \ \  J(\bm{\pi}_\theta) =  \log p(\mathcal{O}_{1:T}^i = 1|\mathcal{O}_{1:T}^{-i}=1) \nonumber \\  & \ge \sum_{t=1}^{T} \mathbb{E}_{s \sim P(\cdot | s, \bm{a}), \bm{a} \sim \bm{\pi}_\theta(s)} \bigg[r^{i}\big(s_{t}, a_{t}^{i}, a_{t}^{-i}\big) + \mathcal{H}\Big( \boldsymbol{\pi_{\theta}} (a_{t}^{i}, a_{t}^{-i} | s_{t}) \Big)\bigg]. 
\end{align} 
One major difference from traditional RL is the additional entropy term\footnote{Soft learning is also called maximum-entropy RL \cite{haarnoja2018soft}. } in  Eq. (\ref{eq:mas_soft}). 
Under this new objective, the value function is written as $
V ^ i ( s )  = \mathbb{E}_{\boldsymbol{\pi}_\theta }  \Big[Q^{i} (s_t, a_t^i, a_t^{-i}) -  \log \big( \boldsymbol{\pi}_\theta ( a ^{i}_ { t }, a ^{-i}_ { t } | s _ { t } ) \big)  \Big]
$, and the corresponding optimal Bellman operator is 
\begin{equation}
	\big(\mathbf{H}^{\text{soft}} Q^{i}\big) \big(s, a^i, a^{-i}\big)  \triangleq r^{i}\big(s, a^i, a^{-i}\big)    +   \gamma \cdot \mathbb{E}_{s' \sim P(\cdot| s, \bm{a})}
\Big[ \log \sum_{\bm{a}}  Q^i\big(s', \bm{a}\big)\Big].   
\end{equation}
This process is called \emph{soft learning} because $\log \sum_{\bm{a}}  \exp { \big( Q ( s   , \bm{a}) \big)}  \approx \max _ {  \bm{a} } Q \big(  s  , \bm{a} \big)$.

One substantial benefit of developing a probabilistic framework for multiagent learning is that it can help model the  \emph{bounded rationality} \cite{simon1972theories}. 
Instead of assuming  perfect rationality and agents reaching NE, bounded rationality accounts for situations in which rationality is compromised; it can be constrained by either the difficulty of the decision problem or the agents' own cognitive limitations.
One intuitive example is the psychological experiment of the Keynes beauty contest \cite{Keynes:1936}, in which all players are asked to guess a number between $0$ and $100$ and the winner is the person whose number is closest to the $1/2$ of the average number of all guesses. Readers are recommended to pause here and think about which number you would guess.  
Although the NE of this game is $0$, the majority of people  guess a number between $13$ and $25$ \cite{coricelli2009neural}, which suggests that   human beings  tend to reason only by $1$-$2$ levels of recursion in strategic games \cite{camerer2004cognitive}, i.e., ``I believe how you believe how I believe''. 

\cite{wen2018probabilistic} developed the first MARL powered reasoning model that  accounts for bounded rationality, which they called \emph{probabilistic recursive reasoning} (PR2). 
The key idea of PR2 is that  a dependency structure is assumed when  splitting the joint policy $\bm{\pi}_\theta$,  written by 
\begin{equation}
	\bm{\pi}_{\theta}\left(a^{i}, a^{-i} | s\right)=\pi_{\theta^{i}}^{i}\left(a^{i} | s\right) \rho_{\theta^{-i}}^{-i}\left(a^{-i} | s, a^i\right)  \ \ \ \ (\textbf{PR2}, \ \ \text{Level-}1),  
\end{equation}
that is, the opponent is considering how the learning agent is going to affect its actions, i.e., a Level-$1$ model. 
The unobserved opponent model is approximated by a best-fit model $\rho_{\theta^{-i}}$ when optimising Eq. (\ref{eq:mas_soft}). In the team game setting, since agents' objectives are fully aligned, the optimal $\rho_{\phi^{-i}}$ has a closed-form solution
$  \rho^{-i}_{\phi^{-i}}(a^{-i}| s, a^{i}) \propto  \exp \left( Q^ { i } (s, a^{i}, a^{-i} ) - Q ^ { i} (s, a^{i} ) \right)  $. 
Following the direction of recursive reasoning, \cite{tian2019regularized} proposed an algorithm named ROMMEO that splits the joint policy by 
\begin{equation}
\bm{\pi}_{\theta}\left(a^{i}, a^{-i} | s\right)=\pi_{\theta^{i}}^{i}\left(a^{i} | s, a^{-i}\right) \rho_{\theta^{-i}}^{-i}\left(a^{-i} | s\right)	  \ \ \ \ (\textbf{ROMMEO}, \ \ \text{Level-}1), 
\end{equation}
in which a Level-$1$ model is built from the learning agent's perspective. \cite{grau2018balancing, shi2019soft} introduced a Level-$0$ model where no explicit recursive reasoning is considered. 
\begin{equation}
\bm{\pi}_{\theta}\left(a^{i}, a^{-i} | s\right) = \pi_{\theta^{i}}^{i}\left(a^{i} | s\right) \rho_{\theta^{-i}}^{-i}\left(a^{-i} | s\right) \ \ \ \ (\text{Level-}0).
\end{equation}
However, they generalised the multiagent soft learning framework to include the zero-sum setting. 
\cite{wen2019modelling} recently proposed a mixture of hierarchy Level-$k$ models in which agents can reason at different recursion levels, and higher-level agents make the best response to lower-level agents (see the blue part in Figure \ref{fig:chk}). They called this method \emph{generalised recursive reasoning} (GR2).
\begin{align}
\label{eq:recur}
&\pi _ {k} ^ {i } ( a_k^ { i } | s )  \propto  \int_{a_ {k-1}^{-i}}  \bigg\{  \pi _ {k} ^ {i } ( a_ {k} ^ { i } | s, a_ {k-1} ^ { -i } ) \  \cdot \nonumber \\ &
 \underbrace{\int_{a_{k-2} ^ {i}}  \Big[ \rho  _ {k-1}  ^ { -i } (  a _ {k-1}^ { -i } | s, a_ {k-2} ^ { i  } )   \pi _ {k-2} ^ {i} ( a_{k-2} ^ {i} | s) \Big] \mathrm{d} a _ {k-2}^ { i }}_{\text{opponents of level k-1 best responds to  agent $i$ of level k-2}} \bigg\}\mathrm{d} a_ {k-1}^{-i}. \ \ (\textbf{GR2}, \ \text{Level-}K). 
\end{align}	
In GR2, practical multiagent soft actor-critic methods with convergence guarantee were introduced to make large-$K$ reasoning tractable.

\section{Dec-POMDPs}
Dec-POMDP is a stochastic team game with partial observability. 
However, optimally solving Dec-POMDPs is a challenging combinatorial problem that is $NEXP$-complete \cite{bernstein2002complexity}.  As the horizon increases, the doubly exponential growth in the number of possible policies quickly makes solution methods  intractable.
Most of the solution algorithms for Dec-POMDPs, including the above VDN/QMIX/QTRAN/Q-DPP, are based on the learning paradigm of centralised training with decentralised execution (CTDE) \cite{oliehoek2016concise}. 
CTDE methods assume a centralised controller that can access  observations  across all agents during training. 
A typical implementation is through a centralised critic with a decentralised actor \cite{lowe2017multi}. 
In representing agents' local policies, stochastic finite-state controllers and a correlation device are commonly applied \cite{bernstein2009policy}.
Through this representation, Dec-POMDP can be formulated as non-linear programmes \cite{amato2010optimizing}; 
this process allows the use of a wide range of off-the-shelf optimisation algorithms.  
 \cite{szer2005maa,dibangoye2016optimally,dibangoye2018learning} introduced the transformation from Dec-POMDP into a continuous-state MDP, named the \emph{occupancy-state MDP (oMDP)}. The occupancy state is essentially a distribution over hidden states and the 
joint histories of observation-action pairs. 
In contrast to the standard MDP, where the agent learns an optimal value function that maps histories (or states) to real values, the learner in oMDP learns an optimal value function that maps occupancy states and joint actions to real values (they call the corresponding policy  a \emph{plan}). 
These value functions in oMDP are piece-wise linear and convex. Importantly,  
the benefit of restricting attention on the occupancy state is that the resulting algorithms are guaranteed to converge to a near-optimal plan for any finite Dec-POMDP with a probability of one, while traditional RL methods, such as REINFORCE, may only converge towards a local optimum. 

 In addition to CTDE methods, famous approximation solutions to Dec-POMDP  include the Monte Carlo policy iteration method \cite{wu2010rollout}, which enjoys linear-time complexity in terms of the number of agents, planning by maximum-likelihood methods  \cite{toussaint2008hierarchical,wu2013monte}, which easily scales up to thousands of agents,    and  a method that decentralises POMDP by maintaining shared  memory among agents \cite{nayyar2013decentralized}.

\section{Networked Multiagent MDPs}
 A rapidly growing  area in the  optimisation domain for addressing decentralised learning  for cooperative tasks is the networked multiagent MDP (networked MDP). 
In the context of networked MDP, agents are considered heterogeneous rather than homogeneous; they have different reward functions but still form a team to maximise the team-average reward $ \mathsf{R}=\frac{1}{N} \sum_{i=1}^{N}R^i(s, \bm{a}, s') $. 
Furthermore, in networked MDP, the centralised controller is assumed to be non-existent; instead, agents can only exchange information with their neighbours in a time-varying communication network defined by $G_t=([N], E_t)$, where $E_t$ represents the set of all communicative links between any two of the $N$ neighbouring agents at time step $t$. The states and joint actions are assumed to be globally observable, but each agent's reward is only locally observable to itself.  
Compared to stochastic team games, this setting is believed to be more realistic for real-world applications and can be scaled up to large-scale systems \cite{qu2020scalable,qu2022scalable,du2022scalable,ma2024efficient,fan2024towards} such as smart grids \cite{dall2013distributed} or transport management \cite{adler2002cooperative}.  

The cooperative goal of the agents in networked MDP is to maximise the team average cumulative discounted reward obtained by  all agents over the network,  that is, 
\begin{equation}
	\max_{\bm{\pi}} \frac{1}{N} \sum_{i=1}^{N}\mathbb{E}\Big[\sum_{t\ge0}\gamma^t R_t^i(s_t, \bm{a}_t) \Big].
	\label{eq:network_obj}
\end{equation}
Accordingly, under the joint policy $\bm{\pi} = \prod_{i\in \{1,...,N \}}\pi^i(a^i|s)$, the Q-function is defined as 
\begin{equation}
Q^{\bm{\pi}}(s, \bm{a}) = \frac{1}{N} \sum_{i =1}^{N} \mathbb{E}_{\bm{a}_t \sim \bm{\pi}(\cdot|s_t), s_t \sim P(\cdot| s_t, \bm{a}_t)} \left[ \sum_{t\ge 0} \gamma^t R_t^i(s_t, \bm{a}_t) \Big| s_0=s, a_0=\bm{a} \right]
\label{eq:network_q}
\end{equation}
To optimise Eq. (\ref{eq:network_q}), the optimal Bellman operator is  written as 
\begin{equation}
\left(\mathbf{H}^{\text{networked MDP}}Q\right)\left(s, \bm{a}\right) = \frac{1}{N}\sum_{i=1}^{N}R^i(s, \bm{a})  + \gamma \cdot \mathbb{E}_{s'\sim P(\cdot|s,\bm{a})}\left[\max_{\bm{a}'\sim \pmb{\mathbb{A}}} Q\left(s', \bm{a}'\right)\right].
\label{eq:network_best_q}
\end{equation}
However, since agents can know only their own reward, they do not share the estimation of the Q function but rather maintain their own copy. Therefore, from each agent's perspective, the individual optimal Bellman operator is  written as 
\begin{equation}
\left({\mathbf{H}}^{\text{networked MDP}, i}Q^i\right)\left(s, \bm{a}\right) = R^i(s, \bm{a})  + \gamma \cdot \mathbb{E}_{s'\sim P(\cdot|s,\bm{a})}\left[ \max_{\bm{a}'\sim \pmb{\mathbb{A}}}  Q^i\left(s', \bm{a}'\right)\right].
\label{eq:network_best_q_ind}
\end{equation}
 To solve the optimal joint policy $\bm{\pi}^*$, the agents must reach \textbf{consensus} over the global optimal policy estimation, that is, if   $Q^1=\cdots=Q^N=Q^*$, we know  
 \begin{equation}
 	\left(\mathbf{H}^{\text{networked MDP}}Q^*\right)\left(s, \bm{a}\right) = \frac{1}{N}\sum_{i=1}^{N}\left({\mathbf{H}}^{\text{networked MDP}, i}Q^i\right). 
 	\label{eq:network_consensus}
 \end{equation} 
 To satisfy Eq. (\ref{eq:network_consensus}), \cite{zhang2018finite}  proposed a method based on  neural fitted-Q iteration (FQI)  \cite{riedmiller2005neural} in the batch RL setting \cite{lange2012batch}. Specifically, let $\mathcal{F}_\theta$ denote the parametric function class of neural networks that approximate Q-functions,  let $\mathcal{D} = \{(s_k, \bm{a}^i_k , s_k')\}$ be the replay buffer that contains all the transition data available to all agents, and let $\{R^i_k\}$ be the local reward known only to each agent. The objective of FQI can be written as  
\begin{equation}
\min_{f\in\mathcal{F}_\theta} \frac{1}{N}\sum_{i=1}^{N}\frac{1}{2K}\sum_{j=1}^{K}\Big[ y^i_k - f(s_k, \bm{a}_k; \theta) \Big]^2, \ \ \  \text{with } y^i_k = R^i_k + \gamma \cdot \max_{\bm{a}\in \pmb{\mathbb{A}}  }Q^i_k(s_k' ,\bm{a}).
\label{eq:network_fqi}
\end{equation}
In each iteration, $K$  samples are drawn from $\mathcal{D}$. Since $y^i_k$ is known only to each agent $i$, Eq. (\ref{eq:network_fqi}) becomes a typical  consensus optimisation problem (i.e., consensus must be reached for $\theta$) \cite{nedic2009distributed}. 
Multiple effective distributed optimisers can be applied to solve this problem, including the \emph{DIGing} algorithm \cite{nedic2017achieving}. 
Let $g^i(\theta^i) = \frac{1}{2K}\sum_{j=1}^{K}\big[ y^i_k - f(s_k, \bm{a}_k; \theta) \big]^2$, $\alpha$ be the learning rate, and $G([N], E_l)$ be the topology of the network in the $l$st iteration;  the \emph{DIGing} algorithm designs  the gradient updates for each agent $i$ as  
{\begin{align}
\theta_{l+1}^{i}&=\sum_{j=1}^{N} E_{l}(i, j) \cdot \theta_{l}^{j}-\alpha \cdot \rho_{l}^{i}, \nonumber \\
\rho_{l+1}^{i}&=\sum_{j=1}^{N} E_{l}(i, j) \cdot \rho_{l}^{j}+\nabla g^{i}\left(\theta_{l+1}^{i}\right)-\nabla g^{i}\left(\theta_{l}^{i}\right).
\label{eq:diging}
 \end{align}
 }
 Intuitively, Eq. (\ref{eq:diging}) implies that if all agents aim to reach a consensus on $\theta$, they must incorporate a weighted combination of their neighbours'  estimates into their own gradient updates. 
 However, due to the usage of neural networks, the agents may not reach an exact consensus. \cite{zhang2018finite} also studied the finite-sample bound in a high-probability sense that quantifies the generalisation error of the proposed neural FQI algorithm.   
 
 The idea of reaching consensus can be directly applied to solving Eq. (\ref{eq:network_obj}) via policy-gradient methods.   \cite{zhang2018fully} proposed an actor-critic algorithm in which the global Q-function is approximated individually by each agent.  
 On the basis of Eq. (\ref{eq:mas_pg}), the critic of $Q^{i, \bm{\pi}_{\theta}}(s, \bm{a})$ is modelled by another neural network parameterised by $\omega^i$, i.e.,  $Q^{i}(\cdot, \cdot; \omega^i)$, and the parameter $\omega^i$ is updated as 
 \begin{equation}
 \omega_{t+1}^i = \sum_{j=1}^{N} E_t(i, j)\cdot \Big(\omega_t^j + \alpha \cdot \delta^j_t \cdot \nabla_\omega Q_t^j(\omega_t^j)	\Big)
 \label{eq:update_w}
 \end{equation}
 where $\delta_t^j = R^j_t + \gamma \cdot \max_{\bm{a}\in \pmb{\mathbb{A}}  }Q^j_t(s_t' ,\bm{a}; \omega^j_t) - Q^j_t(s_t' ,\bm{a}; \omega^j_t)$ is the TD error. 
 Similar to Eq. (\ref{eq:diging}), the update in  Eq. (\ref{eq:update_w}) is a weighted sum of all the neighbouring gradients. 
The same group of authors later extended this approach to cover the continuous-action space in which a deterministic policy gradient method of Eq. (\ref{eq:ma_dpg}) is applied \cite{zhang2018networked}. 
 Moreover, \cite{zhang2018fully} and \cite{zhang2018networked} applied a linear function approximation to achieve an almost sure convergence guarantee.  
 Following this thread, 
 \cite{suttle2019multi} and \cite{zhang2019distributed} extended the actor-critic method to an off-policy setting, rendering more data-efficient MARL algorithms.  
 

\section{Stochastic Potential Games}
The potential game (PG) first appeared in \cite{monderer1996potential}. 
The physical meaning of Eq. (\ref{eq:potential}) is that if any agent changes its policy unilaterally, the changes in reward will be represented on the potential function  shared by all agents. 
A PG is guaranteed to have a pure-strategy NE -- a desirable property that does not generally hold in normal-form games.
Many efforts have since been dedicated to finding the NE of (static) PGs \cite{la2016potential}, among which fictitious play \cite{berger2007brown} and generalised weakened fictitious play \cite{leslie2006generalised} are probably the most common solutions.  

Generally, stochastic PGs (SPGs)\footnote{As with team games, stochastic PG is also called dynamic PG or Markov PG.} can be regarded as the ``single-agent component'' of a multiagent stochastic game \cite{candogan2011flows} since all agents' interests in SPGs are described by a single potential function.  
However, the analysis of SPGs is exceptionally sparse. 
\cite{zazo2015dynamic} studied an SPG with deterministic transition dynamics in which agents consider only \emph{open-loop policies}\footnote{Open loop means that agents' actions are a function of time only. By contrast, close-loop policies take into  account the state. In  deterministic systems, these policies can be optimal and coincide in value. For a stochastic system, an open-loop strategy is unlikely to be optimal since it cannot  adapt to state transitions.}. 
 In fact, generalising a PG to the stochastic setting is further complicated because agents must now execute policies that depend on the state and consider the actions of other players.
In this setting, 
\cite{gonzalez2013discrete} investigated a type of SPG in which they derive a sufficient condition for NE, but it requires each agent's reward function to be a concave function of the state and the transition function to be invertible. 
\cite{macua2018learning} studied a general form of SPG where a  \emph{closed-loop} NE can be found. 
Although they demonstrated the equivalence between solving the closed-loop NE and solving a single-agent optimal control problem, the agents' policies must depend only on disjoint subsets of components of the state. 
 Notably, both \cite{gonzalez2013discrete} and \cite{macua2018learning} proposed centralised methods; optimisation over the joint action space surely results in a combinatorial complexity when solving the SPGs. In addition, they  do not consider an RL setting in which the system is \emph{a priori} unknown.

The work of \cite{mguni2020stochastic} is probably the most comprehensive treatment of SPGs in a model-free setting. 
Similar to \cite{macua2018learning}, the authors revealed that the NE of the PG in pure strategies could be found by solving a dual-form MDP, but they reached the conclusion without the disjoint state assumption: the transition dynamics and potential function must be known. Specifically, they provided an algorithm to estimate the potential function based on the reward samples. 
To avoid combinatorial explosion, they  also proposed a distributed policy-gradient method based on generalised  weakened fictitious play \cite{leslie2006generalised} that has linear-time complexity.  

Recently, \cite{mazumdar2018convergence} studied the dynamics  of gradient-based learning on potential games. They found that in a general superclass of potential games named \emph{Morse-Smale games} \cite{hirsch2012differential}, the limit sets of competitive gradient-based learning with stochastic updates are attractors almost surely, and those attractors are either  local Nash equilibria or non-Nash locally asymptotically stable equilibria but not saddle points.

\chapter{Learning in Zero-Sum Games}
\label{chap:zero-sum}
Zero-sum games are games where gains for one group of players correspond to losses for another group of players, so that the total utilities sum to zero. 
In this chapter we mainly focus on two-player zero-sum games, but briefly review zero-sum games with more than two players at the end of this chapter. 

In two-player zero-sum games, each player wants to choose a strategy such that their opponent's best response against that strategy achieves minimal expected utility. This notion is captured in the following LP which can be used to efficiently solve small matrix games. 


\begin{align}
\label{eq:opt_zero_sum_prime}
\text{min} \quad  U_1^*& \\
\text{s.t.} \quad 
\sum_{a^2 \in \mathbb{A}^2} R^1(a^1, a^2)\cdot \pi^2(a^2) &\le U_1^*,   \forall a^1 \in \mathbb{A}^1 \\ 
\sum_{a^2\in\mathbb{A}^2} \pi^2(a^2) &=1 \\ \pi^2(a^2) &\ge 0,   \forall a^2 \in \mathbb{A}^2 
\end{align}

Eq. (\ref{eq:opt_zero_sum_prime}) is considered from the min-player's perspective. One can also derive a dual-form LP from the max-player's perspective.
In discrete games, the minimax theorem \cite{von1945theory}, shown in Eq. (\ref{eq:minimax_1}) is a consequence of the strong duality theorem of LP\footnote{Solving zero-sum games is equivalent to solving a LP; \cite{dantzig1951proof} also proved the correctness of the other direction, that is, any LP can be reduced to a zero-sum game, though some degenerate solutions need careful treatments \cite{adler2013equivalence}.} \cite{matousek2007understanding}.
\begin{equation}\label{eq:minimax_1}
\min_{\pi^1}\max_{\pi^2}  \mathbb{E}\Big[R\big(\pi^1, \pi^2\big)\Big] = \max_{\pi^2}\min_{\pi^1} \mathbb{E}\Big[R\big(\pi^1, \pi^2\big)\Big]
\end{equation}
As a result of the minimax theorem, playing a Nash equilibrium (NE) is an unambiguous solution concept for a player. By playing a NE they maximize their reward even if the opponent knows their strategy. Playing a NE also guarantees receiving the NE value of the game (for example, tying in tic-tac-toe). As a result, algorithms that seek to find NE in large two-player zero sum games have achieved superhuman performance on Chess\cite{campbell2002deep}, Go\cite{silver2016mastering}, and Poker\cite{brown2018superhuman}. 

 
In games that are larger than those that can be solved with an LP, learning algorithms are used to iteratively converge to a (potentially approximate) NE. 
In the following sections, we will use
the following terms to refer to different convergence rates in literature:
``linear'' indicates a rate of $\tilde{\mathcal{O}}(\log{\frac{1}{\epsilon}})$,
``sublinear'' is slower than ``linear'' and it's rate is $\tilde{\mathcal{O}}(\frac{1}{\epsilon^p})$ with $p\geq 1$, and
``superlinear'' means faster than ``linear'' with a rate of $\tilde{\mathcal{O}}(\log(\log{\frac{1}{\epsilon}}))$, where $\epsilon$ means eventually the algorithm will be $\epsilon-$approximately close to the optimal solution.

\section{Minimax Optimization}
\label{sec:minimax}
Minimax optimization is one of the important and well-studied topic in the optimization fields, which is to optimize the function $f(x,y)$ in a dual-objective manner, for two variables $x\in\mathbb{R}^m$ and $y\in\mathbb{R}^n$:
\begin{equation}
    \min_{x\in\mathbb{R}^m} \max_{y\in\mathbb{R}^n} f(x,y)
    \label{eq:minimax}
\end{equation}
where $f(x,y):\mathbb{R}^m\times \mathbb{R}^n \rightarrow \mathbb{R}$ is not necessarily convex-concave. $f(x,y)$ is convex-concave if $f(\cdot, y)$ is convex for all $y\in\mathbb{R}^n$ and $f(x, \cdot)$ is concave for all $x\in\mathbb{R}^m$. 

Different conventional approaches are proposed to solve this minimax optimization problem:
\emph{Proximal Point} (PP)\cite{martinet1970regularisation, rockafellar1976augmented} is an early proposed method for minimax optimization problem. It has the update rule as following:
\begin{align}
    x_{t+1} = x_t - \eta g_{t+1}, g_{t+1} = \frac{d f(x_{t+1},y_{t+1})}{dx }\\
    y_{t+1} = y_t + \eta h_{t+1}, h_{t+1} = \frac{d f(x_{t+1},y_{t+1})}{dy }
\end{align}
where $\eta$ is the learning rate controlling the step size of the update.
As we can see, it is an implicit algorithm since $f(x_{t+1},y_{t+1})$ is unknown before the update. The implementation of PP requires an inverse of matrix $(\mathbf{I}+\eta \nabla_x f)^{-1}$ and $(\mathbf{I}-\eta \nabla_y f)^{-1}$, which may not be computationally affordable in practice. PP method is proven to converge linearly to NE for bilinear cases like normal-form games or strongly-convex-strongly-concave cases\cite{mokhtari2020unified}.

\emph{Gradient Descent Ascent} (GDA) is a basic type of algorithm where the min-player takes the gradient descent and the max-player takes the gradient ascent, as its name says. GDA has the update rule as following: 
\begin{align}
    x_{t+1} = x_t - \eta g_t, g_t = \frac{d f(x_t,y_t)}{dx }\\
    y_{t+1} = y_t + \eta h_t, h_t = \frac{d f(x_t,y_t)}{dy }
\end{align}
where $\eta$ is the learning rate. GDA achieves average-iterate convergence guarantee with a rate $\mathcal{O}(1/\epsilon^2)$ under $L$-Lipschitz or $l$-smooth convex-concave cases\cite{nemirovski2004prox}; and it has a rate $\mathcal{O}(\kappa^2\log\frac{1}{\epsilon})$ for strongly-convex-strongly-concave cases, $\kappa=\frac{l}{\mu}$ for $l$-smooth and $\mu$-strongly-convex-strongly-concave. However, GDA does not have last-iterate convergence guarantee, there are cases where GDA has cyclic or divergent behaviours instead of converges, and it fails in simple bilinear cases\cite{cheung2019vortices} and non-convex or non-concave cases. 

\emph{Optimistic Gradient Descent Ascent} (OGDA)\cite{daskalakis2017training}: 
\begin{align}
    z_{t+1} = z_t - 2\eta F(z_t) + \eta F(z_{t-1})
\end{align}
where $z_t=(x_t, y_t), F(z_t)=(\frac{d f(x_t,y_t)}{dx}, -\frac{d f(x_t,y_t)}{dy})$, and $\eta$ is the learning rate. OGDA achieves average-iterate convergence guarantee with a rate $\mathcal{O}(1/\epsilon)$ for $l$-smooth convex-concave cases\cite{mokhtari2020convergence}, \citet{golowich2020last} also prove OGDA to have a last-iterate convergence rate of $\mathcal{O}(1/\epsilon^9)$ for convex-concave cases. OGDA is further proved to achieve a linear convergence rate of $\mathcal{O}(\kappa\log{\frac{1}{\epsilon}})$ for $l$-smooth and $\mu$-strongly-convex-strongly-concave cases,  where $\kappa=\frac{l}{\mu}$\cite{mokhtari2020unified}.

\emph{Stochastic GDA} (SGDA) is the algorithm to be applied in the case when oracle gradient is not accessible, but only sample estimation of the gradient is provided. It follows the same setting as stochastic gradient descent versus gradient descent: $\nabla_x f(x,y)\approx \frac{1}{M}\sum_{i=1}^M G(x, y, \xi_i)$, where $\{\xi_i\}_{i\in[M]}$ are data samples and $G$ is the gradient on the sample. SGDA converges in $l$-smooth and nonconvex-$\mu$-strongly-concave case with a rate of $\tilde{\mathcal{O}}(\frac{\kappa^3}{\epsilon^4})$\cite{lin2020gradient}, where $\kappa=\frac{l}{\mu}$. It also converges for nonconvex-concave cases with a rate of $\tilde{\mathcal{O}}(\frac{1}{\epsilon^8})$.

\emph{Two-time-scale GDA}\cite{heusel2017gans} applies an asymmetric learning rate for the the max- and min-player, which can be written as following: 
\begin{align}
    x_{t+1} = x_t - \eta_x g_t, g_t = \frac{d f(x_t,y_t)}{dx }\\
    y_{t+1} = y_t + \eta_y h_t, h_t = \frac{d f(x_t,y_t)}{dy }
\end{align}
where $\eta_x, \eta_y$ are different learning rates for two sides.
It is proven to converge in $l$-smooth and nonconvex-$\mu$-strongly-concave case with a rate of $\tilde{\mathcal{O}}(\frac{\kappa^2}{\epsilon^2})$\cite{lin2020gradient}, where $\kappa=\frac{l}{\mu}$. It also converges for nonconvex-concave cases with a rate of $\tilde{\mathcal{O}}(\frac{1}{\epsilon^6})$. \emph{Two-timescale Stochastic GDA} (two-timescale SGDA)\cite{daskalakis2020independent} is an extension of two-time-scale GDA to two-player SG setting, with reinforcement learning method like REINFORCE for gradient estimation, therefore the gradient is not oracle and suffers from sample estimation variances. However, the method is provably convergent to NE with finite samples.

\emph{$\gamma$-GDA}\cite{jin2020local} can be viewed as a special case of \emph{Two-time-scale GDA}, and it's update rule is as following:
\begin{align}
    x_{t+1} = x_t - \frac{\eta}{\gamma} g_t, g_t = \frac{d f(x_t,y_t)}{dx }\\
    y_{t+1} = y_t + \eta h_t, h_t = \frac{d f(x_t,y_t)}{dy }
\end{align}
where the max-player has the learning rate $\eta$ and the min-player has a slower learning rate $\frac{\eta}{\gamma}$ through dividing it by $\gamma > 1$. It's proven that $\gamma$-GDA does not necessarily converge to local NE. By taking $\gamma \rightarrow \infty$, we have the \emph{$\infty$-GDA} algorithm, which has a infinite slow update of the min-player compared with the max-player. In this case, the max-player will always be approximately close to an oracle at the time scale of the min-player, therefore the algorithm is also called \emph{GD with Max-Oracle} (GDMax). GDMax is proven to converge to stationary points called local minimax, which is a superset of the set of local NE\cite{jin2020local}. 

\emph{Multiplicative Weights Update} (MWU), also known as \emph{Hedge}:
\begin{align}
    x_i^{t+1} =x_i^t\frac{e^{\eta g^t_i}}{\sum_{j\in[m]}\mathbf{x}_j^t e^{\eta g^t_j}}, g^t=\frac{d f(x^t,y^t)}{dx } \label{eq:mwu1}\\
    y_j^{t+1} =y_j^t\frac{e^{-\eta h^t_j}}{\sum_{i\in[n]}\mathbf{y}_i^t e^{-\eta h^t_i}}, h^t=\frac{d f(x^t,y^t)}{dy } \label{eq:mwu2}
\end{align}
where $x_i^t, y_j^t$ are the $i$-th and $j$-th entry of probabilistic density in the row and column player's strategies at time $t$, in spaces of size $m$ and $n$. $\eta$ is the learning rate.
Similar as vanilla GDA, vanilla MWU is found to have a cyclic behaviour and fail to converge even in simple bilinear cases\cite{bailey2018multiplicative}, therefore it does not have a last-iterate convergence guarantee.

\emph{Optimistic Multiplicative Weights Update} (OMWU)\cite{daskalakis2018last}, also known as \emph{Optimistic Hedge}: 
\begin{align}
    x_i^{t+1} =x_i^t\frac{e^{2\eta g^t_i-\eta g^{t-1}_i}}{\sum_{j\in[m]}\mathbf{x}_j^t e^{2\eta g^t_j-\eta g^{t-1}_j}}, \forall i\in[A], \\
    y_j^{t+1} =y_j^t\frac{e^{-2\eta h^t_j+\eta h^{t-1}_j}}{\sum_{i\in[n]}y_i^t e^{-2\eta h^t_i+\eta h^{t-1}_i}}, \forall i\in[B],  \\
    \text{with } g^t=\frac{d f(x^t,y^t)}{dx }, g^{t-1}=\frac{d f(x^{t-1},y^{t-1})}{dx } \\
     h^t=\frac{d f(x^t,y^t)}{dy }, h^{t-1}=\frac{d f(x^{t-1},y^{t-1})}{dy }
\end{align}
where $x_i^t, y_j^t$ are the $i$-th and $j$-th entry of probabilistic density in the row and column player's strategies at time $t$, in spaces of size $m$ and $n$. $\eta$ is the learning rate as in MWU. OMWU is proved to achieve a linear convergence rate for bilinear cases when the equilibrium is unique\cite{daskalakis2018last}.

A unification of OGDA and OMWU is called \emph{Optimistic Mirror Descent Ascent} (OMDA)\cite{wei2020linear}, which can be written as:
\begin{align}
    z_t = \arg\min_{z\in\mathcal{Z}}\{\eta \langle z, F(z_{t-1})\rangle+D_\psi(z, \widehat{z_t})\} \label{eq:omda1}\\
    \widehat{z_{t+1}} = \arg\min_{z\in\mathcal{Z}}\{\eta \langle z, F(z_{t})\rangle+D_\psi(z, \widehat{z_t})\} \label{eq:omda2}
\end{align}
where $z_t=(x_t, y_t), F(z_t)=(\frac{d f(x_t,y_t)}{dx}, -\frac{d f(x_t,y_t)}{dy}), \mathcal{Z}=\mathcal{X}\times\mathcal{Y}$, $D_\psi(u, v)$ is the Bregman divergence in mirror descent. To see that OMDA is a unification of OGDA and OMWU, it can be shown that $D_\psi(u, v)=\frac{1}{2}||u-v||^2$ for OGDA and $D_\psi(u, v)=\text{KL}(u,v)$ for OMWU. The readers can try to prove that by specifying different Bregman divergence, the update rules in Eq. (\ref{eq:omda1}) and (\ref{eq:omda2}) will recover the rules of OMDA and OMWU.

\emph{Extra-Gradient} (EG)\cite{korpelevich1976extragradient} method can be interpreted as an approximation of PP method\cite{mokhtari2020unified}.
\begin{align}
    z_{t+\frac{1}{2}} = z_t - \eta F(z_t), \\
    z_{t+1} = z_t - \eta F(z_{t+\frac{1}{2}})
\end{align}
where $z_t=(x_t, y_t), F(z_t)=(\frac{d f(x_t,y_t)}{dx}, -\frac{d f(x_t,y_t)}{dy})$, and $\eta$ is the learning rate. EG achieves average-iterate convergence guarantee with a rate $\mathcal{O}(1/\epsilon)$ for $l$-smooth convex-concave cases\cite{mokhtari2020convergence}, \cite{daskalakis2018last} also prove EG to have a last-iterate convergence rate of $\mathcal{O}(1/\epsilon^2)$ for convex-concave cases. EG is further proved to achieve a linear convergence rate of $\mathcal{O}(\kappa\log{\frac{1}{\epsilon}})$ for both bilinear cases and $l$-smooth and $\mu$-strongly-convex-strongly-concave cases,  where $\kappa=\frac{l}{\mu}$\cite{mokhtari2020unified}. It is an open problem whether EG can achieve last-iterate convergence guarantee\cite{liang2019interaction, mokhtari2020unified}. As an approximation of PP method, EG is found to have a faster convergence rate than OGDA for bilinear problems\cite{mokhtari2020unified}.

The algorithms introduced in this section are essentially important for solving equilibria in games, especially in two-player zero-sum games. We will see that the minimax optimization is exactly the objective for solving NE when NE exists in a normal-form game. Moreover, the procedure of solving minimax optimization problem will serve as a subroutine for more complex games with multiple steps, including stochastic games (SGs), extensive-form games (EFGs), etc. We will see more about this in the following sections.

\section{Discrete-Action Normal-Form Games}

Now we consider two-player zero-sum games, i.e., fully competitive games. The minimax optimization lies at the core of the solution of this problem.
By plugging $f(\mathbf{x},\mathbf{y})=\mathbf{x}^\intercal \mathbf{A} \mathbf{y}$ in Eq. (\ref{eq:minimax}), finding the NE in normal-form games given the payoff function actually follows the \emph{minimax} optimization formulation:
\begin{equation}
    \min_{\mathbf{x}\in\triangle_\mathbb{A}} \max_{\mathbf{y}\in\triangle_\mathbb{B}} \mathbf{x}^\intercal \mathbf{A} \mathbf{y}
    \label{eq:norm_form_minmax}
\end{equation}
where $\mathbf{A}\in\mathbb{R}^{|\mathbb{A}|}\times \mathbb{R}^{|\mathbb{B}|}$ is the payoff matrix, $\triangle_\mathbb{A}\in \mathbb{R}^{|\mathbb{A}|},\triangle_\mathbb{B}\in\mathbb{R}^{|\mathbb{B}|}$ are probability simplicies on the the discrete action spaces $\mathbb{A}$ and $\mathbb{B}$ for two players. For simplicity, we may denote $A=|\mathbb{A}|$ and $B=|\mathbb{B}|$ in the following paragraphs. We will start with some relatively simple algorithms based on the best response subroutine for finding NE in two-player zero-sum normal-form games.

\emph{Self-Play} (SP)\cite{fudenberg1998theory} is provably convergent for multi-player potential game, but not for zero-sum games. SP has the update rule,
\begin{align}
    x_i^{t+1} = \mathsf{Br}_i(x_{-i}^t), \forall i\in[n]
\end{align}
where $\mathsf{Br}_i(\cdot)$ is the best-response subroutine for a given (joint) strategy, and it returns an optimal strategy for the $i$-th player, $n$ is the number of players in the game. It is easy to see that SP does not even converge for simple two-player zero-sum games like \textit{Rock-Paper-Scissors}. 

\emph{Fictitious Play} (FP)\cite{brown1951iterative} is one of the earliest proposed algorithm for solving games in an iterative procedure, which can be written as:
\begin{align}
    \widehat{x_i^t} &= \mathsf{Br}_i(x_{-i}^t=\frac{1}{t}\sum_{\tau=0}^{\tau=t-1}\Pr(a_{-i}^\tau=a, a\in\mathbb{A}))\\
    x_i^{t+1}&=(1-\frac{1}{t})x_i^t+\frac{1}{t}\widehat{x_i^t}, \forall i\in[n]
\end{align}
where $\mathsf{Br}_i(\cdot)$ is the best-response subroutine for a given (joint) strategy, and it returns an optimal strategy for the $i$-th player, $n$ is the number of players in the game. 
FP achieves $\epsilon$-NE with a convergence rate of $(1/\epsilon)^{\Omega(\mathbb{A})}$ for a normal-form game with action space $\mathbb{A}$\cite{daskalakis2014counter}.

\emph{Double Oracle} (DO)\cite{mcmahan2003planning} is another algorithm with iterative best-response procedure. Different from FP, which is based on the best response of the historical average strategy of opponents, DO iteratively solves the best response of the opponents' Nash equilibrium strategy on a restricted strategy set, and keeps adding new best response strategy to the strategy set. It can be written as,
\begin{align}
    \widehat{x_i^t} &= \mathsf{Br}_i(\mathsf{Nash}_{-i}(\{\widehat{x_{-i}^\tau}|\tau=0,1,\dots,t-1\}))\\
    x_i^{t+1}&=\mathsf{Nash}(\{\widehat{x_{i}^\tau}|\tau=0,1,\dots,t\}), \forall i\in[n]
\end{align}
where $\mathsf{Nash}_{-i}(\cdot)$ will return the Nash strategy on the input restricted set of strategies for players except for $i$, $\mathsf{Br}_i(\cdot)$ is the best-response subroutine for player $i$, and $n$ is the number of players in the game.
The DO has convergence rate of $\mathcal{O}(|\mathbb{A}|)$ for a normal-form game with action space $\mathbb{A}$. 

Apart from these algorithms based on best response, there are some other types of algorithms simultaneously updating the strategies for two players to find the NE in games, which will be detailed as following.

\emph{Replicator Dynamics} (RD) is another type of algorithm with continuous gradient flow. The symmetric version of RD for symmetric games with payoff matrix $\mathbf{A}$ has the update rule as following:  
\begin{align}
    \frac{d x_k}{dt}=x_k[(\mathbf{Ax})_k-\mathbf{x}^\intercal \mathbf{Ax}]
\end{align}
where $x_k$ represents the $k$-th component of the probabilistic distribution of the strategy for each single player in the game. Since the game is 

\emph{Asymmetric Replicator Dynamics} for two-player games with asymmetric payoff matrices $\mathbf{A}, \mathbf{B}, \mathbf{A}^\intercal\neq \mathbf{B}$, 
\begin{align}
    \frac{d x_k}{dt}=x_k[(\mathbf{Ay})_k-\mathbf{x}^\intercal \mathbf{Ay}], \quad
    \frac{d y_k}{dt}=y_k[(\mathbf{x^\intercal B})_k-\mathbf{x}^\intercal \mathbf{By}]
\end{align}
where $x_k, y_k$ represent the $k$-th component of the probabilistic distributions of strategies for row and column players, respectively.
(more see \cite{bloembergen2015evolutionary})

As shown in Eq. (\ref{eq:norm_form_minmax}), finding NE in normal-form games is a sub-problem of minimax optimization, with a bilinear structure (therefore convex-concave). The methods introduced in above Sec.~\ref{sec:minimax} can directly be applied for solving NE in normal-form games. For example, we can derive the update rule for MWU method on normal-form games by plugging $f(\mathbf{x},\mathbf{y})=\mathbf{x}^\intercal \mathbf{A} \mathbf{y}$ into Eq. (\ref{eq:mwu1}) and (\ref{eq:mwu2}):
\begin{align}
    x_i^{t+1} =x_i^t\frac{e^{\eta(\mathbf{A}\mathbf{y}^t)_i}}{\sum_{j\in[m]}\mathbf{x}_j^t e^{\eta(\mathbf{A}\mathbf{y}^t)_j}}, x_i^1=\frac{1}{A}\nonumber\\
    y_j^{t+1} =y_j^t\frac{e^{-\eta(\mathbf{A}^\intercal \mathbf{x}^t)_j}}{\sum_{i\in[n]}\mathbf{y}_i^t e^{-\eta(\mathbf{A}^\intercal \mathbf{x}^t)_i}}, y_j^1=\frac{1}{B}
    \label{eq:mwu}
\end{align}
and the corresponding two-player OMWU algorithm will be:
\begin{align}
    x_i^{t+1} =x_i^t\frac{e^{2\eta(\mathbf{A}\mathbf{y}^t)_i-\eta(\mathbf{A}\mathbf{y}^{t-1})_i}}{\sum_{j\in[m]}\mathbf{x}_j^t e^{2\eta(\mathbf{A}\mathbf{y}^{t})_j-\eta(\mathbf{A}\mathbf{y}^{t-1})_j}}, \forall i\in[A]\nonumber\\
    y_j^{t+1} =y_j^t\frac{e^{-2\eta(\mathbf{A}^\intercal \mathbf{x}^t)_j+\eta(\mathbf{A}^\intercal \mathbf{x}^{t-1})_j}}{\sum_{i\in[n]}y_i^t e^{-2\eta(\mathbf{A}^\intercal \mathbf{x}^t)_i+\eta(\mathbf{A}^\intercal \mathbf{x}^{t-1})_i}}, \forall j\in[B]
    \label{eq:omwu}
\end{align}
Similarly, methods including GDA and OGDA have the corresponding formulas for normal-form games.

As introduced before, the MWU algorithm is also known as \emph{Hedge}\cite{littlestone1994weighted, freund1997decision}, which is originated from online learning for regret minimization. It applies an exponentially weighted  function to derive a new strategy:
\begin{equation}
\pi^{t+1}(a_i) = \dfrac{\pi^{t}(a_i)e^{-\eta R_t(a_i)}}{\sum_{j=1}^{K} \pi^{t}(a_j)e^{-\eta R_t(a_j)}}, \ \  \pi^1(\cdot) = \dfrac{1}{K}. 
\label{eq:hedge}	
\end{equation} 
where $a_i$ is the $i$-th component within the action space. Notice that $R_t$ in above equation is the regret instead of reward at time step $t$.
In computing Eq. (\ref{eq:hedge}), \emph{Hedge} needs access to the full information of the reward values for all actions, including those that are not selected.  
For the above two-player setting, the player at each side optimizes its strategy following the regret-minimization principle. Specifically, the regret is defined as: $R_t=-\mathbf{A}\mathbf{y}^t$ for the first player and $R_t=\mathbf{A^\intercal}\mathbf{x}^t$ for the second player.

The \emph{Hedge} algorithm as Eq. (\ref{eq:mwu}) achieves a regret of $\mathcal{O}(\sqrt{T\log A})$ for multi-player general-sum game, with each player having action space of size $A$\cite{cesa2006prediction}. 
  \emph{EXP3}  \cite{auer1995gambling} extended the \emph{Hedge} algorithm for a  \emph{partial information game} in which the player knows only the reward of the chosen action (i.e., a bandit version) and has to estimate the loss of the actions that it does not select. 
\cite{brown2017dynamic} augmented the \emph{Hedge} algorithm with a tree-pruning technique based on dynamic thresholding. 
 \cite{gordon2007no} developed  \emph{Lagrangian hedging}, which unifies no-regret algorithms, including both regret matching and Hedge, through a class of  potential functions.
\emph{Optimistic Hedge} (or called OMWU) provably achieves $\mathcal{O}(N\log A \log^4 T)$ regret for N-player general-sum games, with each player having action space of size $A$\cite{daskalakis2021near}.

For regret minimization, one can also apply Blackwell's approachability theorem \cite{blackwell1956analog} to minimize the regret independently on each information set, also known as \emph{Regret Matching} (RM) \cite{hart2001reinforcement}. 
As we are most concerned with positive regret, denoted by $\lfloor \cdot\rfloor_+$,  we have $\forall S \in \mathbb{S}^i, \forall a \in \chi(S)$, the strategy of player $i$ at time $t+1$ as Eq. (\ref{eq:regret-match}). 

\begin{align}
\pi^{i}_{t+1}(S, a)=\left\{\begin{array}{ll}\dfrac{ \lfloor \mathsf{Reg}_{t}^i \big(S, a\big) \rfloor_+}{\sum_{a \in \chi(S)} \lfloor \mathsf{Reg}_{t}^i \big(S, a\big) \rfloor_+} & \text {if } \sum_{a \in \chi(S)} \lfloor \mathsf{Reg}_{t}^i (S, a) \rfloor_+>0 \\ \dfrac{1}{|\chi(S)|} & \text {otherwise }\end{array}\right .
\label{eq:regret-match}
\end{align}

In the standard CFR algorithm,  for each information set, Eq. (\ref{eq:regret-match}) is used to compute action probabilities in proportion to the positive cumulative regrets. 

We recommend to read the reference\cite{cesa2006prediction}  for a comprehensive overview of no-regret algorithms.



\section{Continuous-Action Normal-Form Games}
Recently, 
the challenge of training generative adversarial networks (GANs) \cite{goodfellow2014generative} has ignited tremendous research interest in  understanding policy gradient methods in two-player continuous games, specifically, games with a continuous station-action space and nonconvex-nonconcave loss landscape.  
 In GANs, two neural network parameterised models -- the generator G and the discriminator D -- play a zero-sum game. In this game, the generator attempts to generate data that ``look'' authentic such that the discriminator cannot tell the difference from the true data; on the other hand, the discriminator tries not to be deceived by the generator.  The loss function in this scenario is written as  
 \begin{align}
  	\label{eq:gan}
 	\min _{\theta_\text{G}\in\mathbb{R}^d}& \max _{\theta_\text{D}\in\mathbb{R}^d}f\big(\theta_\text{G}, \theta_\text{D}\big) =  \\ &    \bigg[\mathbb{E}_{x \sim p_{\operatorname{data}}} \Big[ \log \text{D}_{\theta_\text{D}}\big(x\big) \Big]+\mathbb{E}_{z \sim p(z)}\Big[ \log \Big(1-\text{D}_{\theta_\text{D}}\big(\text{G}_{\theta_\text{G}}(z)\big)\Big)\Big]\bigg] \nonumber
 \end{align}
where $\theta_\text{G}$ and $\theta_\text{D}$ represent neural networks parameters and $z$ is a random signal, serving as the input to the generator.
In searching for the NE, one naive approach is to update both $\theta_\text{G}$ and  $\theta_\text{D}$ by simultaneously implementing the \emph{Gradient Descent Ascent} (GDA) updates  with the same step size in Eq. (\ref{eq:gan}). This approach is equivalent to a MARL algorithm in which both agents are applying policy-gradient methods.  
With trivial adjustments to the step size \cite{bowling2002multiagent, bowling2005convergence,zhang2010multi}, GDA methods can work effectively in two-player two-action (thus convex-concave) games. 
However, in the nonconvex-nonconcave case, where the minimax theorem no longer holds, GDA with equal stepsize can converge to limit cycles or even diverge in a general setting, and GDA can exhibit oscillation in case of nonconvexity \cite{NEURIPS2020_52aaa62e}. \cite{flokas2021solving} further proved that if the hidden game is strictly convex-concave then vanilla GDA converges not merely to local Nash, but typically to the von-Neumann solution. If the game lacks strict convexity properties, GDA may fail to converge to any equilibrium. Therefore, in order to solve the above problems, more variants of GDA have been developed. \cite{lin2020gradient} presented the convergence results of the two-time-scale GDA algorithm for solving nonconvex-concave minimax problems, showing that the algorithm can find a stationary point efficiently. \cite{kalogiannis2021teamwork} complemented some negative results which are finding a NE in some settings can be shown to be CLS-hard by designing a modified GDA that converges locally to NE. \cite{deng2021local} proposed local Stochastic Gradient Descent Ascent (local SGDA), where the primal and dual variables can be trained locally and averaged periodically to significantly reduce the number of communications. \cite{sharma2022federated} proved that Local SGDA has order-optimal sample complexity for several classes of nonconvex-concave and nonconvex-nonconcave minimax problems, and also enjoys linear speedup with respect to the number of clients. \cite{li2021complexity} showed that for quadratic or nearly quadratic nonconvex-strongly-concave functions under some assumptions, two-time-scale GDA and SGDA with appropriate stepsizes achieve a linear convergence rate. \cite{zhang2022near} proved that alternating gradient descent-ascent (Alt-GDA) achieves a near-optimal local convergence rate for strongly convex-strongly concave (SCSC) problems while simultaneous GDA converges at a much slower rate.

Based on the recent research results above, GDA methods are notoriously flawed from three aspects.
First, GDA algorithms may not converge at all \cite{balduzzi2018mechanics,mertikopoulos2018optimistic,daskalakis2018limit}, resulting in limit cycles\footnote{Limit cycle is a terminology in the study of dynamical systems, which describes oscillatory systems. In game theory, an example of limit cycles in the strategy space can be found in Rock-Paper-Scissor game.} in which even the time average\footnote{In two-player two-action games, \cite{singh2000nash} showed that the time average payoffs would converge to an NE value if their policies do not.} does not coincide with NE \cite{mazumdar2019policy}.   
Second, there exist undesired stable stationary points for the GDA algorithms that are not local optima of the game \cite{adolphs2019local, mazumdar2019policy}.
Third,  there exist games whose equilibria are not the attractors of GDA  methods at all \cite{mazumdar2019policy}. 
These problems are partly caused by the intransitive dynamics (e.g., a typical intransitive game is rock-paper-scissors game) that are inherent in  zero-sum  games  \cite{omidshafiei2020navigating,balduzzi2018mechanics} and  the fact that each agent may have a non-smooth objective function.  
In fact,  
 even in simple linear-quadratic games, the reward function cannot satisfy the smoothness condition\footnote{A differentiable function is said to be smooth if the gradients of the function are continuous.} globally, and the games are  surprisingly not convex either  \cite{fazel2018global, mazumdar2019policy, zhang2019policy}.
 
Three  mainstream  approaches have been  followed to develop algorithms that have at least a local convergence guarantee.  
 One natural idea  is to make the inner loop solvable at a reasonably high level and then focus  on a simpler type of game. In other words, the algorithm tries to find a stationary point of the function $\Phi(\cdot):=\max _{\theta_\text{D} \in \mathbb{R}^d} f\big(\cdot, \theta_\text{D}\big) $, instead of Eq.  (\ref{eq:gan}). 
 For example, by considering games with a nonconvex and (strongly) concave loss landscape,  
\cite{lin2019gradient,nouiehed2019solving,rafique2018non,thekumparampil2019efficient,lu2020hybrid, kong2019accelerated} presented an affirmative answer that  GDA methods can converge to a stationary point in the outer loop of optimising $\Phi(\cdot):=\max _{\theta_\text{D} \in \mathbb{R}^d} f\big(\cdot, \theta_\text{D}\big) $. 
  Based on this understanding, they developed various GDA variants that apply the ``best response'' in the inner loop while maintaining an inexact gradient descent in the outer loop. 
  We refer to \cite{lin2019gradient} [Table 1] for a detailed summary of the time complexity of the above  methods.

 The second mainstream  idea is to shift the equilibrium of interest from the NE, which is induced by simultaneous gradient updates, to the Stackelberg equilibrium, which is a solution concept in leader-follower (i.e., alternating update) games.  
 \cite{jin2019local} introduced the concept of the local Stackelberg equilibrium, named \emph{local minimax}, based on which he established the connection to GDA methods by showing that  all stable limit points of GDA are exactly local minimax points.
\cite{fiez2019convergence} also built connections between the NE and Stackelberg equilibrium  by formulating the conditions under which attracting points of GDA dynamics are Stackelberg equilibria in zero-sum games.
When the loss function is bilinear,   theoretical evidence was found that alternating updates converge faster than simultaneous GDA methods  \cite{zhang2019convergence}.  

The third mainstream idea is to analyse the loss landscape from a game-theoretic perspective and design corresponding algorithms that mitigate oscillatory  behaviour. 
Compared to the previous two mainstream ideas, which helped generate more theoretical insights than applicable algorithms, works within this stream demonstrate strong empirical improvements in training GANs.  
\cite{mescheder2017numerics} investigated the game Hessian and identified that issues on the eigenvalues trigger the limit cycles. 
As a result, they proposed a new type of update rule based on consensus optimisation,  together with a convergence guarantee to a local NE in smooth two-player zero-sum games. 
\cite{adolphs2019local} leveraged the curvature information of the loss landscape to propose algorithms in which all stable limit points are guaranteed to be local NEs. 
Similarly, \cite{mazumdar2019finding} took advantage of the differential structure of the game and  constructed an algorithm for which the local NEs are the only attracting fixed points. 
\cite{NEURIPS2020_52aaa62e} introduced a ``smoothing'' scheme which can be combined with GDA to stabilize the oscillation and ensure convergence to a stationary solution. \cite{yang2022faster} established new convergence results for alternating GDA and smoothed GDA under the mild assumption that the objective satisfies the Polyak-Lojasiewicz (PL) condition about one variable. \cite{he2022gdaam} proposed GDA with anderson mixing (GDA-AM) which can achieve global convergence for bilinear problems under mild conditions. GAD-AM views the GDA dynamics as a fixed-point iteration and solves it using Anderson Mixing to converge to the local minimax. \cite{NEURIPS2019_25048eb6} proposed a multi-step GDA (MGDA) algorithm that finds an ${\epsilon}$–first order stationary point of the game in iterations. \cite{lee2021semi} proposed a new semi-anchoring (SA) technique
for the MGDA method. This makes the MGDA method find a stationary point of a
structured nonconvex-nonconcave composite minimax problem. \cite{barazandeh2021solving} showed that a simple multi-step proximal gradient descent-ascent algorithm converges to ${\epsilon}$–first order Nash equilibrium of the min-max game. Finally, some studies have proved some characteristics of GDA. \cite {mladenovic2021generalized} studied GDA flow and proved global convergence under mild assumptions. \cite{fiez2020gradient} studied the role that a finite timescale separation parameter has on the learning dynamics in GDA. \cite{NEURIPS2019_6c7cd904} proved that GDA dynamics can exhibit a variety of behaviors antithetical to convergence to the game theoretically meaningful min-max solution in some specific instances. While studying the properties of GDA, the results in \cite{doan2022convergence} enhanced the convergence properties of its discrete-time counterpart.
In addition,  \cite{daskalakis2017training, mertikopoulos2018optimistic}
 addressed the issue of limit cycling behaviour in training GANs by proposing to apply the technique of \emph{optimistic mirror descent (OMD)}, which is initially proposed by \cite{rakhlin2013optimization}. OMD achieves the  last-iterate convergent guarantee in bilinear convex-concave games. Specifically, at each time step, OMD adjusts the gradient of that time step by considering the opponent policy at the next time step. Let $M_{t+1}$ be the predictor of the next iteration gradient\footnote{In practice, it is usually set as the last iteration gradient.}; we can write OMD as follows. 
 \begin{align}
 \label{eq:omd}
\theta_{\text{G}, t+1} &=\theta_{\text{G}, t}+\alpha \cdot\left(\nabla_{\theta_{\text{G}}, t} f\big(\theta_\text{G}, \theta_\text{D}\big) +M_{\theta_{\text{G}}, t+1}-M_{\theta_{\text{G}}, t}\right) \nonumber \\ \theta_{\text{D}, t+1} &=\theta_{\text{G}, t}-\alpha \cdot\left(\nabla_{\theta_\text{D}, t} f\big(\theta_\text{G}, \theta_\text{D}\big) +M_{\theta_\text{D}, t+1}-M_{\theta_\text{D}, t}\right) 
 \end{align}
In fact,  the pivotal idea of opponent prediction in OMD, developed in the optimisation domain, resembles the  idea of approximate policy prediction in the MARL domain \cite{zhang2010multi, foerster2018learning}.

Due to the advantages of OMD, many variants continue to improve this method, such as \cite{mertikopoulos2018optimistic} proved that OMD converges in all coherent problems by taking an “extra-gradient” step. \cite{vyas2022competitive} proposed \emph{optimistic competitive gradient optimization (OCGO)}, an optimistic variant, for which they show convergence rate to saddle points in $\alpha$-coherent class of functions. \cite{daskalakis2017training} established that a variant of the widely used Gradient Descent/Ascent procedure, called \emph{Optimistic Gradient Descent/Ascent (OGDA)}, exhibits last-iterate convergence to saddle points in unconstrained convex-concave min-max optimization problems. \cite{b̈ohm2022two} rediscovered a method related to OGDA, for the deterministic and the stochastic problem they showed a convergence rate of $O({1}/{k})$ and $O({1}/\sqrt{k})$, \cite{schafer2020competitive} proposed \emph{competitive mirror descent (CMD)}: a general method for solving such problems based on first order information that can be obtained by automatic differentiation. \cite{huang2021efficient} proposed an \emph{accelerated stochastic mirror descent ascent (VR-SMDA)} method based on the variance reduced technique. \cite{huang2021no} proposed a distributed mirror descent algorithm for computing a Nash equilibrium and proved its convergence with suitably selected diminishing step-sizes for a strictly convex-concave cost function. \cite{wibisono2022alternating} proposed and analyze the alternating mirror descent algorithm, in which each player takes turns to take action following the mirror descent algorithm for constrained optimization. Some work further reveals the properties of OMD, such as \cite{anagnostides2022last} revealed that OMD either reaches arbitrarily close to a Nash equilibrium, or it outperforms the robust price of anarchy in efficiency. \cite{anagnostides2022optimistic} showed that when OMD does not reach arbitrarily close to a NE, the cumulative regret of both players is not only negative, but decays linearly with time. \cite{gao2021second} proposed a second-order extension of the continuous-time game-theoretic mirror descent dynamics which converges to mere variationally stable states. \cite{bohm2022solving} proved novel convergence results which matching the $1/{k}$ rate for the best iterate in terms of the squared operator norm recently shown for the EG for a generalized version of the OGDA. \cite{anagnostides2022frequency} found that OGD and its variants are instances of \emph{proportional integral derivative (PID)} control.
 
The other approach is  extragradient (EG), EG (Korpelevich,1976) is one of the most popular methods for solving saddle point and variational in equalities problems (VIP). there remain important open questions about convergence of EG. \cite{gorbunov2022extragradient} derived the first last-iterate $O({1}/{k})$ convergence rate for EG for monotone and Lipschitz VIP without any additional assumptions on the operator. \cite{enrich2019extragradient} proposed an additional variance reduction mechanism for EG to obtain speed-ups in smooth convex games. \cite{luoextragradient} unified and established a best-convergence rate of two variants of the extragradient method for approximating a solution of a co-monotone inclusion constituted by the sum of two operators. \cite{liao2021local} studied a class of stochastic minimax methods and develop a communication-efficient distributed stochastic extragradient algorithm,
LocalAdaSEG, with an adaptive learning rate suitable for solving convex-concave minimax problems in the Parameter-Server model. \cite{lee2021fast} proposed a two-time-scale variant of the EG, named EG+, with a slow $O({1}/{k})$ rate on the squared gradient norm, where $k$ denotes the number of iterations. \cite{cen2021fast} developed provably efficient extragradient methods to find the quantal response equilibrium (QRE)—which are solutions to zero-sum two-player matrix games with entropy regularization—at a linear rate. \cite{li2021convergence} presented an analysis of the same-sample Stochastic EG (SEG) method with constant step size. \cite{he2021age} proposed AGE, an alternating extra-gradient method with nonlinear gradient extrapolation. It estimates the lookahead step using a nonlinear mixing of past gradient sequences. \cite{du2022optimal} presented a stochastic accelerated gradient EG (AG-EG) that achieves such a relatively mature characterization of optimality in saddle-point optimization. \cite{fasoulakis2021forward} showed that its EG algorithm reaches first an ${\eta}^{1/{\rho}}$-approximate Nash equilibrium, with ${\rho}>1$. Finally, \cite{azizian2020tight} proved that EG achieves the optimal rate for a wide class of algorithms with any number of extrapolations.

In addition to the above three categories of methods, there are many other gradient based methods. \cite{loizou2021stochastic} introduced the expected co-coercivity condition and provided the first last-iterate convergence guarantees of SGDA and stochastic consensus optimization (SCO) under this condition for solving a class of stochastic variational inequality problems that are potentially non-monotone. \cite{loizou2020stochastic} proposed a novel unbiased estimator for the stochastic Hamiltonian gradient descent (SHGD) and showed that SHGD converges linearly to the neighbourhood of a stationary point. \cite{qi2021training} proposed the adaptive Composite Gradients (ACG) method, linearly convergent in bilinear games under suitable settings. \cite{xu2021zeroth} proposed a zeroth-order alternating randomized gradient projection (ZO-AGP) and its iteration complexity to obtain an $\epsilon$-stationary point is bounded by $O(\epsilon^{-4})$. \cite{wu2019logan} improved CS-GAN with natural gradient-based latent optimisation and showed that it improves adversarial dynamics. \cite{balduzzi2018mechanics} decomposed the second-order dynamics into two components. The first is related to potential games, which reduce to gradient descent on an implicit function; the second relates to Hamiltonian games, a new class of games that obey a conservation law, akin to conservation laws in classical mechanical systems. 
\cite{mazumdar2018convergence} proposed a general framework for competitive gradient-based learning, which sheds light on the issue of convergence to non-Nash strategies in general-sum and zero-sum games which have no relevance to the underlying game.

There are some other methods that provide us with a new perspective. For example, \cite{grnarova2021generative} optimized a different objective that circumvents the min-max structure using the notion of duality gap from game theory. The convergence and stability properties of the method proposed in \cite{NEURIPS2019_56c51a39} remain robust under strong interactions between players, even without adapting the step size. \cite{franci2021training} proposed a \textit{stochastic relaxed forward-backward (SRFB)} algorithm for GANs, and demonstrated convergence to an exact solution as more data becomes available. \cite{hsieh2020limits} conducted an in-depth study of a comprehensive class of state-of-the-art algorithms and prevalent heuristics for non-convex / non-concave problems. \cite{perolat2021poincare} investigated how adapting the reward—by adding a regularization term—can yield strong convergence guarantees in monotone games. \cite{letcher2018stability} addressed and solved these games using a new algorithm called \textit{Stable Opponent Shaping (SOS)}, which inherits strong convergence guarantees from lookahead and the shaping capacity from Learning with Opponent-Learning Awareness (LOLA). To accelerate the calculation of equilibrium, \cite{yoon2022accelerated} showed that these accelerated algorithms exhibit what we call the \textit{merging path (MP)} property. \cite{azizian2020accelerating} used matrix iteration theory to characterize acceleration in smooth games.

Contemporary work on learning in continuous games has commonly overlooked the hierarchical decision-making structure present in machine learning problems formulated as games, instead treating them as simultaneous play games and adopting the Nash equilibrium solution concept. \cite{fiez2020implicit}, \cite{fiez2019convergence} deviated from this paradigm and provide a comprehensive study of learning in Stackelberg games.These works provided insights into the optimization landscape of zero-sum games by establishing connections between Nash and Stackelberg equilibria along with the limit points of simultaneous gradient descent.

A central obstacle in the optimization of continuous zero-sum games is the rotational dynamics that hinder their convergence. Existing methods typically employ intuitive, carefully hand-designed mechanisms for controlling such rotations. some works take a novel approach to address this issue by casting min-max optimization as a physical system, such as \cite{hemmat2020lead} leveraged tools from physics to introduce LEAD (Least-Action Dynamics), a second-order optimizer for min-max games. \cite{bailey2019multi} established a formal and robust connection between multiagent systems and Hamiltonian dynamics. \cite{ibrahim2020linear} gave a linear lower bound for n-player differentiable games, by using the spectral properties of the update operator. \cite{domingo2020mean} proved a law of large numbers that relates particle dynamics to mean-field dynamics. \cite{ha2022convergence} presented a spectral analysis and provide a geometric explanation of how and when it actually improves the convergence around a stationary point.
 
Thus far, the most promising results are probably those of \cite{bu2019global} and \cite{zhang2019policy},  which reported the first results in solving zero-sum LQ games with a global convergence guarantee. Specifically, \cite{zhang2019policy} developed the solution through projected nested-gradient methods, while \cite{bu2019global} solved the problem through a projection-free Stackelberg leadership model. Both of the models achieve a sublinear rate for convergence. 

These results are essentially related to game dynamics, recent advancements in learning dynamics for games have addressed critical challenges in achieving efficient regret minimization under various game-theoretic settings. Notably, significant progress has been made in bounding the swap regret in general-sum multiplayer games. A novel combination of optimistic regularized learning with self-concordant barriers enables near-optimal bounds on swap regret, bypassing complex frameworks like higher-order smoothness, and improving existing results to achieve bounds proportional to second-order path lengths \cite{anagnostides2022uncoupled}. In dynamic multi-agent settings, where the underlying repeated games evolve over time, researchers have extended the convergence analysis of optimistic gradient descent (OGD) to time-varying games. This framework provides a refined equilibrium gap and regret bounds, particularly under strong convexity-concavity, with applications to meta-learning and variation-dependent regret bounds \cite{anagnostides2024convergence}. Furthermore, new algorithms address regret minimization in general convex games with arbitrary compact strategy sets, achieving exponential improvements in regret bounds via an optimistic follow-the-regularized-leader approach. These dynamics leverage self-concordant regularizers and proximal oracles to maintain computational efficiency while generalizing previous results \cite{farina2022near}. 
In addition, recent work extends equilibrium computation to continuous-action games via a modified double oracle algorithm suitable for multi-player settings. By maintaining fixed-cardinality pure strategy sets and avoiding the need for exact metagame solving or global best-response computation, this approach significantly reduces memory and computational requirements, even in high-dimensional action spaces. \cite{martin2024simultaneous}
Collectively, these contributions advance our understanding of dynamic learning and equilibrium computation across a broad spectrum of game-theoretic frameworks.

\section{Stochastic Games}
\label{sec:sg_alg}
Stochastic games (SGs)\cite{shapley1953stochastic}, also known as Markov games (MGs)\cite{littman1994markov}, can be defined in the following way. An infinite-horizon discounted version of multi-player general-sum SG/MG is denoted as a tuple $(N, \mathbb{S}, \{\mathbb{A}_i\}_{i=1}^N, \mathbb{P}, \{r_i\}_{i=1}^N, \gamma)$. $N$ is the number of players in the game. $\mathbb{S}$ is the state space, $\{\mathbb{A}_i\}_{i=1}^N$ are the action spaces for each player $i\in[N]$ respectively. $\mathbb{P}$ is the state transition distribution, and $\mathbb{P} ( \cdot | s, a_1, \dots, a_N): \mathbb{S}\times  \{\mathbb{A}_i\}_{i=1}^N \to \Pr(\mathbb{S}) $ is the distribution of the next state given the current state $s$ and joint actions $(a_1, \dots, a_N)$. $r_i\colon \mathbb{S} \times \mathbb{A}_i \to \mathbb{R}$ is the reward function for the $i$-th player. $\gamma\in[0,1]$ is the discount factor. The stochastic policy of the $i$-th player is defined as a probabilistic simplex: $\pi(s)\in \triangle_{\mathbb{A}_i}, s\in\mathbb{S}$.

For the two-player zero-sum SG/MG, it can be defined as a tuple $(N, \mathbb{S}, \mathbb{A}, \mathbb{B}, \mathbb{P}, r, \gamma)$, with the only difference that $\mathbb{A}, \mathbb{B}$ are the action spaces for the max-player and min-player respectively. The transition function $\mathbb{P}$ and $r$ are changed accordingly. In the zero-sum setting, the reward is the gain for the max-player and the loss for the min-player due to the zero-sum payoff structure. Similar to single-agent MDP, value-based methods aim to find an optimal value function, which in the context of zero-sum SGs, corresponds to the minimax NE of the game.  

The MG can be regarded as a direct extension of Markov Decision Process (MDP) for single agent to a multiagent setting, where the Markov property for the agent in MDP is now preserved for the joint transition of all players in MG. An extensive-form game (EFG) can be viewed as the MG with a special tree structure. Although MG can also be viewed in a tree structure perspective (always transferring from a parent node into its child node with the same state and action spaces), it does not require so due to the Markov property. Strictly speaking, MG is a subset of EFG, but this does not hurt to bring up methods specifically designed for MG or EFG to achieve better sample efficiency. We will see this with detailed discussions in the following sections.

\subsection{Tabular Stochastic Games}
From a theoretical view, we survey the existing methods for tabular SG/MG with finite and discrete spaces, for both the state and the action. For simplicity, we denote the action space size $A=|\mathbb{A}|$ (and $B=|\mathbb{B}|$ for second player in the game), and the state space size $S=|\mathbb{S}|$, where $|\cdot|$ denotes the cardinality of the countable set. The horizon of the game is $H$. The tabular case for two-player zero-sum SGs means $A$ and $B$ are finite and small.

In two-player zero-sum SGs with discrete states and actions, we know $V^{1, \pi^1, \pi^2} = -V^{2, \pi^1, \pi^2}$, and by the minimax theorem \cite{von1945theory}, the optimal value function is  $V^* = \max_{\pi^2} \min_{\pi^1} V^{1, \pi^1, \pi^2} = \min_{\pi^1} \max_{\pi^2} V^{1, \pi^1, \pi^2}$. In each stage game defined by $Q^1=-Q^2$, the optimal value can be solved by a matrix zero-sum game through a linear program in Eq. (\ref{eq:opt_zero_sum_prime}). 
\cite{shapley1953stochastic} introduced the first value-iteration method, written as

\begin{equation}
\label{eq:shapley_operator}
(\mathbf{H}^{\text{Shapley}}V)(s) = 	\min_{\pi^1 \in \Delta(\mathbb{A}^1)} \max_{\pi^2 \in \Delta(\mathbb{A}^2)}  \mathbb{E}_{a^1\sim\pi^1, a^2\sim\pi^2, s'\sim \mathbb{P}} \Big[R^1(s, a^1, a^2 ) + \gamma \cdot V(s')  \Big], 
\end{equation}
 and proved $\mathbf{H}^{\text{Shapley}}$  is a contraction mapping (in the sense of the infinity norm) in solving two-player zero-sum SGs. 
In other words, assuming the transitional dynamics and reward function are known, the value-iteration method will  generate a sequence of value functions $\{V_t\}_{t\ge0}$ that asymptotically converges to the fixed point $V^*$, and the corresponding policies will converge to the NE policies $\bm{\pi}^*=(\pi^{1,*}, \pi^{2,*})$. 

If the exact NE can not be found, we will introduce the concept of $\epsilon$-approximate NE, which represent a solution close to true NE by a distance of $\epsilon$. The $\epsilon$-approximate NE for $N$-player is a product policy $\pi=(\pi^1, \pi^2, \dots, \pi^N)$ satisfying:
\begin{align}
    \max_{i\in[N]}\Big(\max_{\widehat{\pi^i}} V^{i, \widehat{\pi^{i}},\pi^{-i}}(s_1) - V^{i,\pi^i, \pi^{-i}} (s_1)\Big) \le \epsilon
\end{align}
This definition covers the case of two players.

In contrast to Shapley's model-based value-iteration method, 
\cite{littman1994markov} proposed a model-free Q-learning method -- \emph{Minimax-Q} -- that extends the classic Q-learning algorithm defined in Eq. (\ref{eq:ma_q_learning}) to solve zero-sum SGs. 
Specifically, in \emph{Minimax-Q}, Eq. (\ref{eq:ma_solve_sg}) can be equivalently written as 
 \begin{align}
		\mathbf{eval}^1 \Big( \big\{Q^1(s_{t+1}, \cdot) \big\} \Big) &= - \mathbf{eval}^2 \Big( \big\{Q^2(s_{t+1}, \cdot) \big\} \Big) \nonumber \\ & = \min_{\pi^1 \in \Delta(\mathbb{A}^1)} \max_{\pi^2 \in \Delta(\mathbb{A}^2)}  \mathbb{E}_{a^1\sim\pi^1, a^2\sim\pi^2} \Big[Q^1(s_{t+1}, a^1, a^2 )  \Big].
		\label{eq:comp_equilibrium}
\end{align}
The Q-learning update rule of \emph{Minimax-Q} is exactly the same as that in Eq. (\ref{eq:ma_q_learning}).  
\emph{Minimax-Q} can be considered an approximation algorithm for computing the fixed point $Q^*$ of the Bellman operator of Eq. (\ref{eq:operator_nash}) through stochastic sampling. Importantly, it assumes no knowledge about the environment. 
\cite{szepesvari1999unified} showed that under similar assumptions to those for Q-learning \cite{watkins1992q},   the Bellman operator of \emph{Minimax-Q} is a contraction mapping operator, and  the stochastic updates made by \emph{Minimax-Q}  eventually lead to  a unique fixed point  that corresponds to the NE value. In a nutshell, \emph{Minimax-Q} is an algorithm provably convergent to NE for two-player zero-sum SGs \cite{szepesvari1999unified}.

In addition to the tabular-form Q-function in \emph{Minimax-Q}, various Q-function approximators have been developed.  For example, \cite{lagoudakis2003learning} studied the factorised linear architectures for Q-function representation. 
\cite{yang2019theoretical} adopted deep neural networks and derived a rigorous finite-sample error bound. 
\cite{zhang2018finite} also derived a finite-sample bound for linear function approximators  in the competitive M-MDPs.   \cite{ding2022deep} shows empirical evidence with a DRL approach for achieving policies that are hard to exploit in two-player Atari games.

\emph{Nash Q-learning}\cite{hu2003nash} is one of the earliest work solving general-sum stochastic games, including zero-sum SG/MG. \emph{Nash Q-learning} provably converges to NE for general-sum games under the assumption that the NE is unique for each stage game during learning process. The update rule of \emph{Nash Q-learning} on a sample $(s, a^1, a^2, r, s')$ follows:
\begin{align}
    Q^*_h(s,a^1, a^2) &= r_h(s,a^1, a^2)+ V^*_{h+1}(s'), \forall h\in[H]\\
    (\pi^{1, *}_h, \pi^{2, *}_h) &= \textsc{Nash}(Q_h(s,\cdot, \cdot))\\
    V^{*}_{h+1}(s') &= \pi^{1, *}_h(\cdot|s)^\intercal Q_{h+1}(s', \cdot,\cdot) \pi_h^{2, *}(\cdot|s) \\
    Q_h(s,a^1, a^2) &= \alpha Q^*_h(s,a^1, a^2) + (1-\alpha) Q_h(s,a^1, a^2)
\end{align}
where $\textsc{Nash}$ is a NE solving subroutine on the payoff matrices as $\textsc{Nash}(Q)=\arg\max_{\pi^1} \arg\min_{\pi^2} (\pi^{1})^{\intercal} Q \pi^2$, $Q^*$ is the target value and $Q$ is the current estimation. It takes a soft update with a learning rate $\alpha$ for each sample $(s, a^1, a^2, r, s')$.
A variant of \emph{Nash Q-learning} called \emph{QVI-MDVSS}\cite{sidford2020solving} with the generative model for querying arbitrary samples and special variance reduction techniques proves $\tilde{\mathcal{O}}(\frac{H^3SAB}{\epsilon^2})$ for two-player zero-sum SG/MG.
For the exploration setting, the convergence rate of an optimistic version of \emph{Nash Q-learning}\cite{bai2020near} for achieving an $\epsilon$-approximate NE is $\tilde{\mathcal{O}}(\frac{H^5SAB}{\epsilon^2})$ for two-player zero-sum SG/MG. As a counterpart, in single-agent RL, \emph{Q-learning} with upper confidence bound achieves a convergence rate of $\tilde{{\Omega}}(\frac{H^4SA}{\epsilon^2})$\cite{jin2018q}, which matches with the lower bound of the model-free single-agent RL setting.

\emph{Nash Value Iteration} (Nash-VI) is the model-based version of \emph{Nash Q-learning}, which combines the Value Iteration (model-based algorithm in single-agent RL) with the NE operator on a matrix. The update rule of \emph{Nash Value Iteration} is similar to \emph{Nash Q-learning}:
\begin{align}
    Q^*_h(s,a^1, a^2) &= r_h(s,a^1, a^2)+ (\mathbb{P}_hV^*_{h+1})(s,a^1, a^2), \forall h\in[H]\\
    (\pi^{1, *}_h(\cdot|s), \pi^{2, *}_h(\cdot|s)) &= \textsc{Nash}(Q_h(s,\cdot, \cdot))\\
    V^*_h(s) &= \pi^{1, *}_h(\cdot|s)^\intercal Q_h(s, \cdot,\cdot) \pi_h^{2, *}(\cdot|s)\\
    Q_h(s,a^1, a^2) &= Q^*_h(s,a^1, a^2)
\end{align}
where $\textsc{Nash}$ is a NE solving subroutine on the payoff matrices, $Q^*$ is the target value and $Q$ is the current estimation. As a model-based algorithms, the transition function $\mathbb{P}_h, \forall h\in[H]$ needs to be estimated before the above update, which is different from the per-sample update in \emph{Nash Q-learning}. In the generative model setting (no consideration for exploration), a variant of Nash-VI\cite{zhang2020model} is proved to achieve $\tilde{\mathcal{O}}(\frac{H^3SAB}{\epsilon^2})$ sample complexity for finding $\epsilon$-approximate NE in two-player zero-sum SG/MG. 

\emph{Value Iteration with Upper/Lower Confidence Bound (VI-ULCB)}\cite{bai2020provable} is an extension of \emph{Upper Confidence Bound Value Iteration} (UCBVI)\cite{azar2017minimax} in single-agent RL to the multiagent setting, with upper confidence bound as exploration bonus. \emph{VI-ULCB} is proven to converge to an $\epsilon$-approximate NE within $\tilde{\mathcal{O}}(\frac{H^4S^2AB}{\epsilon^2})$ steps, which has as higher rate on $S$ but lower rate on $H$ compared with \emph{Nash Q-learning}. The theoretical lower bound for two-player zero-sum SG/MG is ${\Omega}(\frac{H^2S(A+B)}{\epsilon^2})$ for model-based algorithms, and  ${\Omega}(\frac{H^3S\max\{A, B\})}{\epsilon^2})$ for model-free algorithms. Q-learning type algorithms (e.g., Nash Q-learning) are model-free; while value iteration type algorithms, \emph{e.g.} VI-ULCB, are model-based.

As another variant of \emph{Nash-VI} for the exploration setting, \emph{Optimistic Nash Value Iteration (Optimistic Nash-VI)}\cite{liu2021sharp} is proposed to further improved the previous upper bounds. Nash-VI adopts an optimistic value estimation, but with inclusion of an auxiliary bonus apart from the bonus in upper/lower confidence bound for exploration, compared with \emph{VI-ULCB}, to achieve a convergence rate of $\tilde{\mathcal{O}}(\frac{H^3SAB}{\epsilon^2})$, which removes a $S$ term in \emph{UCBVI}. Moreover, Nash-VI finds Coarse Correlated Equalibirum (CCE) for multiplayer general-sum games with a convergence rate of $\tilde{\mathcal{O}}(\frac{H^4S^2\Pi_{i=1}^NA_i}{\epsilon^2})$.

\emph{Nash V-learning}\cite{bai2020near} combines the algorithm \emph{Follow-The-Regularized-Leader} (FTRL) with Q-learning in RL for two-player zero-sum games. With an optimistic value estimation for exploration, \emph{Nash V-learning} is proven to have a convergence rate of $\tilde{\mathcal{O}}(\frac{H^6S(A+B)}{\epsilon^2})$. As a model-free algorithm, \emph{Nash V-learning} is tight compared with the lower bound except for the dependency of the $H$ term. 

\emph{V-learning}\cite{jin2021v} is an algorithm breaking the curse of multiagents, which is seen in above VI-ULCB or Nash-VI algorithms with a dependency of $\Pi_{i=1}^NA_i$ in action spaces. It applies a decentralized learning manner for each agent to estimate $V$ value instead of $Q$ value, since $V$ value only has a dimensional dependency of $\mathcal{O}(S)$ while using the $Q$ value is $\mathcal{O}(S\Pi_{i=1}^NA_i)$. \emph{V-learning} achieves $\epsilon-$NE in two-player zero-sum Markov games for $\tilde{\mathcal{O}}(\frac{H^5S\max\{A,B\}}{\epsilon^2})$ steps. Furthermore, \emph{V-learning} can find $\epsilon-$Coarse Correlated Equilibrium (CCE) in $\tilde{\mathcal{O}}(\frac{H^5S\max\{A,B\}}{\epsilon^2})$ steps. As a concurrent work, CCE-V-learning\cite{song2021can} proves the same rate for the multi-player general-sum SG setting. Meanwhile, the information theoretical lower bound for getting $\epsilon$-NE in two-player zero-sum SGs is $\tilde{\Omega}(\frac{H^3S(A+B)}{\epsilon^2})$\cite{domingues2021episodic, zhang2020model}. More discussions about general-sum settings see Chapter~\ref{chaper:general_sum}.  A summarization of value-based methods for multi-player zero-sum SGs is shown in Table~\ref{tab:zero_sum_sg}.

\begin{table}[H]
\centering
\caption{Convergence guarantee for tabular two-player zero-sum stochastic games.}
\resizebox{\columnwidth}{!}{ 
\begin{tabular}{m{3cm}|m{3cm}|c|c}
\toprule
Type & Algorithm & Training Scheme &  Sample Complexity \\
\hline
\multirow{3}{*}{Model Free} 
& Optimistic Nash Q-learning\cite{hu2003nash, bai2020near}  & centralized & $\tilde{\mathcal{O}}(\frac{H^5SAB}{\epsilon^2})$ \\  \cline{2-4}
& Nash V-learning\cite{bai2020near}  & decentralized & $\tilde{\mathcal{O}}(\frac{H^6S(A+B)}{\epsilon^2})$ \\  \cline{2-4}
& V-learning\cite{jin2021v}  & decentralized &  $\tilde{\mathcal{O}}(\frac{H^5S\max\{A,B\}}{\epsilon^2})$  \\  \hline

\multirow{2}{*}{Model Based} & VI-ULCB\cite{bai2020provable}  & centralized & $\tilde{\mathcal{O}}(\frac{H^4S^2AB}{\epsilon^2})$  \\ \cline{2-4}
& Nash-VI\cite{liu2021sharp}  & centralized & $\tilde{\mathcal{O}}(\frac{H^3SAB}{\epsilon^2})$ \\  \hline
Lower Bound\cite{domingues2021episodic, zhang2020model}  &  &  & $\tilde{\Omega}(\frac{H^3S(A+B)}{\epsilon^2})$  \\
\bottomrule
\end{tabular}
}
\vskip -.2in
\label{tab:zero_sum_sg}
\end{table}

\subsection{Linear Function Approximation}
In recent years, there are also research beyond the tabular settings, which only considers finite and discrete cases. As a relatively simple class in function approximation settings, linear function approximation are studied in both single-agent MDP \cite{jin2020provably} and SGs \cite{xie2020learning, chen2021almost, jin2021power}.

\begin{definition}(Linear function class)
    The linear function class $\mathcal{F}$ is a set of functions linear in the features of state-action pairs. Specifically, for $\forall h\in[H]$,
\begin{align}
    \mathcal{F}_h:=\{\phi_h^\intercal \theta  | \theta \in B_d(R) \}
\end{align}
where $\phi_h:\mathcal{S}\times \mathcal{A}\times \mathcal{B} \rightarrow B_d(1)$ is the feature mapping from the state-action pair (actions for two players) to a $d$-dimensional vector in unit ball. $B_d(R)$ is a $d$-dimensional ball with radius $R$.
\end{definition}

\cite{xie2020learning} proposes an algorithm based on an optimistic version of \emph{Least Square Value Iteration}, which is a theoretically proved algorithm for single-agent MDP \cite{jin2020provably}, and the algorithm is called \emph{Optimistic Minimax Value Iteration} (OMNI-VI) due to the change of single-agent Bellman operator to a Nash/Minimax Bellman operator. The proof shows that the algorithm achieves $\tilde{\mathcal{O}}(\sqrt{d^3H^3 T})$ regret for finding NE in two-player zero-sum MGs, which corresponds to the sample complexity of $\tilde{\mathcal{O}}(\frac{d^3H^4}{\epsilon^2})$

\emph{Nash-UCRL} \cite{chen2021almost} is proposed to apply the ``optimism-in-face-of-uncertainty'' principle in two-player zero-sum MGs as well, and it is proven to achieve $\tilde{\mathcal{O}}(dH\sqrt{T})$ regret under linear function approximation for finding CCE of the game, where $T$ is the total steps in the game and $d$ is the linear function dimension. It corresponds to the sample complexity of $\tilde{\mathcal{O}}(\frac{d^2H^3}{\epsilon^2})$ This result is tighter than the results by \cite{xie2020learning}, and it matches with the information theoretic lower bound $\Omega(dH\sqrt{T})$ for two-player zero-sum MGs with linear structures.

\subsection{General Function Approximation}
To establish understanding towards more general function approximation methods like neural networks, which is proven to have universal function approximation capability with infinitely wide layers, researchers have investigated MGs with general function approximation.

\emph{GOLF-with-Exploiter}\cite{jin2021power} is a model-based algorithm for two-player zero-sum tabular MGs and MGs with linear or kernel function approximation. The results are proved within these function classes since the Bellman Eluder (BE) dimension\cite{jin2020provably} is as low as $\tilde{\mathcal{O}}(d)$. It applies an optimistic estimation for updating policy and exploiter within a confidence set, and the exploiter is adopted in the exploration process. \emph{GOLF-with-Exploiter} is proven to achieve a $\epsilon$-approximate NE in zero-sum MGs with a convergence rate of $\tilde{\mathcal{O}}(\frac{H^2d\log\mathcal{N}_\mathcal{F}}{\epsilon^2})$, where $d$ comes from the BE dimension of the function class, $\mathcal{N}_\mathcal{F}=|\mathcal{F}|$ is the cardinality of function class $\mathcal{F}$, $H$ is the horizon length of an episode by default.

\emph{Optimistic Nash Elimination for Markov Games} (ONEMG)\cite{huang2021towards} is proven to achieve $\tilde{\mathcal{O}}(H\sqrt{d_E K \log \mathcal{N}_\mathcal{F}})$ regret in $K$ episodes of learning, which also has $\tilde{\mathcal{O}}(\frac{1}{\epsilon^2})$ sample complexity. $d_E$ is the proposed Minimax Eluder dimension especially for the zero-sum game setting, as a counterpart of Bellman Eluder dimension in single-agent RL setting, $H$ is the horizon length of an episode by default. $\mathcal{N}_\mathcal{F}$ is the covering number of the general function class $\mathcal{F}$, therefore the rate has a logarithm dependence on the cardinality of the function class. The function class $\mathcal{F}$ here is the function approximation of the $Q^*$, which represents the $Q$ value for Nash equilibrium. For a game with horizon $H$ for each episode, the function class can be disentangled to be $\mathcal{F}=\mathcal{F}_1\times\dots\mathcal{F}_H, \mathcal{F}_h\subset \{f: \mathbb{S}\times \mathbb{A}_1\times \mathbb{A}_2 \rightarrow [0,1]\}$. For example, a parameterized NE value function (with parameters $\theta$) at timestep $h$ is assumed to be an instance within the function class $\mathcal{F}_h$: $Q^*_\theta\in \mathcal{F}_h$. In order to achieve the above regret guarantee, the realizability assumption and the completeness assumption are assumed to be true, which are detailed as following:
\begin{assumption}(Realizability). $Q^*_h\in \mathcal{F}_h, \forall h\in[H]$.
\end{assumption}

\begin{assumption}(Completeness). $\mathcal{T}_hf\in \mathcal{F}_h, \forall f\in\mathcal{F}_{h+1} h\in[H]$, where $\mathcal{T}_h$ is the min-max Bellman operator:
\begin{align}
    \mathcal{T}_h Q_{h+1}(s,a,b)=r(s,a,b)+\gamma \mathbb{E}_{s'\sim \mathbb{P}(\cdot|s,a,b)}[\max_\mu\min_\nu Q_{h+1}(s', \cdot, \cdot)] 
\end{align}
\end{assumption}

\emph{Decentralized Optimistic hypeRpolicy mIrror deScent} (DORIS) \cite{zhan2022decentralized} is a decentralized policy optimization algorithm for zero-sum MGs with general function approximation, and it is proven to achieve $\tilde{\mathcal{O}}(HV_{\max} \sqrt{K d_{BEE}\log\mathcal{N}})$ regret for $K$ episodes, where $V_{\max}$ is the maximum value of $V$-function, $d_{BEE}$ is another type of dimension for quantifying the complexity of function class called Bellman Evaluation Eluder dimension, $H$ is the horizon length. The proof of the rate requires the realizability and generalized completeness assumptions. Similar as ONEMG introduced above, the rate also has a $\sqrt{K}$ dependence on the $K$ and logarithm dependence on the cardinality $\mathcal{N}$ of the function class.

\subsection{Turn-Based Stochastic Games}

An important class of games that lie in the middle of SG and EFG is the two-player zero-sum turn-based SG (2-TBSG). 
In TBSG, the state space is split between two  agents, $\mathbb{S} = \mathbb{S}^1 \cup \mathbb{S}^2, \mathbb{S}^1 \cap \mathbb{S}^2 = \emptyset$, and in every time step, the game is in exactly one of the states, either $\mathbb{S}^1$ or $\mathbb{S}^2$. Two players alternate taking turns to make decisions, and  each state is  controlled\footnote{Note that since the game is turned based,  the Nash policies are deterministic.} by only one of the players $\pi^i:\mathbb{S}^i \rightarrow \mathbb{A}^i, i = 1,2$. The state then transitions into the next state with probability $P: \mathbb{S}^i \times \mathbb{A}^i \rightarrow \mathbb{S}^j, i,j=1,2$. Given a joint policy $\boldsymbol{\pi} = (\pi^1, \pi^2)$, the first player seeks to maximise the value function $V^{\boldsymbol{\pi}(s)}= \mathbb{E}\Big[\sum_{t=0}^{\infty}\gamma^t R\big(s_t, \pi(s_t)\big) |s_0=s \Big]$, while the second player seeks to minimize it, and the saddle point is the NE of the game.

Research on 2-TBSG leads to many important finite-sample bounds, i.e., how many samples one would need before reaching the NE at a given precision, for understanding multiagent learning algorithms.  
\cite{hansen2013strategy} extended \cite{ye2005new,ye2010simplex}'s result from single-agent MDP to 2-TBSG  and proved that the strongly polynomial time complexity of policy iteration algorithms also holds in the context of 2-TBSG if the payoff matrix is fully accessible.   
In the RL setting, in which the transition model is unknown, \cite{sidford2018near,sidford2020solving} provided a near-optimal Q-learning algorithm that computes an $\epsilon$-optimal strategy with high-probability given $\mathcal{O}\big((1-\gamma)^{-3}\epsilon^{-2}\big)$ samples from the transition function for each state-action pair. 
This result of polynomial-time sample complexity is remarkable since it was believed to hold for only single-agent MDPs. 
Recently, \cite{jia2019feature} showed  that if the transition model can be embedded in some state-action feature space, i.e., $\exists \psi_k(s') \text{ such that } P(s'|s,a)=\sum_{k=1}^{K}\phi_k(s,a) \psi_k(s'), \forall s' \in \mathbb{S}, (s, a) \in \mathbb{S} \times \mathbb{A} $,  then the sample complexity of the two-player Q-learning algorithm towards finding an $\epsilon$-NE  is only  linear to the number of features $\mathcal{O}\big(K/(\epsilon^2(1-\gamma)^4)\big)$.

All the above works focus on the offline domain, where they assume  that there exists an \emph{oracle} that can unconditionally provide  state-action transition samples. 
\cite{wei2017online} studied an online setting in an averaged-reward two-player SG. They achieved a polynomial sample-complexity bound if the opponent plays an optimistic best response, and a sublinear regret round against an arbitrary opponent.  

Robust reinforcement learning attempts to learn minimax optimal policies in the face of environment perturbations\cite{morimoto2005robust, pinto2017robust}. Robust RL is usually formulated as a stochastic game, where the adversary is able to perturb something, such as the action or transition dynamics, every timestep. Usually the adversary is assumed to be able to perturb within some pre-defined uncertainty set that stays the same every timestep. For example, the adversary might be able to add noise to a continuous action in an adversarial direction\cite{tessler2019action}.

\section{Extensive-Form Games}
\label{sec:efg_alg}
As briefly introduced in Section \ref{sec:efg-solve}, zero-sum EFG with imperfect information can be efficiently solved via LP in sequence form representations \cite{koller1992complexity,koller1996finding}.
However, these approaches are limited to solving only small-scale problems (e.g., games with $\mathcal{O}(10^7)$ information states). In fact, considerable additional effort is  needed to address real-world games (e.g., limit Texas hold'em, which has $\mathcal{O}(10^{18})$ game states); to name a few, Monte Carlo Tree Search (MCTS) techniques\footnote{Notably, though MCTS methods such as UCT \cite{kocsis2006bandit} work remarkably well in turn-based EFGs, such as GO and chess, they cannot converge to a NE trivially in (even perfect-information) simultaneous-move games \cite{schaeffer2009comparing}. See a rigorous treatment for remedy in \cite{lisy2013convergence}.} \cite{cowling2012information,browne2012survey,silver2016mastering}, isomorphic  abstraction techniques \cite{billings2003approximating, gilpin2006finding}, and iterative (policy) gradient-based approaches
\cite{gordon2007no,gilpin2007gradient, zinkevich2003online}. 

A central idea of iterative methods for EFGs is minimising regret\footnote{One can regard minimising regret as one solution concept for multiagent learning problems, similar to the reward maximisation in single-agent learning.}. 
A learning rule achieves no-regret, also called  \emph{Hannan consistency} in game theoretical terms \cite{hannan1957approximation}, if, intuitively speaking, against any set of opponents it yields a payoff that is no less than the payoff the learning  agent could have obtained by playing any one of its pure strategies in hindsight.
Recall the reward function under a given policy $\boldsymbol{\pi}=(\pi^i, \pi^{-i})$ in Eq. (\ref{eq:efg-reward}); the (average) regret of player $i$ is defined by:  
\begin{equation}
	\mathsf{Reg}_T^i = \frac{1}{T} \max_{\pi^i} \sum_{t=1}^T  \Big[R^i(\pi^i, \pi^{-i}_t) - R^i(\pi^i_t, \pi^{-i}_t) \Big].
	\label{eq:regret}
\end{equation}
A no-regret algorithm satisfies $	\mathsf{Reg}_T^i \rightarrow 0$ as $T \rightarrow \infty$ with probability $1$. 
When Eq. (\ref{eq:regret}) equals  zero,  all agents are acting with their best response to others, which essentially forms a NE. Therefore, one can regard regret as a type of ``distance'' to NE. 
As one would expect, the single-agent Q-learning procedure can be shown to be Hannan consistent in a stochastic game against opponents playing stationary policies \cite{shoham2008multiagent} [Chapter 7] since the optimal Q-function guarantees the best response.
In contrast, the Minimax-Q algorithm in Eq. (\ref{eq:comp_equilibrium}) is not Hannan consistent because if the opponent plays a sub-optimal strategy, Minimax-Q is unable to exploit the opponent due to the over-conservativeness in terms of over-estimating its opponents. 

An important result about regret states is that in a zero-sum game at time $T$, if both players' average regret is less than $\epsilon$, then their average strategy constitutes a $2\epsilon$-NE of the game \cite[Theorem 2]{zinkevich2008regret}. 
In general-sum games, the average strategy of the $\epsilon$-regret algorithm  will reach an $\epsilon$-\emph{coarse correlated equilibrium} of the game  \cite [Theorem 6.3.1]{dinitz2020}. 
This result essentially implies that regret-minimising algorithms (or, algorithms with Hannan consistency) applied in a self-play manner \cite{zhang2024survey} can be used as a general technique to approximate the NE of zero-sum games. 
 Building upon this finding, two families of methods are developed, namely, fictitious play types of methods \cite{berger2007brown} and counterfactual regret minimization \cite{zinkevich2008regret}, which lay the theoretical foundations for modern techniques to solve real-world games. 

It is worth noting that for EFG, a challenging setting involving imperfect information and sequential moves, accelerated dynamics for extensive-form correlated equilibria (EFCE) \cite{farina2022simple} and extensive-form coarse correlated equilibria (EFCCE) have been developed. These methods significantly improve convergence rates, introducing a refined perturbation analysis of Markov chains and efficient fixed-point characterizations \cite{anagnostides2022faster}.

\subsection{Model-Free Approaches}
The four families of model-free techniques for solving extensive-form games—tree-based methods, Double Oracle methods \cite{mcmahan2003planning}, Fictitious Play \cite{berger2007brown}, and policy-gradient-based methods—are closely interconnected, each addressing specific limitations of the others. Tree-based methods, such as Counterfactual Regret Minimization (CFR) \cite{zinkevich2008regret}, serve as a foundation by providing precise solutions for small-scale problems but face scalability challenges due to the exponential growth of the game tree. Double Oracle methods build upon this foundation by iteratively expanding a reduced strategy space, allowing efficient computation in larger games while still leveraging tree structures for solving subgames. 
Fictitious Play \cite{brown1951iterative} introduces a simpler iterative best-response approach that converges to Nash equilibria under specific conditions, offering a balance between computational simplicity and convergence guarantees in certain settings.
Policy-gradient-based methods extend these approaches further by parameterizing strategies and optimizing them directly through gradient ascent, making them suitable for continuous or extremely large strategy spaces where explicit enumeration is infeasible. Together, these techniques represent a progression from exact solutions to scalable approximations, each tailored to different problem complexities and requirements.

\subsubsection{Counterfactual Regret Minimization (CFR)}

Another family of methods achieve Hannan consistency by directly  minimising the regret, in particular,  a special kind of regret named counterfactual regret (CFR) \cite{zinkevich2008regret}. 
Unlike FP methods, which are developed from the stochastic approximation perspective and generally have asymptotic convergence guarantees,  CFR methods are established on the framework of online learning and online convex optimisation \cite{shalev2011online}, which makes analysing the speed of convergence, i.e., the regret bound, to the NE possible.

The key insight from CFR methods is that in order to minimize the total regret in Eq. (\ref{eq:regret}) to approximate the NE, it suffices to  minimize the \emph{immediate counterfactual regret} at the level of each information state. 
Mathematically, \cite{zinkevich2008regret} [Theorem 3] shows that the sum of the immediate counterfactual regret over all encountered information states provides an upper bound for the total regret in Eq. (\ref{eq:regret}),  i.e., 
\begin{equation}
\mathsf{Reg}_T^i \le \sum_{S \in \mathbb{S}^i} \max \Big\{ \mathsf{Reg}_{T, imm}^i (S), 0 \Big\}, \ \ \forall i. 
\label{eq:regret-bound}	
\end{equation}
To fully describe $ \mathsf{Reg}_{T, imm}^i (S)$, we need two additional notations. 
Let $\mu^{\boldsymbol{\pi}}(\boldsymbol{\sigma}_S \rightarrow \boldsymbol{\sigma}_{T})$ denote, given agents' behavioural policies $\boldsymbol{\pi}$, the realisation probability of going from the sequence $\boldsymbol{\sigma}_S$\footnote{Recall that for games of perfect recall, the sequence that leads to the information state, including all the choice nodes within that information state, is unique.}, which leads to the information state $S  \in \mathbb{S}^i$ to its extended sequence $\boldsymbol{\sigma}_{T}$, which continues from $S$ and reaches the terminal state $T$.  Let $\hat{v}^i(\boldsymbol{\pi}, S)$ be the \emph{counterfactual value function}, i.e., the expected reward of agent $i$ in non-terminal information state $S$, which is written as 
\begin{equation}
	\hat{v}^i \big(\boldsymbol{\pi}, S\big) = \sum_{s \in S, T \in \mathbb{T}} \mu^{\pi^{-i}}\big(\boldsymbol{\sigma}_s \big) \mu^{\boldsymbol{\pi}}(\boldsymbol{\sigma}_s \rightarrow \boldsymbol{\sigma}_{T}) R^i(T). 
	\label{eq:counter_value}
\end{equation}
Note that in Eq. (\ref{eq:counter_value}), the contribution from player $i$ in realising $\boldsymbol{\sigma}_s$ is excluded; we treat whatever action current player $i$ needs to reach state $s$ as having a probability of one, that is, $\mu^{\pi^i}(\boldsymbol{\sigma}_s)=1$. 
The motivation is that now one can make the value function $\hat{v}^i \big(\boldsymbol{\pi}, S\big)$ ``counterfactual'' simply by writing the consequence of player $i$ not playing action $a$ in the information state $S$ as $\big(\hat{v}^i(\boldsymbol{\pi}|_{S\rightarrow a}, S) - \hat{v}^i (\boldsymbol{\pi}, S)\big)$, in which $\boldsymbol{\pi}|_{S\rightarrow a}$ is a joint strategy profile identical to $\boldsymbol{\pi}$, except player $i$ always chooses action $a$ when information state $S$ is encountered. 
Finally, based on Eq. (\ref{eq:counter_value}), the immediate counterfactual regret can be expressed as 
\begin{equation}
	\mathsf{Reg}_{T, imm}^i (S) = \max_{a\in \chi(S)} \mathsf{Reg}_{T}^i (S, a)
	\label{eq:imr}
\end{equation}
where 
\begin{equation}
     \mathsf{Reg}_{T}^i (S, a)=  \dfrac{1}{T} \sum_{t=1}^{T} \Big( \hat{v}^i(\boldsymbol{\pi}_t|_{S\rightarrow a}, S) - \hat{v}^i (\boldsymbol{\pi}_t, S) \Big). 
	\label{eq:imr_1}
\end{equation}
Note that the $T$ in Eq. (\ref{eq:counter_value}) is different from that in Eq. (\ref{eq:imr_1}). 

Since minimizing the immediate counterfactual regret minimizes the overall regret, we can find an approximate NE  by choosing a specific behavioral policy $\pi^i(S)$ that minimizes Eq. (\ref{eq:imr_1}). 
To this end, one can apply Blackwell's approachability theorem \cite{blackwell1956analog} to minimize the regret independently on each information set, also known as \emph{regret matching} \cite{hart2001reinforcement}. 
As we are most concerned with positive regret, denoted by $\lfloor \cdot\rfloor_+$,  we have $\forall S \in \mathbb{S}^i, \forall a \in \chi(S)$, the strategy of player $i$ at time $T+1$ as Eq. (\ref{eq:regret-match_1}). 

\begin{align}
\pi^{i}_{T+1}(S, a)=\left\{\begin{array}{ll}\dfrac{ \lfloor \mathsf{Reg}_{T}^i \big(S, a\big) \rfloor_+}{\sum_{a \in \chi(S)} \lfloor \mathsf{Reg}_{T}^i \big(S, a\big) \rfloor_+} & \text {if } \sum_{a \in \chi(S)} \lfloor \mathsf{Reg}_{T}^i (S, a) \rfloor_+>0 \\ \dfrac{1}{|\chi(S)|} & \text {otherwise }\end{array}\right .
\label{eq:regret-match_1}
\end{align}

In the standard CFR algorithm,  for each information set, Eq. (\ref{eq:regret-match}) is used to compute action probabilities in proportion to the positive cumulative regrets. 
In addition to regret matching, another online learning tool that minimizes regret is \emph{Hedge}  
\cite{freund1997decision, littlestone1994weighted}, in which an exponentially weighted  function  is used to derive a new strategy, which is  
\begin{equation}
\pi_{t+1}(a_k) = \dfrac{\pi_{t}(a_k)e^{-\eta R_t(a_k)}}{\sum_{j=1}^{K} \pi_{t}(a_j)e^{-\eta R_t(a_j)}}, \ \  \pi_1(\cdot) = \dfrac{1}{K}. 
\label{eq:hedge_1}	
\end{equation}
In computing Eq. (\ref{eq:hedge_1}), Hedge needs access to the full information of the reward values for all actions, including those that are not selected. 
  \emph{EXP3}  \cite{auer1995gambling} extended the Hedge algorithm for a  \emph{partial information game} in which the player  knows only the reward of the  the chosen action (i.e., a bandit version) and has to estimate the loss of the actions that it does not select. 
\cite{brown2017dynamic} augmented the Hedge algorithm with a tree-pruning technique based on dynamic thresholding. 
 \cite{gordon2007no} developed  \emph{Lagrangian hedging}, which unifies no-regret algorithms, including both regret matching and Hedge, through a class of  potential functions. We recommend \cite{cesa2006prediction}  for a comprehensive overview of no-regret algorithms. 

No-regret algorithms, under the framework of online learning, offer a natural way to study the regret bound (i.e., how fast the regret decays with time). For example, CFR and its variants ensure a counterfactual regret bound of  $\mathcal{O}(\sqrt{T})$\footnote{According to \cite{zinkevich2003online}, any online convex optimization problem can be made to incur $\mathsf{Reg}_T=\Omega(\sqrt{T})$.}, 
as a result of Eq. (\ref{eq:regret-bound}), the convergence rate for the total regret is upper bounded by $\mathcal{O}(\sqrt{T}\cdot|\mathbb{S}|)$, which is linear in the number of information states. In other words, the average policy of applying CFR-type  methods in a two-player zero-sum EFG   generates an $\mathcal{O}(|\mathbb{S}|/\sqrt{T})$-approximate NE after $T$ steps through self-play\footnote{The self-play assumption can in fact be released. \cite{johanson2012finding} shows that in two-player zero-sum games, as long as both agents minimize their regret, not necessarily through the same algorithm,  their time-average policies will converge to NE with the same regret bound $\mathcal{O}(\sqrt{T})$. An example is to let a CFR player play against a best-response opponent. }.

Compared with the LP approach (recall Eq. (\ref{eq:zero-sum_efg})), which is applicable only for small-scale EFGs, the standard CFR method can be applied to limit Texas hold'em with as many as $10^{12}$ states.
CFR$^+$, the fastest implementation of CFR, can solve games with up to $10^{14}$ states \cite{tammelin2015solving}. 
However,  CFR methods still have a bottleneck in that computing Eq. (\ref{eq:counter_value}) requires a traversal of the entire game tree to the terminal nodes in each iteration. 
Pruning the sub-optimal paths in the game tree is a natural solution \cite{brown2015regret, brown2017dynamic,brown2017reduced}. 
 Many CFR variants have been developed to improve computational efficiency further. 
 \cite{lanctot2009monte} integrated Monte Carlo sampling with CFR (MCCFR) to significantly reduce the per iteration time cost of CFR by traversing a smaller sampled portion of the tree. 
\cite{burch2012efficient}
improved MCCFR by sampling only a subset of a player’s actions, which provides even faster convergence rate in games that contain many player actions. 
\cite{gibson2012generalized,schmid2019variance} investigated the sampling variance and proposed MCCFR variants with a variance reduction module. 
\cite{johanson2012efficient} introduced a more accurate MCCFR sampler by considering the set of outcomes from the chance node, rather than sampling only one outcome, as in all previous methods. \cite{xu2024minimizing} extended these developments by exploring the minimization of weighted counterfactual regret using an optimistic Online Mirror Descent (OMD) framework. This led to the proposal of a novel CFR variant, PDCFR+, which unifies the principles of PCFR+ and Discounted CFR (DCFR). By doing so, PDCFR+ effectively mitigates the negative effects of dominated actions while leveraging predictions to consistently accelerate convergence in a principled manner.

\paragraph{Deep CFR}
Apart from Monte Carlo methods, function approximation methods have also been introduced \cite{waugh2014solving, jin2018regret}. The idea of these methods is to predict regret directly, and the no-regret algorithm then uses these predictions in place of the true regret to define a sequence of policies. Deep CFR\cite{brown2019deep} uses neural networks to approximate CFR. Deep CFR iteratively samples trajectories like in MC-CFR and adds the MC-CFR estimators for information sets encountered to a large replay buffer. Then a regret network is trained on the replay buffer to approximate the cumulative counterfactual regret at an information set. This regret network is then used as a policy in the next iteration. 

Although Deep CFR outperforms NFSP on poker games, the variance of the MC-CFR estimator (due to importance sampling) causes the neural network to become unstable. To partially fix this issue, DREAM\cite{steinberger2020dream} uses a learned history-value function as a baseline\cite{schmid2019variance} to reduce variance. However, since the importance sampling term from MC-CFR remains, the variance in DREAM counterfactual regret targets become very large in games with long horizons. To achieve very low variance, ESCHER\cite{mcaleer2022escher} directly estimates counterfactual regret with a history advantage function. ESCHER is guaranteed to converge to an approximate Nash equilibrium with high probability and is shown to outperform DREAM and NFSP in large games. ARMAC\cite{gruslys2020advantage} predicts history advantages conditioned on the policy that generated each iteration and is competitive with NFSP. One potential downside of Deep-CFR-based methods opposed to last-iterate approaches is the need to output an average policy either via checkpointing policies or training an average policy network.



\subsubsection{Policy Gradient}
Self-play is a model-free policy-gradient method where two policies take policy gradient updates against each other simultaneously. Although self-play does not converge to a Nash equilibrium\cite{hennes2019neural}, it has been shown to scale up to large video games\cite{berner2019dota}. 

Interestingly, there exists a hidden equivalence between model-free policy-based/actor-critic MARL methods and the CFR algorithm \cite{jin2018regret, srinivasan2018actor}. In particular, if we consider the counterfactual value function in Eq. (\ref{eq:counter_value}) to be explicitly dependent on the action $a$ that player $i$ chooses at state $S$, in which we have $\hat{v}^i(\boldsymbol{\pi}, S) = \sum_{a\in \chi(S)} \pi^i({S, a}) \hat{q}^i(\boldsymbol{\pi}, S, a)$, then it is  shown in \cite{srinivasan2018actor} [Section 3.2] that  the Q-function in standard MARL $Q^{i, \boldsymbol \pi}(s, \mathbf{a}) = \mathbb{E}_{s'\sim P, \mathbf{a} \sim \boldsymbol{\pi}}\big[\sum_t \gamma^t R^i(s, \mathbf{a}, s')| s, \mathbf{a}\big]$ differs from $ \hat{q}^i(\boldsymbol{\pi}, S, a)$ in CFR only by a constant of the  probability of reaching $S$, that is, 
\begin{equation}
Q^{i, \boldsymbol \pi}\big(s, \mathbf{a}\big) =  \dfrac{\hat{q}^i\big(\boldsymbol{\pi}, S, a\big)}  {\sum_{s \in S} \mu^{\pi^{-i}}\big(\boldsymbol{\sigma}_s \big)}.
\label{eq:cfr-q}	
\end{equation}
Subtracting a value function on both sides of Eq. (\ref{eq:cfr-q}) leads to the fact that the counterfactual regret of $\mathsf{Reg}_T^i(S, a)$ in Eq. (\ref{eq:imr}) differs from the advantage function in MARL, i.e., $Q^{i, \boldsymbol \pi}(s, a^i, a^{-i}) - V^{i, \boldsymbol \pi}(s, a^{-i})$, only by a constant of the realisation probability. 
As a result,  the multiagent actor-critic algorithm \cite{comafoerster} can be formulated as a special type of CFR method, thus sharing a similar convergence guarantee and regret bound in two-player zero-sum games.  

Neural Replicator Dynamics (NeuRD)\cite{hennes2019neural} approximates replicator dynamics in normal form games via a modified policy gradient update. Since replicator dynamics is equivalent to Hedge\cite{kleinberg2009multiplicative}, if one runs NeuRD independently as a counterfactual-regret minimizer in CFR, this algorithm approximates CFR with Hedge, which is known to converge to a Nash equilibrium in self play. The NeuRD update can also be used as a policy gradient update in self play. The resulting algorithm does not have theoretical guarantees but has demonstrated approximate convergence in some games. 

Friction-Follow-the-Regularized-Leader (F-FoReL)\cite{perolat2021poincare, perolat2022mastering} modifies the reward function of a game by adding a relative entropy term comparing the current policy and a reference policy. By modifying the reward in this way, any no-regret algorithm achieves last-iterate convergence to a modified equilibrium. The full algorithm carries out this procedure multiple times, where the reference policy is updated every iteration to be the output of the last iteration. This procedure converges to an approximate Nash equilibrium in the last iterate in normal-form games. In extensive-form games, the NeuRD update is used to update the policy and the reward is modified at every information set. Although there are not convergence guarantees for F-FoReL in extensive-form games, F-FoReL has achieved the impressive result of beating top humans at the very large game of Stratego\cite{perolat2022mastering}. This is a model-free deep reinforcement learning method, without search, that learns to master Stratego through self-play from scratch. This regularization method called "reward transformation" has given rise to a series of modern EFG solving methods that do not require search \cite{meng2023efficient,}.

Similar to F-FoReL, Magnetic Mirror Descent (MMD)\cite{sokota2022unified} is a policy-gradient method that uses regularization to converge in self-play. MMD is similar to mirror descent except that is adds an additional policy regularization term. MMD is guaranteed to converge in the last iterate to a QRE in normal-form games. Using the sequence form and dilated entropy as the divergence, MMD also converges in the last iterate to a QRE in extensive-form games. In deep RL experiments, the authors show that a behavioral-strategy update without convergence guarantees outperforms NFSP on large games. 

Another policy-gradient approache is Actor Critic Hedge (ACH)\cite{fu2021actor}. ACH is an actor-critic method where the critic is an information-state advantage function and the policy is given by Hedge. Experiments show ACH is competitive with NeuRD, RPG, and QPG. 

\subsubsection{Double Oracle}
\label{sec:metagame-review}

\begin{algorithm}[t!]
 \caption{Policy Space Response Oracles (PSRO)\cite{mcmahan2003planning, lanctot2017unified}}\label{psro_alg-1}
 \begin{algorithmic}[1]
 \State \textbf{Initialise:} the ``high-level'' policy set $\mathbb{S}=\prod_{i \in \mathcal{N}}\mathbb{S}^i$,  the meta-game payoff $\tM, \forall S \in \mathbb{S}$, and meta-policy $\vpi^{i} = \operatorname{UNIFORM}(\mathbb{S}^i)$.
 \State \textbf{for} iteration $t \in \{1,2, ... \}$ \textbf{do}:
  \State  \hspace{1em}  \textbf{for} each player $i \in \mathcal{N}$ \textbf{do}:
\State \hspace{2em}  Compute the meta-policy $\vpi_t$ by meta-game solver $\mathcal{S}(\tM_t)$.
\State \hspace{2em} Find a new  policy against others by Oracle: $S^i_t = \mathcal{O}^i(\vpi_t^{-i})$.
\State \hspace{2em} Expand $\mathbb{S}_{t+1}^i  \leftarrow \mathbb{S}_t^i \cup \{S_t^i\}$ and update meta-payoff $\tM_{t+1}$.
\State \hspace{1em} \textbf{terminate if:} $\mathbb{S}^i_{t+1} = \mathbb{S}^i_t, \forall i \in \mathcal{N}$.  
\State \textbf{Return:} $\vpi$ and $\mathbb{S}$. 
\end{algorithmic}
 \end{algorithm}

In solving real-world zero-sum games, such as Go or StarCraft, since the number of pure strategies can be very large, one feasible approach instead is to focus on \emph{meta-games}. 
A meta-game  is  constructed by considering sets of policies for each player, called populations. Populations of policies implicitly define a normal-form meta game where the actions correspond to policies in a population and the payoffs correspond to the expected value when the corresponding policies are played against each other. Ideally, small populations can be found such that some mixture of them will approximate a Nash equilibrium in the full game. 

The double oracle algorithm (DO)\cite{mcmahan2003planning} iteratively builds a meta-game by computing a NE of the meta-game and then adding best responses for both players to the opponent's meta-distribution. 
Policy Space Response Oracles (PSRO)\cite{neu2017unified} is a direct extension of  double oracle \cite{mcmahan2003planning} that uses an RL subroutine as an approximate best response. 
Specifically, one can write PSRO and its variations in Algorithm \ref{psro_alg-1}, which essentially involves an iterative two-step process of solving for the meta-policy first (e.g., Nash over the meta-game), and then based on the meta-policy, finding a new better-performing policy, against the opponent's current meta-policy, to augment the existing population.  
The meta-policy solver, denoted as $\mathcal{S}(\cdot)$,  computes a joint meta-policy profile  $\vpi$ based on the current payoff $\tM$ where different solution concepts can be adopted (e.g., NE). 
Finding a new policy is equivalent to solving a single-player optimisation problem given opponents' policy sets $\mathbb{S}^{-i}$ and meta-policies $\vpi^{-i}$, which are fixed and known. 
One can regard a new policy as given by an \emph{Oracle}, denoted by $\mathcal{O}$. In two-player zero-sum cases, an oracle represents $\mathcal{O}^1(\vpi^2) =\{S^1: \sum_{S^2\in \mathbb{S}^2} \vpi^2(S^2) \cdot \phi (S^1, S^2) > 0 \}$. 
Finally, after a new policy is found, the payoff table $\tM$ is expanded, and the missing entries are filled by running new game simulations. 
The above two-step process loops over each player at each iteration, and it terminates if no new policies can be found for any players.  

Algorithm \ref{psro_alg-1} is a general framework, with appropriate choices of meta-game solver $\mathcal{S}$ and oracle $\mathcal{O}$, it can represent solvers for different types of meta-games. 
For example, it is trivial to see that FP/GWFP is recovered when $\mathcal{S}=\operatorname{UNIFORM}(\cdot)$ and $\mathcal{O}^i=\mathbf{Br}^i(\cdot)/\mathbf{Br}_\epsilon^i(\cdot)$. 
The double oracle \cite{mcmahan2003planning} and PSRO methods \cite{lanctot2017unified} refer to the cases when the meta-solver computes NE.

PSRO has advantages and disadvantages compared to other approaches such as CFR-based approaches and policy-gradient-based approaches. One advantage is that PSRO can potentially terminate a lot faster than other approaches. If PSRO happens to add pure strategies that support a Nash equilibrium of the game in few iterations then PSRO can terminate very early. Intuitively, games that are highly transitive or that require many non-strategic skills might be good candidates for PSRO. For example, Starcraft requires low-level skills that do not need to be mixed over, and a PSRO-based method was able to beat expert humans\cite{vinyals2019alphastar}. Also, PSRO provides a natural measure of approximate exploitability, which can be useful when evaluating progress in large games. A few downsides of PSRO are as follows. First, PSRO requires training best responses sequentially, which can be time-consuming. Second, PSRO can arbitrarily increase exploitability from one iteration to the next, which can be a problem when terminating early. Third, PSRO might need to expand all pure strategies in the game, which is exponential in the number of information sets. 

Pipeline PSRO\cite{mcaleer2020pipeline} solves the first problem by pre-training multiple best responses simultaneously while maintaining PSRO convergence guarantees. Anytime PSRO\cite{mcaleer2022anytime} is a version of PSRO that does not increase exploitability. It does so by adaptively modifying the restricted distribution via a no-regret algorithm against the opponent best response while the opponent best response is training. The resulting restricted distribution approximates the least-exploitable restricted distribution, which only gets less exploitable as more policies are added to the population. To address the final problem, Neural Extensive-Form Double Oracle (NXDO)\cite{mcaleer2021xdo} constructs an extensive-form restricted game instead of a normal-form restricted game. Instead of sampling one policy at the root of the game and then continuing to play from that policy, NXDO allows players to switch between policies in the population from one information set to another. As a result, convergence is guaranteed in a number of iterations that is linear in the number of information sets, as opposed to exponential for PSRO. 
Despite these improvements, PSRO and NXDO still lack regret bounds or rates of convergence besides the trivial bound when all strategies are added. Also, after the restricted distribution becomes strong enough, training the best response from scratch via RL can be challenging.
Regret-Minimizing Double Oracle provides a unified theoretical framework to analyze these methods \cite{tang2023regret, tang2024sample}, including their convergence rates and sample complexity. This framework enables extending ODO to extensive-form games while determining its sample complexity and highlights the exponential sample complexity of XDO \cite{mcaleer2021xdo} due to the decaying stopping threshold of restricted games.

 \subsubsection{Fictitious Play}

 Fictitious play (FP) \cite{berger2007brown}  is one of the oldest learning procedures in game theory that is provably convergent for zero-sum games, potential games, and two-player n-action games with generic payoffs. Although the algorithm guarantees convergence to the NE, the convergence is not efficient in the worst case \cite{daskalakis2014counter}.
 In FP, each player maintains a belief about the empirical mean of the opponents' average policy, based on which the player selects the best response.  
 With the best response defined in Eq. (\ref{eq:best_response}), we can write the FP updates as:  \par
 {\small
 \begin{align}
 a^{i,*}_{t} \in \mathbf{Br}^{i}\Big( \pi^{-i}_t =  \dfrac{1}{t} \sum_{\tau=
 0}^{t-1} \mathds{1}\left\{a^{-i}_{\tau}=a, a\in \mathbb{A}\right\}\Big), \pi^i_{t+1} = \Big(1 - \dfrac{1}{t}\Big)\pi^i_{t} + \dfrac{1}{t} a^{i,*}_{t}, \forall i.
 \label{eq:fp}	
 \end{align}}
 
 In the FP scheme, each agent is oblivious to the other agents' reward; however, they need full access to their own payoff matrix in the stage game. 
In the continuous case with an infinitesimal learning rate of $1/t \rightarrow 0$,  Eq. (\ref{eq:fp}) is equivalent to $d \boldsymbol{\pi}_t / dt \in \mathbf{Br}\big(\boldsymbol{\pi}_t \big) - \boldsymbol{\pi}_t $ in which $\mathbf{Br}(\boldsymbol{\pi}_t) = \big(\mathbf{Br}(\pi^{-1}_t),..., \mathbf{Br}(\pi^{-N}_t) \big) $.
 \cite{viossat2013no} proved that continuous FP leads to no regret and is thus Hannan consistent. 
 If the empirical distribution of each $\pi^i_t$ converges in FP, then it converges to a NE\footnote{Note that the convergence in Nash strategy does not necessarily mean the agents will receive the expected payoff value at NE. In the example of Rock-Paper-Scissor games, agents' actions are still miscorrelated after convergence, flipping between one of the three strategies, though their average policies do converge to $(1/3, 1/3, 1/3)$. }.

 Although standard discrete-time FP is not Hannan consistent \cite[Exercise 3.8]{cesa2006prediction}, various extensions have been proposed that guarantee such a property; see a full list summarised in \cite{hart2013simple} [Section 10.9].
 Smooth FP \cite{fudenberg1993learning,fudenberg1995consistency} is a stochastic variant of FP (thus also called stochastic FP) that considers a smooth $\epsilon$-best response in which the probability of each action is a softmax function of that action's utility/reward  against the historical frequency of the opponents' play. 
 In smooth FP, each player's strategy is a genuine mixed strategy. Let $R^i(a^i_1, \pi^{-i}_t)$ be the expected reward of player $i$'s action $a^i_1 \in \mathbb{A}^i$ under opponents' strategy $\pi^{-i}$;  the probability of playing $a^i_1$ in the best response is written as 
 \begin{equation}
\mathbf{Br}^{i}_{\lambda}(\pi_t^{-i}):=\frac{\exp \left(\frac{1}{\lambda} R^i\left(a^{i}_{1}, \pi_t^{-i}\right)\right)}{\sum_{k=1}^{|\mathbb{A}^i|} \exp \left(\frac{1}{\lambda} R^i\left(a^{i}_{k}, \pi_t^{-i}\right)\right)}	.
\label{eq:smooth-fp}
 \end{equation}
 \cite{benaim2013consistency} verified the Hannan consistency of the  smooth best response with the smoothing parameter $\lambda$ being time dependent and vanishing  asymptotically. 
 In potential games, smooth FP is known to converge to a neighbourhood of the set of NE \cite{hofbauer2002global}. Recently, \cite{swenson2019smooth} showed a generic result that in almost all $N \times 2$ potential games, smooth FP converges to the neighbourhood of a pure-strategy NE with a probability of one.

 In fact, ``smoothing'' the cumulative payoffs before computing the best response is crucial to designing learning procedures that achieve Hannan consistency \cite{kaniovski1995learning}. 
 One way to achieve such smoothness is through stochastic smoothing or adding perturbations\footnote{The physical meaning of perturbing the cumulative payoff is to consider the incomplete information about what the opponent has been playing, variability in their payoffs, and unexplained trembles.}.
 For example, the smooth best response in Eq. (\ref{eq:smooth-fp}) is a closed-form solution   if one perturbs the cumulative reward by an additional entropy function, that is,   
 \begin{equation}
 	\pi^{i,*} \in   \mathbf{Br}(\pi^{-i}) = \Big\{ \arg\max_{\hat{\pi}\in\Delta(\mathbb{A}^i)} \mathbb{E}_{\hat{\pi}^{i}, \pi^{-i}}\big[R^i + \lambda \cdot \log(\hat{\pi})\big] \Big\}.
 	\label{eq:smoothfp}
 \end{equation}
Apart from smooth FP,  another way to add perturbation is the \emph{sampled FP} in which during each round, the player samples historical time points using a randomised sampling scheme,  and plays the best response  to the other players' moves, restricted to the set of sampled time points. Sampled FP is shown to be Hannan consistent when used with Bernoulli sampling \cite{li2018sampled}.

Among the many extensions of FP, the most important is probably \emph{generalised weakened FP (GWFP)} \cite{leslie2006generalised}, which releases the standard FP by allowing both approximate best response and perturbed average strategy updates. 
Specifically, if we write the $\epsilon$-best response of player $i$ as
 \begin{equation}
 R^i\Big(\mathbf{Br}_{\epsilon}(\pi^{-i}), \pi^{-i}\Big) \ge \sup_{\pi\in \Delta(\mathbb{A}^i)}R^i\Big(\pi, \pi^{-i} \Big) - \epsilon	.
 \label{eq:br_e}
 \end{equation}
 then the GWFP updating steps change from Eq. (\ref{eq:fp}) to 
 \begin{equation}
\pi^i_{t+1} = \Big(1 - \alpha^{t+1}\Big)\pi^i_{t} + \alpha_{t+1} \Big({\color{blue}\mathbf{Br}^i_{\epsilon}}(\pi^{-i}) + {\color{red}{M^i_{t+1}}}  \Big),  \ \ \ \ \forall i .
 \label{eq:gwfp}	
 \end{equation}
GWFP is Hannan consistent if $\alpha_t \rightarrow 0, \epsilon_t \rightarrow 0, \sum_{\alpha_t}=\infty$ when $t \rightarrow \infty$, and $\{M_t\}$ meets $\lim _{t \rightarrow \infty} \sup _{k}\big\{\big\|\sum_{i=t}^{k-1} \alpha^{i+1} M^{i+1} \big\| \text { s.t. } \sum_{i=t}^{k-1} \alpha^{i+1} \leq T \big\}=0$. 
It is trivial to see that GWFP recovers FP when $\alpha_t=1/t, \epsilon_t=0, M_t=0$. 
GWFP is an important extension of FP in that it provides two key components for  bridging game theoretic ideas with RL techniques.  
With the approximate best response (highlighted in blue, also named as the ``weakened'' term), this approach allows one to adopt a model-free RL algorithm, such as deep Q-learning, to compute the best response. Moreover, the perturbation term (highlighted in red, also named as the ``generalised'' term) enables one to incorporate policy exploration; if one applies an entropy term as the perturbation in addition to the best response  (in which the smooth FP in Eq. (\ref{eq:smoothfp}) is also recovered), the scheme of  maximum-entropy RL methods \cite{haarnoja2018soft} is recovered.
In fact, the generalised term also accounts for the perturbation that comes from the fact the beliefs are not updated towards the exact mixed strategy $\pi^{-i}$ but instead towards the observed actions \cite{benaim1999mixed}.  
As a direct application,  \cite{perolat2018actor} implemented the GWFP process through an actor-critic framework \cite{konda2000actor} in the MARL setting. 
 
Brown's original version of FP \cite{berger2007brown} describes alternating updates by players; yet, the modern usage of FP involves players updating their beliefs simultaneously \cite{berger2007brown}.
 In fact,  \cite{heinrich2015fictitious} only recently proposed the first FP algorithm for EFG using the sequence-form representation.
   The extensive-form FP is essentially an adaptation of GWFP from NFG to EFG based on the insight that a mixture of normal-form strategies can be implemented by a weighted combination of behavioural strategies that have the same realisation plan  (recall Section \ref{sec:sequence-form}). 
 Specifically, let $\pi$ and $\beta$ be two behavioural strategies, $\Pi$ and $B$ be the two realisation-equivalent mixed strategies\footnote{Recall that in games with perfect recall, Kuhn's theorem \cite{kuhn1950extensive} suggests that the behavioural strategy and mixed strategies are equivalent in terms of the realisation probability of different outcomes.}, and $\alpha \in \mathbb{R}^+$; then, for each information state $S$, we have 
 \begin{equation}
 \tilde{\pi}(S) = \pi(S) + \dfrac{\alpha \mu^{\beta}(\sigma_S) }{(1-\alpha) \mu^{\pi}(\sigma_S) + \alpha \mu^{\beta}(\sigma_S)} \Big(\beta(S) - \pi(S)  \Big)    , \ \ \forall S \in \mathbb{S}, 
 \label{eq:efg-fp}	
 \end{equation}
where $\sigma_S$ is the sequence leading to $S$,  $\mu^{\pi/\beta}(\sigma_S)$ is the realisation probability of $\sigma_S$ under a given policy,  and
 $ \tilde{\pi}(S)$ defines a new behaviour  that is realisation equivalent to the mixed strategy $(1-\alpha)\Pi + \alpha B$.
 The extensive-form FP essentially iterates between Eq. (\ref{eq:br_e}), which computes the $\epsilon$-best response, and Eq. (\ref{eq:efg-fp}), which updates the old behavioural strategy with a step size of $\alpha$. 
 Note that these two steps must iterate over all information states of the game in each iteration. 
  Similar to the normal-form FP in Eq. (\ref{eq:fp}), extensive-form FP generates a sequence of $\{\boldsymbol \pi_t\}_{t\ge1}$ that provably converges to the NE of a zero-sum game under self-play if the step size $\alpha$ goes to zero asymptotically. 
  As a further enhancement,  \cite{heinrich2016deep} implemented  neural fictitious self-play (NFSP),  in which the best response step is computed by deep Q-learning \cite{mnih2015human} and the policy mixture step is computed through supervised learning.  
  NFSP requires the storage of large replay buffers of past experiences; 
   \cite{lockhart2019computing} removes this requirement by obtaining the policy mixture  for each player through an independent policy-gradient step against the respective best-responding opponent. 
All these amendments help make extensive-form FP  applicable to  real-world games with large-scale information states.  

\subsection{Search}
In this section we briefly outline techniques for performing search in two-player zero-sum games. The previous sections in this chapter covered model-free techniques. These techniques are useful for obtaining a blueprint policy which roughly approximates a Nash equilibrium. However, usually these blueprint policies are not exact enough to achieve superhuman performance. To achieve superhuman performance on games such as chess, go, and poker, search was required. 

Search in two-player zero-sum extensive-form games refers to any technique that improves its blueprint policy at test time against an opponent. At a given information set, a search technique will perform \emph{subgame solving} to approximately solve the remaining subgame. Hopefully, the resulting strategy at that information set will be better than the blueprint policy. 

\subsubsection{Perfect-Information Search}
In perfect-information games, to make a decision at a state, all that is needed is to consider all future states reachable from that state. As a result, most perfect-information search techniques have a recursive structure. The simplest example is min-max exhaustive search\cite{russell2010artificial}. Min-max search recursively solves all subgames nested inside a given subgame. The simplest subgame and the base case are subgames that only contain the terminal nodes. These subgames are solved by assuming that the max player will play the maximal action or the min player will play the minimal action. In subgames that are not terminal nodes, the value computed by lower level subgames are passed up, and the player then chooses the best action of the available actions. The resulting policy is the Nash equilibrium of that subgame, because at each decision point, each player acted min-max optimally. 

Of course, exhaustive search can only work in games as small as tic-tac-toe. Many techniques exist for scaling up this idea to large games. Instead of performing exhaustive search, these methods truncate the search by substituting the value of a subgame with a heuristic based on that state. For example, in chess this heuristic could be based on the number of pieces and the position. These techniques have lead to superhuman performance at chess and checkers\cite{campbell2002deep, schaeffer1996chinook}. 

More recently, deep learning techniques have been used to find heuristics for even larger games. AlphaGo\cite{silver2016mastering} uses a deep neural network to approximate the value of a state for a given player. This heuristic is used in Monte-Carlo Tree Search (MCTS)\cite{browne2012survey} as a substitute for the value of a subgame at a particular state. AlphaGo trained the value network and policy using a large amount of human data. AlphaZero\cite{silver2017mastering} was able to learn a value and policy network from scratch by training it on rollouts from self-play games of the prior version of itself. 

\subsubsection{Imperfect-Information Search}
Imperfect-information search is considerably harder than perfect-information search. This is because the information needed to act in an information set does not just depend on states that can be reached from the current state, but also on everywhere else in the game tree. 

As a result, subgame solving in imperfect information games must either construct \emph{gadget games} or reason about \emph{public belief states}. Gadget games are games that are constructed at a subgame so that once solved, the strategy found from solving the gadget game can be used for that subgame in the original game. Unsafe subgame solving methods\cite{ganzfried2015endgame, gilpin2006texas, gilpin2007better, billings2003approximating} have no guarantees on whether the exploitability of the new strategy will be lower than the original blueprint strategy. Safe subgame solving\cite{burch2014solving} constructs gadget games in a way that the opponent can opt out and receive the utility of the blueprint strategy. As a result, by solving this type of subgame, the new strategy is guaranteed to be less exploitable than before. Max-margin subgame solving\cite{moravcik2016refining} maximizes the difference in exploitability between the blueprint and the new strategy. Safe and nested subgame solving improves max-margin subgame solving further by considering all subgames in conjunction and was shown to achieve superhuman performance in poker\cite{brown2017safe}.  
However, current subgame-solving techniques typically analyze the entire common-knowledge closure of the player's current information set, encompassing all nodes within which it is common knowledge that the current node lies. While effective in games with relatively simple information structures like poker, these methods are impractical for games with complex information structures, where the common-knowledge closure becomes infeasibly large to enumerate or approximate. To address this limitation, recent work introduced $k$-KLSS (order-$k$ knowledge-limited subgame solving), a novel subgame-solving approach that leverages low-order knowledge instead of relying on the common-knowledge closure \cite{zhang2021subgame}. This method enables agents to prune unreachable nodes upon encountering an information set, significantly reducing the game tree size relative to traditional methods. 

Methods based on public belief states such as DeepStack\cite{moravvcik2017deepstack} and ReBeL\cite{brown2020combining} keep an explicit distribution over all possible hidden states and perform search in this space.

\subsubsection{Imperfect-Information Subgame Solving}
Imperfect-information subgame solving has undergone significant advancements, especially through the development of core algorithms, real-time techniques, and scalable refinements. Key theoretical foundations include decomposition methods such as CFR-D \cite{burch2013cfr}, which offers memory-efficient solutions with full-game consistency guarantees, addressing computational limits in large imperfect-information games (IIGs) \cite{burch2014solving,burch2013cfr}. Real-time solving methods, particularly nested solving \cite{brown2017safe}, allow dynamic refinement of strategies during games, significantly reducing exploitability in real-time settings like poker. Techniques such as depth-limited solving \cite{brown2018depth} provide a balance between decision quality and memory usage, crucial for handling large state-space domains while ensuring safety against adversary exploitation. Additionally, methods like subgame margin help manage exploitability during transitions between precomputed strategies and refined subgames \cite{moravcik2016refining}, further improving the robustness of subgame-solving algorithms.

Despite substantial progress, scalability and abstraction remain significant challenges. Abstraction refinement techniques, such as lossless or dynamic refinements \cite{moravcik2016refining,ganzfried2014potential}, ensure that simplified representations of games do not compromise the accuracy of subgame solutions. Approaches like Earth Mover’s Distance in clustering help optimize state aggregation but require careful balancing to avoid over-simplification \cite{ganzfried2014potential}. Furthermore, frameworks like Generative Subgame Solving \cite{ge2024efficient} and Opponent-Limited Subgame Solving \cite{liu2023opponent} provide scalability by reducing subgame tree sizes, a crucial step for handling high-dimensional games and adversarial setups. These frameworks have shown potential for large-scale games like adversarial team scenarios, but their applicability beyond poker remains underexplored, with generalizability to other domains like economic models and security games still facing hurdles.

The real-world application of these techniques, particularly in poker, has validated their effectiveness, with systems like Libratus \cite{sandholm2010state} and DeepStack \cite{moravvcik2017deepstack} showcasing superhuman performance by combining subgame refinement and recursive CFR. However, when extending these techniques to non-poker domains, such as adversarial team games or sequential auctions \cite{sandholm2010state}, scalability and the control of exploitability remain unresolved challenges. Further advancements in imperfect-recall solutions \cite{ganzfried2014potential, vsustr2020sound}, opponent modeling \cite{milec2021continual}, and methods to control safety and exploitability are crucial for broadening the applicability of subgame-solving algorithms \cite{liu2022safe, vsustr2020sound}. As these methods evolve, the challenge will be to integrate them into more diverse, real-world settings while managing the computational complexity and ensuring robust performance across domains.

\subsubsection{Game Abstraction in Imperfect-Information Games}
Game abstraction plays a vital role in solving large-scale, imperfect-information games by simplifying the computational complexity involved in reasoning about vast strategy spaces \cite{Sandholm15:Abstraction}. The key objective is to reduce the game’s complexity while preserving enough information to make effective decisions. Early research focused on three main types of abstraction: state \cite{li2006towards}, action \cite{gilpin2009algorithms}, and information abstractions \cite{kroer2014extensive}. State abstraction groups similar states together to reduce the number of possibilities agents need to consider, making it computationally feasible to analyze large games like poker \cite{gilpin2008expectation}. Action abstraction reduces the set of available actions to the most strategic moves, helping agents focus on key decisions. Information abstraction, such as imperfect-recall methods \cite{kroer2014extensive}, simplifies hidden information, making it more manageable for agents to compute strategies. While temporal abstractions \cite{xu2024strategy} remain less explored, they hold promise for simplifying decision-making over extended horizons.

Modern approaches in game abstraction have focused on balancing between lossless and lossy abstractions \cite{gilpin2009algorithms}. Lossless abstractions maintain the full strategic fidelity of the game \cite{gilpin2007lossless}, but they are often computationally infeasible for large-scale settings due to the vastness of strategy spaces. On the other hand, lossy abstractions reduce the game’s complexity at the expense of some strategic accuracy. Despite this trade-off, lossy abstractions \cite{sandholm2012lossy} have been equipped with rigorous theoretical guarantees, such as bounds on error propagation and exploitability \cite{Kroer18:Unified_ai_3}, ensuring that the abstraction does not overly compromise the quality of the strategies derived from it. Additionally, dynamic and adaptive refinement techniques \cite{brown2015simultaneous, bard2014asymmetric} have been introduced to improve abstraction methods in real-time. These techniques, such as simultaneous abstraction and equilibrium finding, dynamically refine abstractions as the game progresses, leading to more efficient computations and higher-quality solutions.

The scalability of these abstraction methods has been significantly enhanced through unified frameworks that combine various types of abstraction into a single, scalable system \cite{li2006towards}. These frameworks ensure that different abstraction methods can be applied cohesively across different domains, providing both computational efficiency and theoretical guarantees. Notable examples include systems that integrate Nash equilibrium solvers with abstraction pipelines, enabling the solution of massive strategy spaces in complex games like poker \cite{Sandholm15:Abstraction, Kroer14:Extensivea}. Furthermore, abstraction methods have shown practical success in real-world applications, particularly in poker, where techniques like Libratus have demonstrated superhuman performance by continuously refining abstractions throughout games \cite{sandholm2010state}. Despite these advances, challenges remain, particularly in multi-agent and non-zero-sum games, where abstraction methods struggle to provide guarantees on solution quality. Future research will likely focus on addressing these limitations and extending abstraction techniques to more dynamic, multi-agent environments, as well as exploring under-researched areas like temporal abstractions and long-horizon decision-making.

\section{Online Markov Decision Processes}

A common situation in which online learning techniques are applied is in stateless games, where the learning agent faces an identical decision problem in each trial (e.g., playing a multi-arm bandit in the casino). 
However,  real-world decision problems often occur in a dynamic and changing environment.
  Such an environment is commonly captured by a state variable which, when incorporated into online learning, leads to an online MDP.
Online MDP \cite{even2009online,yu2009markov,auer2009near}, also called adversarial MDP\footnote{The word ``adversarial'' is inherited from the online learning literature, i.e., \emph{stochastic bandit} vs \emph{adversarial bandit} \cite{auer2002nonstochastic}. Adversary means there exists a virtual adversary (or, nature) who has complete control over the reward function and transition dynamics, and the adversary does not necessarily maintain a fully competitive relationship with  the learning agent.}, focuses on the problem in which the reward and transition dynamics can change over time, i.e., they are non-stationary and time-dependent.

In contrast to an ordinary stochastic game, the opponent/adversary in an online MDP is not necessarily rational or even self-optimising. 
The aim of studying online MDP is to provide the agent with policies that perform well against every possible opponent (including but not limited to adversarial opponents), and the objective of the learning agent is to minimize its average loss during the learning process. Quantitatively, the loss is measured  by how worse off the agent is compared to the best stationary policy in retrospect. The \emph{expected regret} is thus different from Eq. (\ref{eq:regret}) (unless in  repeated games)  and is written as 
\begin{equation}
	\mathsf{Reg}_T = \frac{1}{T} \sup_{\pi \in \Pi} \mathbb{E}_{\pi}\Big[ \sum_{t=1}^T  R_t\big(s_t^*, a_t^*\big) - R_t\big(s_t, a_t\big) \Big]
	\label{eq:regret-omdp}
\end{equation}
where  $\mathbb{E}_{\pi}$ denotes the expectation  over the sequence of $(s_t^*, a_t^*)$ induced by the stationary policy $\pi$.
Note that the reward function sequence and the transition kernel sequence are given by the adversary, and they are not influenced by the retrospective sequence $(s_t^*, a_t^*)$. 
 
 The goal is to find a no-regret algorithm that can satisfy $	\mathsf{Reg}_T \rightarrow 0$ as $T \rightarrow \infty$ with probability $1$. 
A sufficient condition that ensures the existence of no-regret algorithms for online MDPs is the \emph{oblivious} assumption -- both the reward functions and transition kernels are fixed in advance, although they are unknown to the learning agent.  This scenario is in contrast to the stateless setting in which no-regret is achievable, even if the opponent is allowed to be \emph{adaptive/non-oblivious}: they can choose the  reward function and transition kernels in accordance to  $(s_0, a_0, ..., s_t)$ from the learning agent.
In short, \cite{yu2009markov,mannor2003empirical} demonstrated that in order to achieve sub-linear regret, it is essential that the changing rewards are chosen obliviously. 
Furthermore, \cite{yadkori2013online} showed with the example of an online shortest path problem that there does not exist a polynomial-time solution (in terms of the size of the state-action space) where both the reward functions and transition dynamics are adversarially chosen, even if the adversary is \emph{oblivious} (i.e., it cannot adapt to the other agent's historical actions). 
Most recently, \cite{ortner2020variational,cheung2020reinforcement} investigated online MDPs where the transitional dynamics are allowed to change slowly (i.e., the total variation does not exceed a specific budget). 
Therefore, the majority of existing no-regret algorithms  for online MDP focus on an oblivious adversary for the reward function only. 
The nuances of different algorithms   lie in whether the transitional kernel is assumed to be known to the learning agent and whether the feedback reward that the agent receives is in the full-information setting or in the bandit setting (i.e., one can only observe the reward of a taken action). 

Two design principles can lead to no-regret algorithms that solve online MDPs with an oblivious adversary controlling the reward function.
One is to leverage the local-global regret decomposition result \cite{even2005experts,even2009online} [Lemma 5.4], which demonstrates that one can in fact achieve  no regret globally by running a local regret-minimisation algorithm at each state; a similar result is observed for the CFR algorithm described in Eq. (\ref{eq:imr}). Let $\mu^*(\cdot)$ denote the state occupancy induced by policy $\pi^*$; we then obtain the decomposition result by 
\begin{equation}
	\mathsf{Reg}_T = \sum_{s \in \mathbb{S}} \mu^{*}(s) \sum_{t=1}^{T} \underbrace{\sum_{a \in \mathbb{A}}\Big(\pi^{*}(a \mid s)-\pi_{t}(a \mid s)\Big) Q_{t}\big(s, a\big)}_{\text{local regret in state $s$ with reward function  $Q_t(s, \cdot)$}}.
	\label{eq:regret-global-local}
\end{equation}
Under full knowledge of the transition function  and full-information feedback about the reward, \cite{even2009online} proposed the famous \emph{MDP-Expert (MDP-E)} algorithm, which adopts   \emph{Hedge} \cite{freund1997decision} as the regret minimizer  and achieves $\mathcal{O}( \sqrt{\tau^3 T \ln |\mathbb{A}|})$ regret, where $\tau$ is the bound on the mixing time  of MDP \footnote{Roughly, it can be considered as the time that a policy needs to reach the stationary status in  MDPs. See a precise definition in \citet{even2009online} [Assumption 3.1].  }. 
For comparison,  the theoretical lower bound for regret in a fixed MDP (i.e., no adversary perturbs the reward function)  is  $\Omega(\sqrt{|\mathbb{S}||\mathbb{A}|T})$\footnote{This lower bound  has recently  been achieved by \cite{azar2017minimax} up to a logarithmic factor.} \cite{auer2009near}.
Interestingly, \cite{neu2017unified} showed that there in fact exists an equivalence between TRPO methods \cite{schulman2015trust} and MDP-E methods. 
Under bandit feedback, \cite{neu2010online} analysed \emph{MDP-EXP3}, which achieves a regret bound of $\mathcal{O}( \sqrt{\tau^3 T |\mathbb{A}| \log |\mathbb{A}|/\beta})$, where $\beta$ is a lower bound on the probability of reaching a certain state under a given policy.
Later, \cite{neu2014online} removed the dependency on $\beta$ and achieved  $\mathcal{O}(\sqrt{T}\log T)$ regret. 
One major advantage of local-global design principle is that it can work seamlessly with function approximation methods \cite{bertsekas1996neuro}. For example, 
\cite{yu2009markov} eliminated the requirement of knowing the transition kernel by incorporating Q-learning methods; their proposed \emph{Q-follow the perturbed leader (Q-FPL)} method achieved $\mathcal{O}(T^{2/3})$ regret. 
\emph{OPPO} algorithm  achieves $\mathcal{O}({\sqrt{d^2H^3T}})$ regret \cite{cai2019provably}, where . $d$ is the feature dimension, $H$
 is the episode horizon, and $T$ is the total number of steps. Note that \emph{OPPO} requires the full information of the reward functions to be available, but the transition dynamics are unknown. In contrast, \cite{jin2019learning} first addresses the problem of learning online MDPs with unknown transition dynamics and bandit feedback.

Apart from the local-global decomposition principle, another design principle is to formulate the regret minimisation problem as an online linear optimisation (OLO) problem and then apply gradient-descent type methods. 
Specifically, since the regret in Eq. (\ref{eq:regret-global-local}) can be further  written as the inner product of 	$\mathsf{Reg}_T = \sum_{t=1}^{T} \langle \mu^* - \mu_t, R_t \rangle $, one can run the gradient descent method by
\begin{equation}
	\mu_{t+1} = \arg \max_{\mu \in \mathcal{U}} \Big\{\big\langle \mu, R_t  \big\rangle - \dfrac{1}{\eta} \mathcal{D} \big(\mu | \mu_t\big)  \Big\},
	\label{eq:regert-descent}
\end{equation}
where $\mathcal{U}=\big\{\mu\in \Delta_{\mathbb{S}\times \mathbb{A}}: \sum_a\mu(s, a)= \sum_{s', a'}P(s|s', a') \mu(s',a') \big\}$ is the set of all valid stationary distributions\footnote{In the online MDP literature, it is generally assumed that every policy reaches its stationary distribution immediately; see the policy mixing time assumption in \cite{yu2009markov} [Assumption 2.1].}, where $\mathcal{D}$ denotes a certain form of divergence 
and the policy can be extracted by  $\pi_{t+1}(a|s) = \mu_{t+1}(s,a)/\mu(s)$. 
One significant advantage of this type of method is that it can flexibly handle different model constraints and extensions. 
If one uses Bregman divergence as $\mathcal{D}$, then online mirror descent is recovered \cite{nemirovsky1983problem} and is guaranteed to achieve a nearly optimal regret for OLO problems  \cite{srebro2011universality}. 
\cite{zimin2013online} and \cite{dick2014online} adopted a relative entropy for $\mathcal{D}$; the subsequent \emph{online relative entropy policy search (O-REPS)} algorithm achieves an $\mathcal{O}(\sqrt{\tau T \log (|\mathbb{S}| |\mathbb{A}|}))$ regret in the full-information setting and an $\mathcal{O}(\sqrt{T |\mathbb{S}| |\mathbb{A}| \log (|\mathbb{S}| |\mathbb{A}|}))$ regret in the bandit setting. For comparison, the aforementioned MDP-E algorithm achieves  $\mathcal{O}( \sqrt{\tau^3 T \ln |\mathbb{A}|})$  and  $\mathcal{O}( \sqrt{\tau^3 T |\mathbb{A}| \log |\mathbb{A}|/\beta})$, respectively. 
When the transition dynamics are unknown to the agent, 
\cite{rosenberg2019online} extended O-REPS by incorporating the classic idea of \emph{optimism in the face of uncertainty} in \cite{auer2009near}, and the induced \emph{UC-O-REPS} algorithm achieved $\mathcal{O}(|\mathbb{S}|\sqrt{|\mathbb{A}|T})$ regret.

\section{Team Games}
Team games are games in which members of the same team share the same utility functions. In two-team zero-sum games, two teams compete in a zero-sum game. One solution concept in zero-sum team games is called TMECor\cite{Celli18:Computational,Farina18:Ex, zhang2021team, zhang2022team}. In TMECor, players on the same team are allowed to coordinate before playing, and since it involves cooperation within the team, cooperative MARL algorithms can be used to approximate this solution concept \cite{mcaleer2023team}. A related solution concept is an equilibrium where team members are not able to coordinate, which is known as a \emph{team-maxmin equilibrium} (TME) strategy\cite{von1997team,Basilico17:Team,Zhang20:Computing,Zhang20:Converging, kalogiannis2021teamwork}. A TME yields the maximum expected utility for the team players against a best-responding team opponent. The TMECor solution concept has several advantages over TME. First, the team is guaranteed at least as much expected utility under TMECor than under TME\cite{Celli18:Computational}. Second, the TMECor objective is convex while the TME objective is not convex. Lastly, in general, finding a TMECor strategy is \textsf{NP}-hard and inapproximable\cite{Celli18:Computational}. However, many real-world scenarios might not permit communication.



















\chapter{Learning in General-Sum Games}
\label{chaper:general_sum}
Solving general-sum stochastic games (SGs) entails an entirely different level of difficulty than solving team games or zero-sum games. 
In a static two-player normal-form game, finding the Nash equilibrium (NE) is known to be $PPAD$-complete \cite{chen2006settling}.
\section{Solutions by Mathematical Programming}
To solve a two-player general-sum discounted stochastic
game with discrete states and discrete actions,  \cite{filar2012competitive} [Chapter 3.8] formulated the problem as a nonlinear program; the matrix form is written as follows:
\begin{equation}
\label{eq:opt_two_gen}
\begin{array}{l}{\min _{\mathbf{V}, \bm{\pi}} f(\mathbf{V}, \bm{\pi})=\sum_{i=1}^{2} \boldsymbol{1}_{|\mathbb{S}|}^{T}\bigg[V^{i}-\Big(\mathbf{R}^{i}(\bm{\pi}) + \gamma \cdot \mathbf{P}(\pi) V^{i}\Big)\bigg] } \\ \text { s.t. } {\begin{array}{ll}{\text { (a) } \pi^{2}(s)^{T}\Big[\mathbf{R}^{1}(s)+\gamma \cdot \sum_{s' } \mathbf{P}(s' | s) V^{1}(s')\Big] \leq V^{1}(s) \boldsymbol{1}_{|\mathbb{A}^1|}^{T}, \ \ \ \  \forall s \in \mathbb{S}} \\ {\text { (b) }\Big[\mathbf{R}^{2}(s)+\gamma \cdot \sum_{s'} \mathbf{P}(s' | s) V^{2}(s')\Big] \pi^{1}(s) \leq V^{2}(s) \boldsymbol{1}_{|\mathbb{A}^2|}, \ \ \ \  \forall s \in \mathbb{S}} \\ {\text { (c) } \pi^{1}\left( s \right) \geq \mathbf{0}, \ \   \pi^{1}(s)^{T} \boldsymbol{1}_{|\mathbb{A}^1|}=1, \ \ \ \  \forall s \in \mathbb{S}} \\ {\text { (d) } \pi^{2}\left(  s \right) \geq \mathbf{0}, \ \   \pi^{2}(s)^{T} \boldsymbol{1}_{|\mathbb{A}^2|}=1, \ \ \ \  \forall s \in \mathbb{S}}\end{array}}\end{array}
\end{equation}
where 
\renewcommand{\theenumi}{(\roman{enumi})}%
\begin{itemize}
  \item $\mathbf{V}=\left\langle V^{i}: i=1,2\right\rangle$ is the vector of agents' values over all states,  $V^{i}=\left\langle V^{i}(s): s \in \mathbb{S}\right\rangle$ is the value vector  for the $i$-th agent. 
  \item $\pi=\left\langle\pi^{i}: i=1,2\right\rangle$ and $\pi^{i}=\left\langle\pi^{i}(s): s \in \mathbb{S}\right\rangle$, where the policy $\pi^{i}(s)=\left\langle\pi^{i}(a| s): a \in \mathbb{A}^{i}\right\rangle$ is the vector representing the stochastic policy  in state $s \in \mathbb{S}$ for the $i$-th agent. 
  \item $\mathbf{R}^{i}(s)=\left[R^{i}\left(s, a^{1}, a^{2}\right): a^{1} \in \mathbb{A}^{1}, a^{2} \in \mathbb{A}^2\right]$ is the reward matrix for the $i^{\text{th}}$ agent  in state $s \in \mathbb{S}$. The rows correspond to the actions of the second agent, and the columns correspond to those of the first agent. With a slight abuse of notation, we use 
 $\mathbf{R}^{i}(\bm{\pi})=\mathbf{R}^{i}\left(\left\langle\pi^{1}, \pi^{2}\right\rangle\right)=\left\langle\pi^{2}(s)^{T} \mathbf{R}^{i}(s) \pi^{1}(s): s \in \mathbb{S}\right\rangle$ to represent the expected reward vector over all states under joint policy $\bm{\pi}$.
  \item $\mathbf{P}(s' | s)=\left[P(s' | s, \bm{a}): \bm{a}=\left\langle a^{1}, a^{2}\right\rangle, a^{1} \in \mathbb{A}^{1}, a^{2} \in \mathbb{A}^{2}\right]$ is a matrix representing the probability of transitioning from the current state $s \in \mathbb{S}$ to the  next state $s' \in \mathbb{S}$. The rows represent the actions of the second agent, and the columns represent those of the first agent.
With a slight abuse of notation, we use $\mathbf{P}(\bm{\pi})=\mathbf{P}\left(\left\langle\pi^{1}, \pi^{2}\right\rangle\right)=\left[\pi^{2}(s)^{T} \mathbf{P}(s' | s) \pi^{1}(s): s \in \mathbb{S}, s' \in \mathbb{S}\right]$ to represent the expected transition probability over all  state pairs under joint policy $\bm{\pi}$.  \end{itemize}

This is a nonlinear programme because the inequality constraints in the optimisation problem are quadratic in $\mathbf{V}$ and $\bm{\pi}$. The objective  function in Eq. (\ref{eq:opt_two_gen}) aims to minimise the TD error for a given policy $\bm{\pi}$ over all states, similar to the policy evaluation step in the traditional policy iteration method, and the constraints of $(a)$ and $(b)$ in Eq. (\ref{eq:opt_two_gen}) act as the policy improvement step, which satisfies the equation when the optimal value function is achieved. 
Finally, constraints $(c)$ and $(d)$ ensure the policy is properly defined. 

Although the NE is proved to exist in general-sum SGs in the form of stationary strategies, 
solving Eq. (\ref{eq:opt_two_gen}) in the two-player case  is notoriously challenging. First, Eq. (\ref{eq:opt_two_gen}) has a non-convex feasible region; second, only the global optimum\footnote{Note that in the zero-sum case, every local optimum is global.} of Eq. (\ref{eq:opt_two_gen}) corresponds to the NE of SGs, while the common gradient-descent type of methods can only  guarantee convergence to a local minimum. 
Apart from the efforts by \cite{filar2012competitive},  \cite{breton1986computation} [Chapter 4] developed a formulation that has  nonlinear objectives but linear constraints. Furthermore, \cite{dermed2009solving} formulated the NE solution as multi-objective linear program.  
\cite{herings2004stationary,herings2010homotopy} proposed an algorithm in which a \emph{homotopic path} between the equilibrium points of $N$ independent MDPs and the $N$-player SG is traced numerically. This approach yields a NE point of the stochastic game of interest.
However,  all these methods are tractable only in small-size SGs with at most tens of states and  only two players.

Apart from the concept of NE defined before, we will also use another two solution concepts in the general-sum SGs, which are correlated equilibrium (CE) and coarse correlated equilibrium (CCE), as generalization of NE. These concepts will be defined as following.  

\begin{definition}(Correlated Equilibrium) A correlated equilibrium (CE) in multi-player general-sum SGs is defined as a joint (correlated) policy $\boldsymbol{\pi}:=\{\pi_h(s)\in\triangle_{\mathbb{A}_i}, s\in\mathbb{S}\}$, and it satisfies the following relationship at each time-step:
\begin{align}
    \max_{i\in[N]}\max_{\phi\in\Phi_i}V^{i, \phi\diamond \boldsymbol{\pi}}(s)\le V^{i, \boldsymbol{\pi}}, \forall s\in\mathbb{S},
\end{align}
where the \emph{strategy modification} $\phi:=\{\phi_s:\mathbb{A}_i \rightarrow \mathbb{A}_i\}_{s\in\mathbb{S}}\in\Phi_i$ is a function set for player $i$ at each time-step, changing the action of the player. Specifically, by $\phi\diamond \pi$ it changes the policy $\boldsymbol{\pi}$ from choosing $\boldsymbol{a}=(a_1, \dots, a_N)$ at state $s$ and a certain time-step, to actions $(a_1, \cdots, a_{i-1}, \phi_s(a_i), a_{i+1}, \dots, a_N)$ instead\cite{liu2021sharp}.  
    
\end{definition}

\begin{definition}(Coarse Correlated Equilibrium) A coarse correlated equilibrium (CCE) in multi-player general-sum SGs is defined as a joint (correlated) policy $\boldsymbol{\pi}:=\{\pi_h(s)\in\triangle_{\mathbb{A}_i}, s\in\mathbb{S}\}$, and it satisfies the following relationship at each time-step:
\begin{align}
    \max_{i\in[N]}\max_{\widehat{\pi_i}}V^{i, \widehat{\pi_i}, \pi_{-i}}(s)\le V^{i, \boldsymbol{\pi}}, \forall s\in\mathbb{S}
\end{align}.
    
\end{definition}
It can be seen that CCE is a generalized version of NE, and NE is just the case when the correlated policy in CCE can be factorized as the product of individual policy for each player. 

There is a relationship between NE, CE and CCE as shown in Theorem~\ref{thm:relation_ne_ce}, for which the proof will be omitted.

\begin{theorem}
\label{thm:relation_ne_ce}
For general-sum SGs, \{NE\}$\subseteq$\{CE\}$\subseteq$\{CCE\}.
\end{theorem}

Mathematical programming methods, particularly linear programming, have been crucial for solving equilibria in general-sum extensive-form games. Solutions such as Extensive-Form Correlated Equilibria (EFCE) \cite{farina2022simple} and Stackelberg Extensive-Form Correlated Equilibria (SEFCE) offer strong theoretical guarantees but face scalability challenges due to the exponential complexity of game trees \cite{cermak2016using}. Recent advancements include innovations like correlation-directed acyclic graphs (DAGs) \cite{zhang2022optimal}, which help manage the computational complexity of EFCE. Despite these improvements, the high computational demands of these techniques limit their applicability in large-scale games, requiring further research to improve scalability without sacrificing solution quality.

\section{Solutions by Value-Based Methods}

A series of value-based methods have been proposed to address general-sum SGs. A majority of these methods adopt classic Q-learning \cite{watkins1992q} as a centralised controller, with the differences being what solution concept the central Q-learner should  apply to guide the agents  to converge in each iteration. 
For example,  the Nash-Q learner in Eqs. (\ref{eq:ma_nash_q} \&  \ref{eq:operator_nash}) applies NE as the solution concept, the correlated-Q learner adopts correlated equilibrium \cite{greenwald2003correlated}, and the friend-or-foe learner considers both cooperative (see Eq. (\ref{eq:coop_equilibrium})) and competitive equilibrium (see Eq. (\ref{eq:comp_equilibrium}))  \cite{littman2001friend}.  
Although many  algorithms come with convergence guarantees,  the corresponding  assumptions  are often overly restrictive to be applicable in general. 
When Nash-Q learning was first proposed \cite{hu1998multiagent}, it required the NE of the SG be unique such that the convergence property could  hold. 
Though strong, this assumption was still noted by  \cite{bowling2000convergence}   to be insufficient to justify the convergence of  the Nash-Q algorithm. Later, \cite{hu2003nash} corrected her  convergence proof by tightening the assumption even further;  the uniqueness  of the NE must hold for every single stage game encountered during state transitions.
Years later,  a strikingly negative result by \cite{zinkevich2006cyclic} concluded that the entire class of value-iteration methods could be excluded from consideration for computing stationary equilibria, including both NE and correlated equilibrium, in general-sum SGs. 
Unlike those in single-agent RL, the Q values in the multiagent case are  inherently defective for reconstructing the equilibrium policy.  

There is a variant of \emph{Nash-VI} algorithm for the exploration setting in multi-player SGs, called \emph{Optimistic Nash Value Iteration (Optimistic Nash-VI)}\cite{liu2021sharp}, which is provably finding Coarse Correlated Equalibirum (CCE) for multiplayer general-sum SGs with a convergence rate of $\tilde{\mathcal{O}}(\frac{H^4S^2\Pi_{i=1}^NA_i}{\epsilon^2})$. Similar as in single-agent RL, value iteration (VI) is a model-based type algorithm as a counterpart of model-free Q-learning. In multiagent setting, Nash-VI and its variants are also model-based, which first estimates the transition model $\mathbb{P}(s'|s,\mathbf{a})$ and leveraging the NE, CE or CCE solving subroutine on payoff matrices to achieve the corresponding equilibria on multi-step SGs. However, \emph{Nash-VI} has a product dependency on the action space as $\Pi_{i=1}^NA_i$, which is referred as the ``curse of multiagents''. This will bring large computation when the number of players is huge. To break the ``curse of multiagents'',  \emph{V-learning}\cite{jin2021v} is proposed. In \emph{V-learning}, each agent estimates the $V$ value instead of $Q$ value, since $V$ value only has a dimensional dependency of $\mathcal{O}(S)$ while using the $Q$ value is $\mathcal{O}(S\Pi_{i=1}^NA_i)$. \emph{V-learning} finds $\epsilon-$CCE in multi-player general-sum SGs in $\tilde{\mathcal{O}}(\frac{H^5S\max_{i\in[N]}A_i}{\epsilon^2})$ steps, and finds $\epsilon-$CE in $\tilde{\mathcal{O}}(\frac{H^5S(\max_{i\in[N]}A_i)^2}{\epsilon^2})$ steps. There are some concurrent work proposed at this time, \emph{Optimistic V-learning with Stabilized Online Mirror Descent} (V-learning OMD)\cite{mao2022provably}, \emph{CCE-V-learning} and \emph{CE-V-learning}\cite{song2021can} prove the similar rates as \emph{V-learning} with slightly worse dependence on the horizon $H$ for the multi-player general-sum SGs. The information theoretical lower bound of multi-player general-sum SGs is proved to be $\tilde{\Omega}(\frac{H^3S\max_{i\in[N]}A_i}{\epsilon^2})$ for NE, CE and CCE\cite{jin2021v, mao2022provably}. A summarization of value-based methods for multi-player general-sum SGs is shown in Table~\ref{tab:general_sum_sg}. Among them, \cite{jin2021v} achieves tighter sample complexity bounds, which is why we only include its complexity results in this table.

Value-based methods have been slower to mature in the context of general-sum games. Recent work focuses on adapting neural networks to approximate enforceable payoff frontiers (EPF) \cite{song2022sample}, particularly for Stackelberg problems. These methods aim to reduce computational costs while maintaining effective value propagation in games with complex strategies. However, significant work remains to refine these methods, particularly in handling the complexity of value functions in non-cooperative settings and improving their practical efficiency in dynamic environments.

\begin{table}[H]
\centering
\caption{Convergence guarantee for tabular multi-player general-sum stochastic games.}
\resizebox{\columnwidth}{!}{ 
\begin{tabular}{m{5cm}|c|c}
\toprule
Algorithm & Solution Concept &  Sample Complexity  \\
\hline
\multirow{3}{*}{} 
 Nash-VI\cite{liu2021sharp}  & NE, CE, CCE & $\tilde{\mathcal{O}}(\frac{H^4S^2\Pi_{i=1}^NA_i}{\epsilon^2})$  \\  \hline
\multirow{2}{*}{V-learning\cite{jin2021v}} & CCE &  $\tilde{\mathcal{O}}(\frac{H^5S\max_{i\in[N]}A_i}{\epsilon^2})$ \\  \cline{2-3}
& CE &  $\tilde{\mathcal{O}}(\frac{H^5S(\max_{i\in[N]}A_i)^2}{\epsilon^2})$ \\  \hline
Lower Bound\cite{jin2021v, mao2022provably}  & NE, CE, CCE &  $\tilde{\Omega}(\frac{H^3S\max_{i\in[N]}A_i}{\epsilon^2})$ \\

\bottomrule
\end{tabular}
}
\vskip -.2in
\label{tab:general_sum_sg}
\end{table}

\section{Solutions by Two-Timescale Analysis}

In addition to the centralised Q-learning approach, decentralised Q-learning algorithms have recently received considerable attention because of their potential for scalability. 
Although independent learners have been accused of having convergence issues \cite{tan1993multi}, decentralised methods have made substantial progress
with the help of two-timescale  stochastic analysis \cite{borkar1997stochastic} and its application in RL \cite{borkar2002reinforcement}.    

Two-timescale stochastic analysis   is a set of tools certifying that, in a system with two coupled stochastic processes that evolve at different speeds,  if the fast process converges to a unique limit  point for any particular fixed value of the slow process, we can, quantitatively, analyse the asymptotic behaviour of the algorithm as if the fast process is always fully calibrated to the current value of the slow process \cite{borkar1997stochastic}. 
As a direct application,  \cite{leslie2003convergent, leslie2005individual} noted that independent Q-learners with agent-dependent learning rates could break the symmetry that leads to the non-convergent limit cycles; as a result, they can converge almost surely to the NE in two-player collaboration games,  two-player zero-sum games, and multi-player matching pennies. 
Similarly, \cite{prasad2015two}
introduced a two-timescale update rule that ensures the training dynamics  reach a stationary local NE in general-sum SGs if the critic learns faster than the actor. 
Later, 
\cite{perkins2015mixed} proposed a distributed actor-critic algorithm that enjoys provable convergence  in solving static potential games with continuous actions. 
Similarly, 
\cite{arslan2016decentralized} developed a two-timescale variant of Q-learning that is guaranteed to converge to an equilibrium in SGs with weakly acyclic characteristics, which generalises potential games. 
Other applications include developing two-timescale update rules for training GANs \cite{heusel2017gans} and  developing a two-timescale algorithm with guaranteed asymptotic convergence to the  Stackelberg equilibrium  in general-sum Stackelberg games. 

Two-timescale analysis has emerged as an important tool for improving the convergence speed and sample efficiency in general-sum extensive-form games. This method differentiates between fast and slow timescales in the learning process, making it particularly useful in settings with bandit feedback, such as the $K$-EFCE framework \cite{song2022sample}. By decoupling the update rates of different game variables, these methods enable more efficient learning processes, especially in large, multi-agent environments. However, balancing the timescales for optimal performance and ensuring convergence across different agent types remain ongoing challenges.

%

\section{Solutions by Policy-Based Methods}
Convergence to NE via direct policy search has been extensively studied; however, early results were limited mainly by  stateless two-player two-action games \cite{singh2000nash,bowling2002multiagent, bowling2005convergence, conitzer2007awesome,abdallah2008multiagent,zhang2010multi}.
Recently, GAN training has posed a new challenge, thereby rekindling interest  in understanding the policy gradient dynamics of continuous games \cite{nagarajan2017gradient, heusel2017gans, mescheder2018training, mescheder2017numerics}. 

Analysing gradient-based algorithms through dynamic systems \cite{shub2013global} is a natural approach  to yield more significant insights into convergence behaviour. 
However, a fundamental difference is observed when one attempts to apply the same analysis  from the single-agent case to the multiagent case
because the combined dynamics of gradient-based learning schemes in multiagent games do not necessarily correspond to a proper \emph{gradient flow} -- a critical premise for almost sure convergence to a local minimum.  
In fact, the difficulty of solving general-sum continuous games is exacerbated by the usage of deep networks with stochastic gradient descent. 
In this context, a key equilibrium concept of interest is the \emph{local NE} \cite{ratliff2013characterization} or \emph{differential NE} \cite{ratliff2014genericity}, defined as follows. 
\begin{definition}[Local Nash Equilibrium]
	For an $N$-player continuous game denoted by $\{\ell_i:\mathbb{R}^d \rightarrow R \}_{i\in \{1,...,N\}}$ with each agent's loss $\ell_i$ being twice continuously differentiable, the parameters are $\bm{w}=(\bm{w}_1, ..., \bm{w}_n)\in \mathbb{R}^d$, and each player controls $\bm{w}_i \in \mathbb{R}^{d_i}, \sum_i{d_i}=d$. Let $\boldsymbol{\xi}(\mathbf{w})=\left(\nabla_{\mathbf{w}_{1}} \ell_{1}, \ldots, \nabla_{\mathbf{w}_{n}} \ell_{n}\right) \in \mathbb{R}^{d}$ be  the simultaneous gradient of the losses w.r.t. the parameters of the respective players, and let $\mathbf{H}(\mathbf{w}):=\nabla_{\mathbf{w}} \cdot \boldsymbol{\xi}(\mathbf{w})^{\top}$ be the $(d\times d)$ Hessian matrix of the gradient, written as \par
	{\small\[
	\mathbf{H}(\mathbf{w})=\left(\begin{array}{cccc}\nabla_{\mathbf{w}_{1}}^{2} \ell_{1} & \nabla_{\mathbf{w}_{1}, \mathbf{w}_{2}}^{2} \ell_{1} & \cdots & \nabla_{\mathbf{w}_{1}, \mathbf{w}_{n}}^{2} \ell_{1} \\ \nabla_{\mathbf{w}_{2}, \mathbf{w}_{1}}^{2} \ell_{2} & \nabla_{\mathbf{w}_{2}}^{2} \ell_{2} & \cdots & \nabla_{\mathbf{w}_{2}, \mathbf{w}_{n}}^{2} \ell_{2} \\ \vdots & & & \vdots \\ \nabla_{\mathbf{w}_{n}, \mathbf{w}_{1}}^{2} \ell_{n} & \nabla_{\mathbf{w}_{n}, \mathbf{w}_{2}}^{2} \ell_{n} & \cdots & \nabla_{\mathbf{w}_{n}}^{2} \ell_{n}\end{array}\right)
	\]}
	where $\nabla_{\mathbf{w}_{i}, \mathbf{w}_{j}}^{2} \ell_{k}$ is the $(d_i \times d_j)$  block of $2$nd-order derivatives. 
	 A differentiable NE for the game is $\bm{w}^*$  if $\boldsymbol{\xi}(\mathbf{w}^*)=0$ and $\nabla_{\mathbf{w}_{i}}^{2} \ell_i \succ 0, \  \forall i \in \{1,...,N\}$; furthermore, this result is a local NE if $\det \mathbf{H}(\mathbf{w}^*) \neq 0$. 
\end{definition}
A recent result by \cite{mazumdar2018convergence} suggested that gradient-based algorithms can almost surely avoid a subset of local NE in general-sum games; even worse, there exist non-Nash stationary points.  
As a tentative treatment, \cite{balduzzi2018mechanics} applied \emph{Helmholtz decomposition}\footnote{This approach is similar in ideology to the work by \cite{candogan2011flows}, where they leverage the combinatorial Hodge decomposition to decompose any multi-player normal-form game into a potential game plus a harmonic game. However, their equivalence is an open question.} to decompose the game Hessian $\mathbf{H}(\bm{w})$ into a potential part plus a Hamiltonian part. Based on the decomposition, they designed a gradient-based method to address each part and  combined them into \emph{symplectic gradient adjustment (GDA)}, which is able to find all local NE for zero-sum games and a subset of local NE for general-sum games. 
More recently, 
\cite{chasnov2019convergence} separately considered  the cases of 1) agents with oracle access to the exact gradient $\boldsymbol{\xi}(\mathbf{w})$ and 2) agents with only an unbiased estimator for $\boldsymbol{\xi}(\mathbf{w})$. In the first case, they provided asymptotic and finite-time convergence rates for the gradient-based learning process to reach the differential NE. 
In the second case, they derived concentration bounds guaranteeing with high probability that agents will converge to a neighbourhood of a stable local NE in finite time. 
In the same framework, \cite{fiez2019convergence} studied Stackelberg games in which agents take turns to conduct the gradient update rather than acting simultaneously and established the connection under which the equilibrium points of simultaneous gradient descent are Stackelberg equilibria in zero-sum games.  
\cite{mertikopoulos2019learning} investigated the local convergence of no-regret learning and found local NE is attracting under gradient play  if and only if a NE satisfies a property known as \emph{variational stability}. This idea is   inspired by the seminal notion of \emph{evolutionary stability} observed in animal populations  \cite{smith1973logic}.

Finally, it is worth highlighting that the above theoretical analysis of the performance of gradient-based methods on stateless continuous games cannot be taken for granted in SGs.  The main reason is that the assumption on the differentiability of the loss function required in continuous games may not hold in general-sum SGs. As clearly noted by \cite{mazumdar2019policy, fazel2018global, zhang2019policy}, even in the extreme setting of linear-quadratic games, the value functions are not guaranteed to be globally smooth (w.r.t. each agent's policy parameter).

\chapter{\texorpdfstring{Learning in Games when $N \rightarrow +\infty$}{Learning in Games when N -> infinity}}

\label{sec:mfs}
As detailed in Section \ref{sec:challenge}, 
designing learning algorithms in a multiagent system with $N \gg 2$ is a challenging task. One major reason is that the solution concept, such as Nash equilibrium, is difficult to compute in general due to the curse of dimensionality of the multiagent problem itself.  
However, if one considers a continuum of agents with $N \rightarrow +\infty$, then the learning problem becomes surprisingly tractable.
The intuition is that one can effectively transform a  many-body interaction problem into a two-body interaction problem (i.e., agent vs the population mean)  via mean-field approximation. 

The idea of mean-field approximation, which considers the behaviour of large numbers of particles where individual particles have a negligible impact on the system, originated from physics.  Important applications include solving  Ising models\footnote{An Ising model is a model used to study magnetic phase transitions under different system temperatures. In a 2D Ising model, one can imagine the magnetic spins are laid out on a lattice, and each spin can have one of two directions, either up or down. When the system temperature is high,  the direction of  the spins is chaotic, and when the temperature is low, the directions of the spins tend to be aligned. Without the mean-field approximation, computing the probability of the spin direction is a combinatorial hard problem; for example, in a $5\times 5$ 2D lattice, there are  $2^{25}$  possible spin configurations. A successful approach to solving the Ising model is to observe the phase change under different temperatures and compare it against the ground truth.} \cite{weiss1907hypothese,kadanoff2009more}, or more recently, understanding the learning dynamics of over-parameterised deep neural networks \cite{lu2020mean,song2018mean,sirignano2020mean,hu2019mean}. 
In the game theory and MARL context, mean-field approximation essentially enables one to think of the interactions between every possible permutation of agents as an interaction between each agent itself and the aggregated mean effect of the population of the other agents, such that the $N$-player game ($N \rightarrow +\infty$) turns into a ``two''-player game. Moreover, under  \emph{the law of large numbers} and \emph{the theory of propagation of chaos} \cite{gartner1988mckean, mckean1967propagation,sznitman1991topics}, the aggregated version of the optimisation problem in Eq. (\ref{eq:cumu_marl})   asymptotically approximates the original $N$-player game. 

\begin{figure}[t!]
     \centering
\includegraphics[width=.85\textwidth]{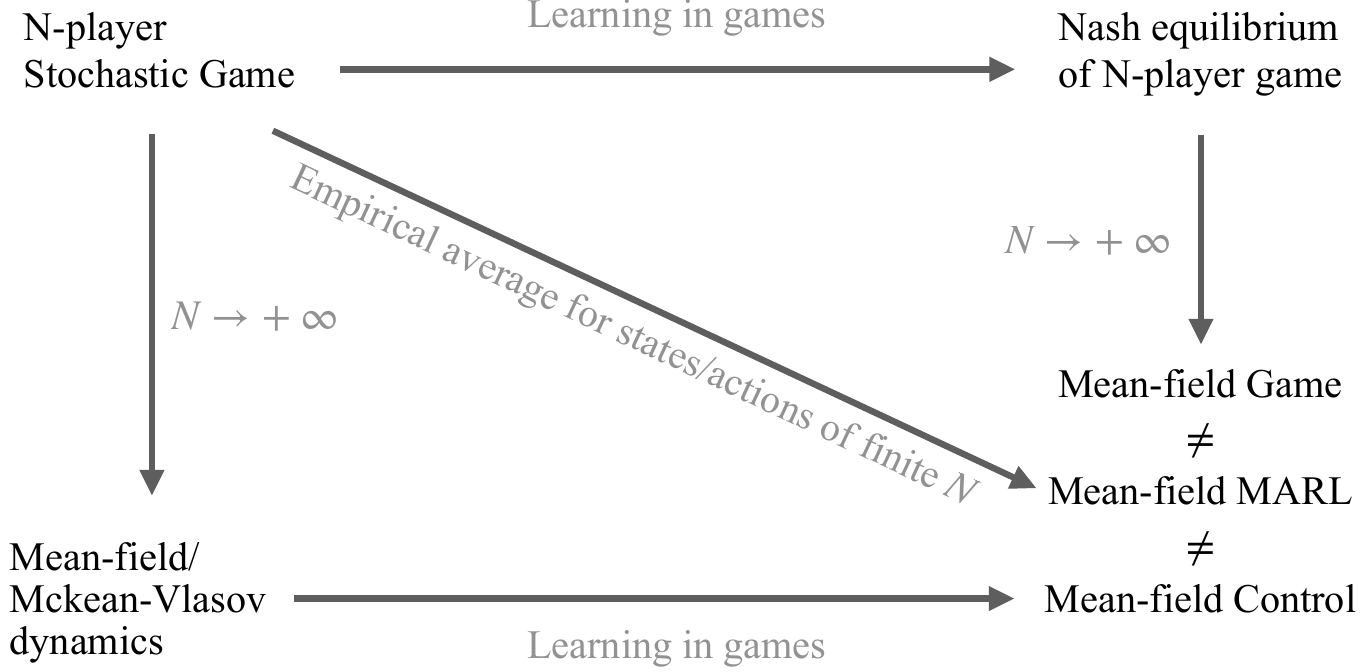}
     \caption{Relations of mean-field  learning algorithms in games with large $N$.}
\label{fig:mfs}
\end{figure}

The assumption in the mean-field regime that each agent responds only to the mean effect of the population may appear rather limited initially; however, for many real-world applications,  agents often cannot access  the information of all other agents but can instead know the global information about the population. For example, in high-frequency trading in finance \cite{cardaliaguet2018mean, lehalle2019mean},   each trader cannot know every other trader's position in the market, although they have access to the aggregated order book from the exchange. 
Another example is real-time bidding for online advertisements \cite{guo2019learning, iyer2014mean}, in which  participants can only observe, for example, the second-best prize that wins the auction but not the individual bids from other participants. 

There is a subtlety associated with types of games in which one applies the  mean-field theory. 
If one applies the mean-field type theory in non-cooperative\footnote{Note that the word ``non-cooperative'' does not mean agents cannot collaborate to complete a task, it means agents cannot collude to form a coalition: they have to behave independently. } games, in which agents act independently to maximize their own individual reward, and the solution concept is NE, then the scenario is usually referred to as a \emph{mean-field game (MFG)} \cite{jovanovic1988anonymous, huang2006large,lasry2007mean, gueant2011mean}. 
If one applies mean-field theory in cooperative games in which there exists a central controller to control all agents cooperatively to reach some Pareto optimality, then the situation is usually referred to as  \emph{mean-field control (MFC)} \cite{bensoussan2013mean, andersson2011maximum}, or \emph{McKean-Vlasov dynamics (MKV)} control. 
If one applies the mean-field approximation to solve a standard SG through MARL, specifically, to factorize each agent's reward function or the joint-Q function,  such that they depend only on the agent's local state and the mean action of others, then it is called \emph{mean-field MARL (MF-MARL)} \cite{yang2018mean,subramanian2020multi,zhou2019factorized}. 

Despite the difference in the applicable game types, 
technically, the differences among MFG/MFC/MF-MARL can be elaborated from the perspective of  the order in which the equilibrium is learned (optimised)  and the limit as $N \rightarrow +\infty$ is taken \cite{carmona2013control}. 
MFG learns  the equilibrium of the game first and then takes the limit as $N \rightarrow +\infty$, while MFC takes the limit first and optimises the equilibrium later. 
MF-MARL is somewhat in between. 
The mean-field in MF-MARL refers to the empirical average of the states and/or actions of a \emph{finite} population; $N$ does not have to reach infinity, though the approximation converges asymptotically to the original game when $N$ is large.
This result is in contrast to the mean-field in MFG and MFC, which is essentially a  probability distribution of states and/or actions of an \emph{infinite} population (i.e., the Mckean-Vlasov dynamics).   
Before providing more details, 
we summarise the relationships of MFG, MFC, and MF-MARL in Figure \ref{fig:mfs}. 
Readers are recommended to revisit their differences after finishing reading the below subsections.

\section{Non-cooperative Mean-Field Game}
MFGs have been widely studied in different domains,  including physics, economics, and stochastic control \cite{carmona2018probabilistic, gueant2011mean}. 
An intuitive example to quickly illustrate  the idea of MFG is the problem of  \emph{when does the meeting start} \cite{gueant2011mean}.
For a  meeting in the real world, people often schedule a calendar time $t$ in advance, and the actual start time $T$ depends on when the majority of participants (e.g., $90\%$) arrive. Each participant plans to arrive at $\tau^i$, and the actual arrival time, $\tilde{\tau}^i = \tau^i +\sigma^i \epsilon^i$, is often influenced by some uncontrolled factors $\sigma^i \epsilon^i, \epsilon^i\sim \mathcal{N}(0,1)$, such as weather or traffic.  Assuming all players are rational, they do not want to be later than either $t$ or $T$; moreover, they do not want to arrive too early and have to wait. The cost function of each  individual  can be written as $c^i({t, T, \tilde{\tau}^i})=\mathbb{E}\big[\alpha \lfloor\tilde{\tau}^i - t \rfloor_+ + \beta \lfloor \tilde{\tau}^i - T \rfloor_+ + \gamma \lfloor T - \tilde{\tau}^i\rfloor_+ \big],
$ where $\alpha, \beta, \gamma$ are constants. The key question to ask is when  is the best time for an agent to arrive, as a result, when will the meeting actually start, i.e., what is $T$?
  
  The challenge of the above problem lies in the coupled relationship between $T$ and $\tau^i$; that is, in order to compute $T$, we need to know $\tau^i$, which is based on $T$ itself.
  Therefore,  
solving the time $T$ is essentially equivalent to finding the fixed point, if it exists, of the stochastic process that generates $T$. 
In fact, $T$ can be effectively computed through a two-step iterative process, and we denote as $\Gamma^1$ and $\Gamma^2$. At $\Gamma^1$,  given the current\footnote{At time step $0$, it can be a random guess. Since the fixed point exists, the final convergence result is irrelevant to the initial guess.} value of  $T$, each agent solves their optimal arrival time $\tau^i$ by minimising their cost  $R^i({t, T, \tilde{\tau}^i})$.  At $\Gamma^2$, agents calibrate the new estimate of $T$ based on all $\tau^i$ values that were  computed in $\Gamma^1$.  $\Gamma^1$ and $\Gamma^2$ continue iterating until $T$ converges to a  fixed point, i.e., $\Gamma^2 \circ \Gamma^1 (T^*) = T^*$.
The key insight is that the interaction with other agents is captured simply by the mean-field quantity. Since the meeting starts only when $90\%$ of the people arrive, if one considers a continuum of players with $N \rightarrow +\infty$,  $T$ becomes the $90th$ quantile of a distribution, and each agent can easily find the best response. This result contrasts to the  cases of a finite number of players, in which the ordered statistic   is intractable, especially when $N$ is large (but still finite).  

Approximating an $N$-player SG by letting $N \rightarrow +\infty$ and letting each player choose an optimal strategy in response to the population's macroscopic information (i.e., the mean field), though analytically friendly, is not cost-free. 
In fact, MFG makes two major assumptions: 1) the impact of each player's action on the outcome is infinitesimal, resulting in all agents being identical,  interchangeable, and indistinguishable; 2) each player maintains \emph{weak interactions} with others only through a mean field, denoted by $L^i \in \Delta^{|\mathbb{S}||\mathbb{A}|}$, which is essentially a population state-action joint distribution  
\begin{equation}
L^i =\Big( \mu^{-i}(\cdot), \alpha^{-i}(\cdot) \Big) =\lim_{N \rightarrow +\infty} \bigg(\dfrac{\sum_{j \neq i}\mathds{1}(s^j=\cdot)}{N-1} , \dfrac{\sum_{j \neq i}\mathds{1}(a^j=\cdot)}{N-1}  \bigg) 	
\label{eq:meanfield}
\end{equation}
where $s^j$ and $a^j$ player $j$'s local state\footnote{Note that in mean-field learning in games, the state is not assumed to be global. This is different from Dec-POMDP, in which there exists an observation function that maps the global state to the local observation for each agent. } and local action. 
Therefore, for SGs that do not share the homogeneity assumption\footnote{In fact, the homogeneity in MFG can be relaxed to allow agents to have (finite) different types \cite{lacker2019mean}, though within each type, agents must be homogeneous. } and weak interaction assumption, MFG is not an effective approximation. 
Furthermore, since agents have no identity  in MFG, one can choose a representative agent (the agent index is thus omitted) and  write the formulation\footnote{MFG is more commonly formulated in a continuous-time setting in the domain of optimal control, where it is typically composed by a backward \emph{Hamilton-Jacobi-Bellman equation} (e.g., the Bellman equation in RL is its discrete-time counterpart) that describes the optimal control problem of an individual agent and a forward \emph{Fokker-Planck equation} that describes the dynamics of the aggregate distribution (i.e., the mean field) of the population.  }
 of the MFG as
\begin{align}  V\Big(s, \pi,\big\{L_{t}\big\}_{t=0}^{\infty}\Big)&:=\mathbb{E}\left[\sum_{t=0}^{\infty} \gamma^{t} R\left(s_{t}, a_{t}, L_{t}\right) \Big| s_{0}=s\right] \nonumber  \\ \text { subject to }  s_{t+1}& \sim P\left(s_{t}, a_{t}, L_{t}\right), a_{t} \sim \pi_{t}\left(s_{t}\right). 
\label{eq:mfg_obj}
\end{align}
Each agent applies a local policy\footnote{A  general non-local policy  $\pi(s, L): \mathbb{S} \times \Delta^{|\mathbb{S}||\mathbb{A}|} \rightarrow \Delta(\mathbb{A})$ is also valid for MFG, and it makes the learning easier by assuming $L$ is fully observable.} $\pi_t: \mathbb{S} \rightarrow \Delta(\mathbb{A})$, which assumes the population state is not observable. 
Note that both the reward function and the transition dynamics depend on the sequence of the mean-field terms $\{L_{t}\}_{t=0}^{\infty}$. 
From each agent's perspective, the MDP is time-varying and is determined by  all other  agents.

 The solution concept in MFG is a variant of the (Markov perfect) NE named the mean-field equilibrium, which is a pair of $\{\pi_t^*, L_t^*\}_{t\ge 0}$ that satisfies two conditions: 1) for fixed $L^*=\{L^*_t\}$, $\pi^*=\{\pi_t^*\}$ is the optimal policy, that is, $V(s, \pi^*, L^*) \ge V(s, \pi, L^*), \forall \pi, s$; 2) $L^*$ matches with the generated mean field when agents follow $\pi^*$.  
The two-step iteration process in the meeting start-time example applied in MFG is then expressed as $\Gamma^1(L_t) = \pi^*_t$ and $\Gamma^2(L_t, \pi^*_t) = L_{t+1}$, and it terminates when $\Gamma^2\circ \Gamma^1(L) = L=L^*$. 
Mean-field equilibrium is essentially a fixed point of MFG, its existence for discrete-time\footnote{The existence of equilibrium in continuous-time MFGs is widely studied in the area of stochastic control  \cite{huang2006large,lasry2007mean,carmona2013probabilistic,carmona2015probabilistic,carmona2016mean, cardaliaguet2015master,fischer2017connection,lacker2018convergence,lacker2015mean}, though it may be of less interest to RL researchers. } discounted MFGs has been verified by \cite{saldi2018markov} in the infinite-population limit $N \to +\infty$ and also in the partially observable setting \cite{saldi2019approximate}. 
However, these works consider the case where the mean field in MFG includes only the population state.  
Recently, \cite{guo2019learning} demonstrated the existence of NE in MFG, taking into account both the population states and actions distributions. In addition, they proved that if $\Gamma^1$ and $\Gamma^2$ meet \emph{small parameter conditions} \cite{huang2006large}, then the NE is unique in the sense of $L^*$.   
In terms of uniqueness, a common result is based on assuming monotonic cost functions \cite{lasry2007mean}. In general, MFGs admit multiple equilibria \cite{nutz2020convergence}; the reachability of multiple equilibria is studied when the cost functions are anti-monotonic \cite{cecchin2019convergence} or quadratic \cite{delarue2020selection}.   

Based on the two-step fixed-point iteration  in MFGs, 
various model-free RL algorithms have been proposed for learning the NE.  
The idea is that  in the step $\Gamma^1$, one can approximate the optimal $\pi_t$ given $L_t$ through single-agent RL algorithms\footnote{Since agents in MFG are homogeneous, if the representative agent reaches convergence, then the joint policy is the NE. Additionally, given $L_t$, the MDP to the representative agent is stationary.} such as (deep) Q-learning \cite{guo2019learning,anahtarci2019fitted,anahtarci2020q}, (deep) policy-gradient methods \cite{guo2020general,elie2020convergence,subramanian2019reinforcement,uz2020approximate}, and actor-critic methods \cite{yang2019provably,fu2019actor}. Then, in step $\Gamma^2$, one can compute the forward $L_{t+1}$ by sampling the new $\pi_t$ directly or via fictitious play \cite{cardaliaguet2017learning,hadikhanloo2019finite,elie2019approximate}. 
A surprisingly good result is that the sample complexity of both value-based and policy-based learning methods for  MFG  shares the same order of magnitude as those of single-agent RL algorithms \cite{guo2020general}.
However, one major subtlety of these learning algorithms for MFGs is how to obtain stable samples for $L_{t+1}$. 
For example, \cite{guo2020general} discovered that applying a softmax policy for each agent and projecting the mean-field quantity on an $\epsilon$-net with finite cover help to significantly stabilize the forward propagation of $L_{t+1}$.   

\section{Cooperative Mean-Field Control}

MFC maintains the same homogeneity assumption and weak interaction assumption as MFG. However,   
unlike MFG, in which each agent behaves independently, there is a central controller that coordinates all agents' behaviours in the context of MFC. 
In cooperative multiagent learning, assuming each agent observes only a local state, the central controller maximises the aggregated accumulative reward: 
\begin{equation}
\sup_{\boldsymbol{\pi}} \dfrac{1}{N} \sum_{i=1}^N \mathbb{E}_{\mathbf{s}_{t+1}\sim P, \mathbf{a}_t \sim \boldsymbol{\pi}} \bigg[ \sum_t \gamma^t R^i(\mathbf{s}_t, \boldsymbol{a}_t) \Big| \mathbf{s}_0 = \mathbf{s} \bigg].  
\label{eq:cumu_marl}	
\end{equation}
Solving Eq. (\ref{eq:cumu_marl}) is a combinatorial problem. Clearly, the sample complexity of applying the Q-learning algorithm grows exponentially in $N$ \cite{even2003learning}. 
To avoid the curse of dimensionality in $N$, MFC \cite{carmona2018probabilistic, gu2019dynamic} pushes $N \rightarrow +\infty$, and under the law of large numbers and the theory of propagation of chaos \cite{gartner1988mckean, mckean1967propagation,sznitman1991topics}, the optimisation problem in Eq. (\ref{eq:cumu_marl}), in the view of a representative agent, can be equivalently written as 
 \begin{align}
\sup_{\pi} \  \mathbb{E} & \bigg[ \sum_t \gamma^t \tilde{R}(s_t, a_t, \mu_t, \alpha_t) \Big| s_0 \sim \mu \bigg] \nonumber \\ 
 \text { subject  to } &   s_{t+1} \sim P\left(s_{t}, a_{t}, \mu_{t}, \alpha_t \right), a_{t} \sim \pi_{t}\left(s_{t}, \mu_t \right). 
\label{eq:mfc_obj}	
\end{align}
in which $(\mu_t, \alpha_t)$ is the respective state and action marginal distribution of the mean-field quantity, $\mu_{t}(\cdot)=\lim_{N \rightarrow +\infty}\sum_{i=1}^{N} {\mathds{1}(s_{t}^i=\cdot)}{/N}$, $\alpha_{t}(\cdot)=\sum_{s \in \mathbb{S}} \mu_{t}(s) \cdot  \pi_{t}\left(s, \mu_{t}\right)(\cdot)$,  and $\tilde{R} = \lim_{N \rightarrow +\infty} \sum_iR^i/N.$
The MFC approach is attractive not only because the dimension of MFC is independent of $N$, but also because MFC has shown to approximate the original cooperative game in terms of both game values and optimal strategies \cite{lacker2017limit, motte2019mean}.

Although the MFC formulation in Eq. (\ref{eq:mfc_obj}) appears similar to the MFG formulation in Eq. (\ref{eq:mfg_obj}), their underlying physical meaning is fundamentally different. 
As is illustrated in Figure \ref{fig:mfs}, the difference is which operation is performed first: learning the equilibrium of the $N$-player game or taking the limit as $N  \rightarrow +\infty$. 
In the fixed-point iteration of MFG, one first assumes $L_t$ is given and then lets the (infinite) number of agents find the best response to $L_t$, while in MFC, one assumes an infinite number of agents to avoid the curse of dimensionality in cooperative MARL and then finds the optimal policy for each agent from a central controller perspective.
 In addition, compared to mean-field NE in MFG, 
the solution concept of the central controller in MFC is the Pareto optimum\footnote{The Pareto optimum is a subset of NE.}, an equilibrium point where no individual can be better off without making others worse off. It is worth noting that when there are multiple equilibria and the agent may have different preferences between them, learning to play Pareto optimal strict Nash equilibrium is helpful \cite{wang2003learning}. Finally, other differences between MFG and MFC can be found in \cite{carmona2013control}.

In MFC, since the marginal distribution of states serves as an input in the agent's policy and is no longer assumed to be known in each iteration (in contrast to MFG),  the dynamic programming principle no longer holds in MFC due to its non-Markovian nature \cite{andersson2011maximum, buckdahn2011general, carmona2015forward}. That is, MFC problems are inherently time-inconsistent.  
A counter-example of the failure of standard Q-learning in MFC can be found in \cite{gu2019dynamic}.  
 One solution is to learn MFC  by adding common noise to the underlying dynamics such that all existing theory on
 learning MDP with stochastic dynamics can be applied, such as Q-learning \cite{carmona2019model}.  
 In the special class of linear-quadratic MFCs, \cite{carmona2019linear} studied the policy-gradient method and its  convergence, and \cite{luo2019natural} explored an  actor-critic  algorithm.
 However, this approach of adding common noise still suffers from high sample complexity and weak empirical performance \cite{gu2019dynamic}. Importantly, applying  dynamic programming in this setting  lacks rigorous verifications, leaving aside the measurability issues and the existence of a stationary optimal policy. 
 
 Another way to address the time inconsistency in MFCs is to consider an  \textbf{enlarged} state-action space  \cite{lauriere2014dynamic,pham2016discrete,pham2018bellman,pham2017dynamic,djete2019mckean,gu2019dynamic}.
 This technique is also called ``lift up'', which essentially means to lift up the state space and the action space into their corresponding probability measure spaces in which dynamic programming principles hold.   
For example, \cite{gu2019dynamic, motte2019mean} proposed to lift the finite state-action space $\mathbb{S}$ and $\mathbb{A}$ to a compact state-action space embedded in Euclidean space denoted by $\mathcal{C}:= \Delta(\mathbb{S}) \times \mathcal{H}$ and $\mathcal{H}:=\big\{h: \mathbb{S} \rightarrow \Delta({\mathbb{A}})\big\}$, and the optimal Q-function associated with the MFC problem in Eq. (\ref{eq:mfc_obj}) is\par 
{\small \begin{equation}
 Q_{\mathcal{C}}(\mu, h)=\sup _{\pi} \  \mathbb{E}\left[\sum_{t=0}^{\infty} \gamma^{t} \tilde{R}\big(s_{t}, a_{t}, \mu_{t}, \alpha_t \big) \Big| s_{0} \sim \mu, u_{0} \sim \alpha, a_{t} \sim \pi_{t}\right]	, \forall (\mu, h) \sim \mathcal{C}. 
 \end{equation}}
 The physical meaning of $\mathcal{H}$ is the set of all possible local policies $h: \mathbb{S} \rightarrow \Delta(\mathbb{A})$ over all different states. 
 Note that after lift up, the mean-field term $\mu_t$  in $\pi_t$ of Eq. (\ref{eq:mfc_obj}) no longer exists as an input to $h$. 
 Although the support of each $h$ is $|\Delta(\mathbb{A})|^{|\mathbb{S}|}$, it proves to be the minimum space under which the Bellman equation can hold. 
The Bellman equation for $Q_{\mathcal{C}}: C \rightarrow \mathbb{R}$ is 
 \begin{align}
 Q_{\mathcal{C}}(\mu, h)=R\big(\mu, h\big)+\gamma \sup _{\tilde{h} \in \mathcal{H}} Q_{\mathcal{C}}\Big(\Phi(\mu, h), \tilde{h}\Big)
 \end{align}
where $R$ and $\Phi$ are the reward function and transition dynamics written as
\begin{align}
 R\big(\mu, h\big) &=\sum_{s \in \mathbb{S}} \sum_{a \in \mathcal{A}} \tilde{R}\big(s, a, \mu, \alpha(\mu, h)\big) \cdot  \mu(s) \cdot h(s)(a) \\ \Phi \big(\mu, h\big) &=\sum_{s \in \mathbb{S}} \sum_{a \in \mathbb{A}} P\big(s, a, \mu, \alpha(\mu, h)\big) \cdot \mu(s) \cdot h(s)(a) 
\end{align}
with $\alpha(\mu, h)(\cdot):=\sum_{s \in \mathbb{S}}  \mu(s) \cdot h(s)(\cdot)$ representing the marginal distribution of  the mean-field quantity in action. 
 The optimal value function is  $V^*(\mu) = \max_{h \in \mathcal{H}} Q_{\mathcal{C}}\big(\mu, h\big)$. 
Since both $\mu$ and $h$ are probability distributions, the difficulty of learning MFC then changes to how to deal with continuous state and continuous action inputs to $Q_{\mathcal{C}}(\mu, h)$, which is still an open research question. 
\cite{gu2020q} tried to discretise the lifted space $\mathcal{C}$ through $\epsilon$-net and then adopted the kernel regression on top of the discretisation; impressively, the sample complexity of the induced Q-learning algorithm is independent of the number of agents $N$. 

\section{Mean-Field MARL}
\label{sec:mf-marl-related}
The scalability issue of multiagent learning in non-cooperative general-sum games can also be alleviated by applying the mean-field approximation directly to each agent's Q-function \cite{zhou2019factorized,yang2018mean,subramanian2020multi}.  
In fact, \cite{yang2018mean} was the first to combine mean-field theory with the MARL algorithm. 
The idea  is to first factorise the Q-function using only the local pairwise  interactions between agents (see Eq. (\ref{eq:mf-pairh})) and then apply the mean-field approximation; specifically, one can write the neighbouring agent's action $a^k$ as the sum of the mean action $\bar{a}^j$ and a fluctuation term $\delta{a^{j,k}}$, i.e., $a^{k}=\bar{a}^{j}+\delta a^{j, k}, \  \bar{a}^{j}=\frac{1}{N^{j}} \sum_{k} a^{k}$,  
in which $\mathcal{N}(j)$ is the set of neighbouring agents of the learning agent $j$ with its size being $N^j= |\mathcal{N}^j|$. 
With the above two processes,  we can reach the  mean-field Q-function  $Q^j(s, a^j, \bar{a}^j)$ that approximates $Q^j(s, \mathbf{a})$ as follows \par
{\small \begin{align}
&Q^j\big(s, \mathbf{a}\big) = \frac{1}{N^j} \sum_{k} Q^j\big(s, a^j, a^k\big) \label{eq:mf-pairh} \\
&= \frac{1}{N^j} \sum_{k} \bigg[ Q^j\big(s, a^j, \bar{a}^j\big) + \nabla_{\bar{a}^j} Q^j\big(s, a^j, \bar{a}^j\big) \cdot \delta{a^{j,k}} \nonumber \\ 
& \qquad \qquad \qquad + \dfrac{1}{2}\,\delta{a^{j,k}} \cdot \nabla^2_{\tilde{a}^{j,k}} Q^j\big(s, a^j, \tilde{a}^{j,k}\big) \cdot \delta{a^{j,k}} \bigg] \\
&= Q^j\big(s, a^j, \bar{a}^j\big) + \nabla_{\bar{a}^j} Q^j\big(s, a^j, \bar{a}^j\big) \cdot \bigg[ \dfrac{1}{N^j} \sum_{k} \delta{a^{j,k}} \bigg] \nonumber \\ 
& \qquad \qquad \qquad + \dfrac{1}{2N^j} \sum_{k} \bigg[ \delta{a^{j,k}} \cdot \nabla^2_{\tilde{a}^{j,k}} Q^j\big(s, a^j, \tilde{a}^{j,k}\big) \cdot \delta{a^{j,k}} \bigg] \label{mean-field-first-1}  \\
&= Q^j\big(s, a^j, \bar{a}^j\big) + \dfrac{1}{2N^j} \sum_{k} R^j_{s, a^j}\big(a^k\big) \notag \\ & \approx Q^j\big(s, a^j, \bar{a}^j\big)~. \label{mean-field-final}
\end{align}}
The second term in Eq. (\ref{mean-field-first-1}) is zero by definition, and the third term can be bounded if the Q-function is smooth, and it is neglected on purpose. 
The mean-field action $\bar{a}^{j}$ can be interpreted as the empirical distribution of the actions taken by agent $j$'s neighbours. However, unlike the mean-field quantity in MFG or MFC, this quantity does not have to assume an infinite population of agents, which is more friendly for many real-world tasks, although a large $N$ can reduce the approximation error between $a^k$ and $\bar{a}^j$ due to the law of large numbers.  
In addition, the mean-field term in MF-MARL does not include the state distribution, unlike MFG or MFC. 

Based on the mean-field Q-function, one can write the Q-learning update as\par 
{\small\begin{align}
Q_{t+1}^{j}\big(s, a^{j}, \bar{a}^{j}\big)=\big(1-\alpha \big) Q_{t}^{j}\big(s, a^{j}, \bar{a}^{j}\big)+\alpha \Big[R^{j}+\gamma v_{t}^{j, \text{MF}}\big(s^{\prime}\big)\Big] \nonumber \\
v_{t}^{j, \text{MF}}\big(s^{\prime}\big)=\sum_{a^{j}} \pi_{t}^{j}\big(a^{j} \mid s^{\prime}, \bar{a}^{j}\big) \cdot \mathbb{E}_{\bar{a}^{j}\left(\boldsymbol{a}^{-j}\right) \sim \boldsymbol{\pi}_{t}^{-j}}\Big[Q_{t}^{j}\big(s^{\prime}, a^{j}, \bar{a}^{j}\big)\Big]. 
\label{eq:mfq-upate}
\end{align}}
The mean action $\bar{a}^j$ depends on $a^j, j\in \mathcal{N}(j)$, which itself depends on the mean action. 
The chicken-and-egg problem is essentially the time inconsistency that also occurs in MFC. 
To avoid coupling between $a^j$ and $\bar{a}^j$, \cite{yang2018mean} proposed a filtration such that in each stage game $\{\boldsymbol{Q}_t\}$, the mean action $\bar{a}^j$ is computed first using each agents' current policies, i.e., $ \bar{a}^{j}=\frac{1}{N^{j}} \sum_{k} a^{k}, a^{k} \sim \pi_{t}^{k}$, and then given  $\bar{a}^j$,  each agent finds the best response by \par
{\small\begin{equation}
\pi_{t}^{j}\left(a^{j} \mid s, \bar{a}^{j}\right)=\dfrac{\exp \left(\beta Q_{t}^{j}\left(s, a^{j}, \bar{a}^{j}\right)\right)}{\sum_{a^{j} \in \mathbb{A}^{j}} \exp \left(\beta Q_{t}^{j}\left(s, a^{j^{\prime}}, \bar{a}^{j}\right)\right)}.
\label{eq:boltz_policy} 	
\end{equation}}
For large $\beta$, the Boltzmann policy in Eq. (\ref{eq:boltz_policy}) proves to be a contraction mapping, which means the optimal action $a^j$ is unique given $\bar{a}^j$; therefore, the chicken-and-egg problem is resolved\footnote{Coincidentally, the techniques of fixing the mean-field term first and adopting the Boltzmann policy for each agent were discovered by \cite{guo2019learning} in learning MFGs at the same time.}.

MF-Q can be regarded as a modification of the Nash-Q learning algorithm \cite{hu2003nash}, with the solution concept changed from NE to mean-field NE (see the definition in MFG). 
As a result, under the same conditions, which include the strong assumption that there exists a unique NE at every stage game encountered, $\mathbf{H}^{\mathrm{MF}} \boldsymbol{Q}(s, \boldsymbol{a})=\mathbb{E}_{s^{\prime} \sim p}\left[\boldsymbol{R}(s, \boldsymbol{a})+\gamma \boldsymbol{v}^{\mathrm{MF}}\left(s^{\prime}\right)\right]$ proves to be a contraction operator. Furthermore, the asymptotic convergence of the MF-Q learning update in Eq. (\ref{eq:mfq-upate}) has also been established.

Considering only pairwise interactions in MF-Q may appear rather limited. However, 
it has been noted that  the pairwise approximation of the agent and its neighbours, while significantly reducing the complexity of the interactions among agents, can still preserve global interactions between any pair of agents   \cite{blume1993statistical}. In fact, such an approach is widely adopted in other machine learning domains, for example, factorisation machines \cite{rendle2010factorization} and learning to rank \cite{cao2007learning}. 
Based on MF-Q, \cite{li2019efficient} solved the real-world taxi order dispatching task for Uber China and demonstrated strong empirical performance against humans.  
\cite{subramanian2019reinforcement} extended MF-Q to include multiple types of agents and applied the method to a large-scale predator-prey simulation scenario. 
\cite{ganapathi2020partially} further relaxed the assumption that agents have access to exact cumulative metrics regarding the mean-field behaviour of the system, and proposed partially observable MF-Q that maintains a distribution to model the uncertainty regarding the mean field of the system.

\chapter{Future Directions}
\label{chapter:conclusion}

\paragraph{MARL Theory.} 

In contrast to the remarkable empirical success of MARL methods, developing theoretical understandings of MARL techniques are very much under-explored in the literature. Although many early works have been conducted on understanding the convergence property and the finite-sample bound of single-agent RL algorithms \cite{bertsekas1996neuro}, extending those results into multiagent, even many-agent, settings seem to be non-trivial.  
Furthermore, it has become a common practice nowadays to use DNNs to represent value functions in RL and multiagent RL. In fact, many recent remarkable successes of multiagent RL benefit from the success of deep learning techniques \cite{vinyals2019alphastargrandmaster,openai-hide-and-seek,pachocki2018openai}. Therefore, there are pressing needs to develop theories that could explain and offer insights into the effectiveness of deep MARL methods. Overall, we believe there is an approximate ten-year gap between the theoretical developments of single-agent RL and multiagent RL algorithms. 
	Learning the lessons from single-agent RL theories and extending them into multiagent settings, especially understanding the incurred difficulty due to  involving multiple agents, and then generalising the theoretical results to include DNNs could probably act as a practical road map in developing MARL theories. Along this thread, we recommend the work of  \cite{zhang2019multi} for a comprehensive summary of existing MARL algorithms that come with theoretical convergence guarantee.

\paragraph{Safe and Robust MARL.}	

Although RL provides a general framework for optimal decision making, it has to incorporate certain types of constraints when RL models are truly to be deployed in the real-world environment. We believe it is critical to firstly  account for MARL with  robustness and  safety constraints; one direct example is on autonomous driving. 
 At a very high level, robustness refers to the property that an algorithm can generalise and maintain robust performance in settings that are different from the training environment \cite{morimoto2005robust,abdullah2019wasserstein}. And safety refers to the property that an algorithm can only act in a pre-defined safety zone with minimum times of violations even during training time \cite{garcia2015comprehensive}. 
In fact, the community is still at the early stage of developing theoretical frameworks to encompass either robust or safe constraint in single-agent settings. 
In the multiagent setting, the problem could only become more challenging because the solution now requires to take into account the coupling effect between agents, especially those agents that have conflict interests 
\cite{li2019robust}.  In addition to opponents, one should also consider robustness towards the uncertainty of environmental dynamics \cite{zhang2020robust}, which in turn will change the behaviours of opponents and pose a more significant challenge.

\paragraph{Model-Based  MARL.} 
 
Most of the algorithms we have introduced in this monograph are \emph{model-free}, in the sense that the RL agent does not need to know how the environment works and it can learn how to behave optimally through purely interacting with the environment.  In the classic control domain, \emph{model-based} approaches have been extensively studied in which the learning agent will first build an explicit state-space ``model''  to understand how the environment works in terms of state-transition dynamics and reward function, and then learn from the ``model''. The benefit of model-based algorithms lies in the fact that they often require much fewer data samples from the environment \cite{deisenroth2011pilco}. The MARL community has initially come up with model-based approaches, for example the famous R-MAX algorithm \cite{brafman2002r}, nearly two decades ago. Surprisingly, the developments along the model-based thread halted ever since. Given the impressive results that model-based approaches have demonstrated on single-agent RL tasks \cite{schrittwieser2020mastering,hafner2019dream,hafner2019learning}, model-based MARL approaches deserves more attention from the community. 

\paragraph{Multiagent Meta-RL.} 

Throughout this monograph, we have introduced many MARL applications;  each task  needs a bespoke MARL model to solve.  A natural question to ask  is whether we can use one model that can generalise across multiple tasks. For example, \cite{terry2020pettingzoo} has put together almost one hundred MARL tasks, including Atari, robotics,  and various kinds of board games and pokers into a Gym API. An ambitious goal is to develop algorithms that can solve all of the tasks  in one or a few shots. This requires multiagent meta-RL techniques.   

Meta-learning aims to train a generalised model on a variety of learning tasks, such that it can solve new learning tasks with few or without additional training samples. In the context of equilibrium finding and learning to play games, a promising direction was explored by \cite{harris2022meta}, who introduced meta-learning for game-theoretic settings. They provided the first meta-learning guarantees for fundamental classes of games, such as two-player zero-sum games, general-sum games, and Stackelberg games. Their framework leverages natural notions of similarity between sequences of games, achieving faster convergence rates to equilibria compared to solving games in isolation, while recovering single-game guarantees for arbitrary sequences. Notably, their experiments with poker endgames demonstrated significant efficiency gains, often by an order of magnitude, when applying meta-learning techniques to games with varying stack sizes.
Additionally, \cite{anagnostides2024convergence} extended the meta-learning framework to the analysis of no-regret learning algorithms, such as optimistic gradient descent (OGD), in dynamic game environments. Their results highlighted how meta-learning could refine variation-dependent regret bounds and enable sharper convergence guarantees for time-varying zero-sum and general-sum games. This work underscores the potential of meta-learning to adapt to dynamically changing games, providing novel insights into learning equilibria efficiently in settings where game parameters evolve over time.

Fortunately, \cite{finn2017model} has proposed  a general meta-learning framework -- MAML -- that is compatible with any model trained with gradient-descent based methods. Although MAML works well on supervised learning tasks, developing  meta-RL algorithms seems to be highly non-trivial \cite{rothfuss2018promp}, and introducing the meta-learning framework on top of MARL is even  an uncharted territory. 
We expect multiagent meta-RL to be a challenging yet fruitful research topic, since  
making a group of agents master multiple games necessarily requires  agents to automatically discover their identities and roles when playing different games; this itself is a hot research idea. Besides, the meta-learner in the outer loop would need to figure out how to compute the gradients with respect to the entire inner-loop subroutine, which must be a MARL algorithm such as multiagent policy gradient method or mean-field Q-learning, and, this would probably lead to exciting enhancements to the existing meta-learning framework. 

\paragraph{Foundation Models for MARL.}
Foundation models for decision making, especially within the scope of multiagent reinforcement learning, symbolize the confluence of foundational models and sequential decision-making concepts\cite{yang2023foundation}. Foundation models, typically pretrained on vast datasets through self-supervised learning, exhibit remarkable knowledge transfer capabilities to an array of downstream tasks. These models are being increasingly employed for tasks requiring sophisticated control, long-term reasoning, search, and planning - all essential components of sequential decision-making processes. However, unique challenges arise when these models interact with external entities or agents, such as learning from external feedback, adapting to unique modalities, and performing long-term reasoning and planning.

Sequential decision-making, including reinforcement learning, imitation learning, planning, search, and optimal control, has typically relied on task-specific or tabula rasa environments with limited prior knowledge. While these methods have achieved commendable feats, they often grapple with issues related to generalization and sample efficiency due to their start-from-scratch approach, which omits broad knowledge from vision, language, or other datasets.

Recently, a growing interest has been seen in merging the principles of foundation models and sequential decision making. This shift is fueled by the recognition that broad datasets can also greatly benefit sequential decision making models, similar to how they've benefitted foundation models. Large-scale vision and language models' success has inspired the creation of bigger datasets for learning multi-model, multi-task, and generalist interactive agents. Simultaneously, foundation models have started branching into complex problems involving long-term reasoning or multiple interactions. This mutually beneficial integration has immense potential to leverage foundational models' world knowledge to enhance sequential decision-making processes, ultimately leading to faster task-solving and better generalization. The burgeoning research area holds promising solutions for existing challenges and continues to open up new avenues for exploration and innovation.

\backmatter  

\bibliography{main} 

\end{document}